\documentclass[twocolumn]{openjournal}

\usepackage{xcolor}
\usepackage{textgreek}
\usepackage[utf8]{inputenc}
\usepackage[english]{babel}
\usepackage{threeparttable}
\usepackage{hyperref}
\hypersetup{
    unicode, 
    colorlinks=true,
    linkcolor=linkcolor,
    citecolor=linkcolor,
    filecolor=linkcolor,
    urlcolor=linkcolor,
}
\usepackage{color,colortbl}
\definecolor{linkcolor}{rgb}{0.0,0.3,0.5}
\usepackage{tensind}
\tensordelimiter{?}
\DeclareGraphicsExtensions{.bmp,.png,.jpg,.pdf}
\usepackage{verbatim}
\usepackage[normalem]{ulem}
\usepackage{orcidlink}
\usepackage{soul}
\usepackage{amsmath}

\usepackage{graphicx}
\usepackage{subcaption}

\graphicspath{ {./figs/} }

\begin{document}
\journalinfo{The Open Journal of Astrophysics}
\title{Wide Area VISTA Extragalactic Survey (WAVES): redshift-limited selection of targets for the Wide survey using decision-tree classification}

\author{G.~Kaur\altaffilmark{1,*}}

\author{M.~Bilicki\altaffilmark{1,*}}

\author{S.~Bellstedt\altaffilmark{2}}

\author{E.~Tempel\altaffilmark{3}}

\author{W.A.~Hellwing\altaffilmark{1}}

\author{I.~Baldry\altaffilmark{4}}

\author{B.~Bandi\altaffilmark{5}}

\author{S.~Barsanti\altaffilmark{6,7}}

\author{S.~Driver\altaffilmark{2}}

\author{N.~Guerra-Varas\altaffilmark{8}}

\author{B.~Holwerda\altaffilmark{9}}

\author{C.~Lagos\altaffilmark{2}}

\author{J.~Loveday\altaffilmark{5}}

\author{A.~Robotham\altaffilmark{2}}

\altaffiltext{1}{Center for Theoretical Physics, Polish Academy of Sciences, Al. Lotników 32/46, 02-668, Warsaw, Poland}
\altaffiltext{2}{ICRAR, The University of Western Australia, 35 Stirling Highway, Crawley, WA 6009, Australia}
\altaffiltext{3}{Tartu Observatory, University of Tartu, Observatooriumi 1, 61602 Tõravere, Estonia}
\altaffiltext{4}{Astrophysics Research Institute, Liverpool John Moores University, IC2, Liverpool Science Park, 146 Brownlow Hill, Liverpool, L3 5RF, UK}
\altaffiltext{5}{Astronomy Centre, University of Sussex, Falmer, Brighton BN1 9QH, UK}
\altaffiltext{6}{Sydney Institute for Astronomy, School of Physics, University of Sydney, NSW 2006, Australia}
\altaffiltext{7}{Research School of Astronomy and Astrophysics, Australian National University, Canberra, ACT 2611, Australia}
\altaffiltext{8}{European Southern Observatory, Karl-Schwarzschild-Str. 2, 85748 Garching near Munich, Germany}
\altaffiltext{9}{Physics \& Astronomy Department, University of Louisville, Louisville, KY 40292, USA}
\altaffiltext{*}{email: gursharanjitk@cft.edu.pl, bilicki@cft.edu.pl}

\begin{abstract}

The Wide-Area VISTA Extragalactic Survey (WAVES) on the 4-metre Multi-Object Spectroscopic Telescope (4MOST) includes two flux-limited subsurveys with very high (95\%) completeness requirements: Wide over $\sim\!1200$ deg$^2$ and Deep over $\sim\!65$ deg$^2$. Both are $Z$-band selected, and additionally redshift-limited, while the true redshifts are not known a priori but will only be measured by 4MOST.
Here, we present a classification-based method to select the targets for WAVES-Wide. Rather than estimating individual redshifts for the input photometric galaxies, we assign probabilities of them being below $z=0.2$, the redshift limit of the subsurvey.
This is done with the supervised machine learning approach of eXtreme Gradient Boosting (XGB), trained and tested on a comprehensive spectroscopic sample overlapping with WAVES fields and using nine-band  VST+VISTA photometry for the feature space.
We check the performance of our classifier both for the fiducial WAVES-Wide limits, as well as for a range of neighboring redshift and magnitude thresholds,  finding purity and completeness at the level of 90-95\% depending on the test set.  The metrics deteriorate for sources close to the selection limits and for test sets selected from DESI, not used in training. We apply the classifier trained on the full spectroscopic sample to the 8.6 million galaxies from the WAVES photometric catalog. This gives us about 1.2 million target candidates, reducing to 1 million if we apply an XGB probability threshold $P_\mathrm{XGB}(z\leq0.2) > 0.72$. Our work demonstrates that a machine-learning classifier could be used to select a flux- and redshift-limited sample from deep photometric data. 

\end{abstract}


\maketitle

\section{Introduction}
\label{intro}

\setcounter{footnote}{0}

The Wide-Area VISTA Extragalactic Survey \citep[WAVES,][]{WAVES} is one of the observational campaigns to be undertaken on the  4-metre Multi-Object Spectroscopic Telescope \citep[4MOST,][]{4MOST} -- a major endeavor of the European Southern Observatory (ESO) designed to spectroscopically map the southern sky, which began operations in 2026. WAVES aims to study the growth of structure, mass, and energy on scales from 1 kpc to 10 Mpc over the last 7 Gyr. At the largest scales (1-10 Mpc), it will examine structures like groups, filaments, and voids, and their recent development, comparing results with numerical simulations to evaluate the Cold Dark Matter paradigm. At intermediate scales (10 kpc - 1 Mpc), WAVES will assess sizes, mass distribution, and merger rates of galaxy groups to understand dark matter halo and stellar mass assembly. At the smallest scales (1-10 kpc), it will provide precise distance and environmental data to complement high-resolution space imaging for studying the evolution of galaxy bulges, discs, and bars. Overall, WAVES will create a comprehensive spectroscopic dataset of 1.6 million galaxies, connecting the low- ($z < 0.2$) and intermediate- ($z < 0.8$) redshift Universe.

To achieve its goals, WAVES is designed in two main subsurveys: Wide and Deep. WAVES-Wide (WW) aims to obtain spectra of a highly complete ($>90\%$) sample of galaxies, limited by the $Z$-band magnitude $Z\lesssim21.1$ and redshift $z\lesssim0.2$, over most ($\sim1200$ deg$^2$) of the joint area of the Kilo-Degree Survey \citep[KiDS,][]{KiDS} and  VISTA Kilo-Degree Infrared
Galaxy Survey \citep[VIKING,][]{VIKING}. WAVES-Deep (WD) is planned to reach $Z\lesssim21.25$ mag and $z\lesssim0.8$ over a $\sim65$ deg$^2$ area composed of the Galaxy And Mass Assembly (GAMA) G23 field and four `Deep Drilling Fields' (DDF) of the Legacy Survey of Space and Time (LSST) at the Rubin Observatory \citep{2019ApJ...873..111I}.

In both  Wide and Deep cases, the WAVES target selection is flux- and redshift-limited. The former condition is straightforward to achieve, as fluxes are readily available from internal WAVES processing of the KiDS+VIKING (KV) data (Bellstedt et al. submitted). The challenge lies in assigning redshift limits in surveys, which by design aim at measuring the redshifts themselves. To accomplish this, WAVES uses multiband KV photometry as a starting point.

The selection of WAVES targets has two stages. First, extragalactic sources are separated from stars. Then the selection of the redshift-limited galaxy samples follows. The star-galaxy separation was originally performed in a similar way as for the GAMA fields \citep{Bellstedt20}, i.e.\ by locating the objects on color/magnitude/size planes. For WAVES purposes, this was originally replaced by an unsupervised machine-learning (ML) approach described by \cite{Cook24}. For the final target catalogs, however, this is updated by using flags from the most recent runs of photo-z estimation codes (Bellstedt et al. submitted; Suelves et al. in prep.). In this work, we will assume that we are given a photometric catalog of galaxies and we will not be dealing with possible star contamination or missing galaxies.

The second stage of WAVES selection is to choose the photometric galaxy targets for spectroscopic observations within specific redshift ranges. This is typically based on multi-band magnitudes and colors \citep[e.g.][]{SDSS-III, VIPERS, DESI}, which is especially efficient if specific galaxy types, such as luminous red galaxies, are required \citep[e.g.][]{SDSS-IV-LRG}. The scientific goals of WAVES require, however, all possible galaxies to be observed, without any additional color- or galaxy-type pre-selections, as long as they are within the assigned magnitude and redshift limits. What is more, the science goals of the survey put stringent requirements of over 90\% spectroscopic completeness to be achieved \citep{WAVES}. Especially for WW, this would be very difficult to meet with color selection, unless one allows for many observed targets to be beyond the $z\simeq0.2$ limit. Due to limited fibre-hour allocation, such contamination cannot be excessive without affecting both WAVES and the entire 4MOST survey efficiency. We note, however, that, as WAVES is targeting galaxies, the ‘false positives' will still give extragalactic spectra, although at higher redshifts than required.

Multi-color information could be gathered into the prediction of photometric redshift \citep[photo-z, e.g.][]{NG22}, which could then be used for more efficient redshift-dependent target selection. In 4MOST, this is employed for one of the WAVES subsurveys, Optical, Radio Continuum and HI Deep Spectroscopic Survey \citep[ORCHIDSS,][]{2023Msngr.190...25D} in addition to radio selection. Another 4MOST community survey,  CHileAN Cluster galaxy Evolution Survey \citep[CHANCES,][]{2023Msngr.190...31H}, makes use of photo-z based selections for its low-z survey. In WW, photo-zs are estimated by various approaches for target selection: Gaussian processes for photometric redshifts \citep[GPz,][]{2022MNRAS.512.3662D}, scaled flux matching \citep[SFM,][]{2021MNRAS.500.1557B}, and Tartu Observatory Photo-zs \citep[TOPz,][]{TOPz, TOPz_2025}, the results of which are then combined, together with our method, into the final target probability (Suelves et al. in prep.). 

However, even the best photo-z methodologies using broad bands, as we have here, do not give sufficient precision to meet the requirements of WW in terms of completeness while preserving purity. Indeed, state-of-the-art results employing KV magnitudes by \cite{Bilicki21} give $\sigma_z\sim0.018(1+z)$ at a median $z\sim0.23$ albeit at brighter magnitudes ($r<20$) than in WW. Deeper flux-limited samples would likely have even larger photo-z scatter \citep[e.g.][]{Porredon21}. This could be improved by employing deep learning using galaxy images for photo-z predictions \citep[e.g.][]{Treyer24, Anjitha25}, but the overall scatter is unlikely to be much smaller than $\sigma_z\sim 0.015(1+z)$. At the WW limiting redshift of $z=0.2$, this amounts to some 10\% relative error, which could be crudely translated to $\sim90\%$ completeness and purity if such photo-zs are employed for target selection.

In this work, we take a different approach towards assigning WW targets. As the problem reduces to answering the question of whether a source is above or below the redshift threshold, this is basically a classification problem. It could be regarded as binary classification, or just one-class assignment, as what is relevant is to correctly assign a galaxy to the class where $z<z_\mathrm{lim}$. Our method to select the WW targets therefore classifies photometric galaxies as either meeting the redshift threshold criterion or not. We solve this classification problem by using an empirical approach of supervised ML. A similar idea was applied in \cite{Wu2022}, where convolutional neural networks were employed on DESI Legacy images to identify $z<0.03$ galaxies, and more recently in \cite{Payerne2024}, where a random forest (RF) was used to select high-redshift Lyman-Break Galaxies from current and future broadband wide photometric surveys. 

Our goal is to assign class “0" to the desired WW targets (below the chosen $Z$-magnitude and redshift thresholds) and class “1" otherwise. For that purpose, we use a tree-based algorithm, which is fast and not very demanding computationally, while at the same time well-suited to such binary classification problems. Moreover, our approach allows us to estimate the output probability of an object being assigned to a given class. Working with a flux-limited sample, we build a supervised ML classifier that is trained on galaxies having both the KV photometry and redshift labels from external spectroscopy. The classifier indirectly learns the color-redshift relation, but rather than providing point redshift estimates, it gives the probability that a given galaxy lies below the desired redshift cut.

Tree-based methods are abundantly used for classification and regression tasks in a similar context. Some examples include \cite{Gerdes2010}, who designed a photo-z algorithm based on boosted decision trees; \cite{2013MNRAS.432.1483C}, who employed RF to estimate photo-zs and their probability density functions; \cite{Zhou2021}, where RF was used to estimate photo-zs for the DESI Legacy galaxy sample; \cite{2024A&A...685A.127S}, where the reddest high-redshift galaxies in the Euclid Deep Fields were identified with gradient-boosted trees; or already mentioned \cite{Payerne2024}. 
Here we employ the XGBoost algorithm \citep{chen2016xgboost}  trained on data with 9-band photometric magnitudes together with the corresponding 36 colors (45 features in total) and true (spectroscopic) redshift labels that are turned into two classes (targets or non-targets). Our choice follows after testing various other ML approaches, such as RF, k-nearest neighbors (kNN), or support vector machines. Similarly, as with some other methods, e.g., kNN, an advantage of tree-based algorithms is their ability to provide not only a discrete class assignment but also its probability. This could be folded into the target selection, for instance, by giving a higher chance of observation to those galaxies where such probabilities are the largest \citep{Tempel2020a}. Another reason for the choice of a tree classifier was the need to get target probability assignments for different selection criteria, as well as short training time, making it computationally efficient. 

XGBoost we employ is a supervised classifier. A general drawback of such methods is their dependence on the training data with true labels. If the training data are incomplete, then in the areas of the parameter space which they do not cover well, the classification might be unreliable. This does affect us as well, and we will discuss the implications below, in relation to the probabilities assigned by the final classifier. In particular, if a too permissive cut on classification probability is applied, the number of resulting target candidates becomes too large to be practically observed. This is likely driven by particular contamination from higher-redshift galaxies whose colors are similar to those of true targets, but their color-redshift relation is not well mapped in the training set. Such contaminants can, however, be removed by raising the classification probability threshold, which leads to lowering the proposed target number.

The XGBoost-based classification pipeline we present here has been adopted by WAVES for the selection of WW targets within a joint framework employing also other approaches \citep{2021MNRAS.500.1557B,2022MNRAS.512.3662D, TOPz, TOPz_2025}. The details will be presented in a separate paper by Suelves et al. (in prep.), while here we focus only on our particular method without comparing it with the others.

In Sec .~\ref {data}, we describe the WAVES photometric data along with the spectroscopic compilation, which is used to train the XGBoost classifier. In Sec.~\ref{Sec:problem}, we outline the challenges associated with target selection for WAVES Wide. The gradient tree-based target selection pipeline and the classifier training are presented in Sec.~\ref{sec:method}. The performance of the classification pipeline is shown in Sec.~\ref{sec:performance}. In Sec.~\ref{sec:inference}, we apply the trained XGBoost classifier to label all the galaxies in the WAVES photometric catalog and assign target probabilities.  
Finally, Sec.~\ref {sec:conclusions} provides a summary and conclusions.

\section{Data}\label{data}
\subsection{WAVES photometric data}
\label{sec:photo}

Our feature space is built from 9-band optical+near-infrared (NIR) magnitudes available in the WAVES photometric catalog. While for the final (inference) dataset we use only the WW coverage, for the training phase we supplement that with photometric measurements in the WD fields. This is necessary because most of the key deep spectroscopic surveys required for our calibration are located only in the Deep fields, while the Wide ones are covered mostly by brighter-end samples. 

The WW catalog is based on ESO VST KiDS plus ESO VISTA VIKING public surveys covering the $ugriZYJHK_s$ bandpasses. 
The input imaging is basically the same as was used to produce the KiDS final Data Release 5 \citep{2024A&A...686A.170W}. However, the image processing and source detection for WAVES are separate from KiDS. The general framework and tools are the same as those used in GAMA \citep{Bellstedt20}, with some modifications to masking and fragmenting procedures. Detection images, deeper than those used in \cite{Bellstedt20}, are created through an $r+i+Z+Y$ stack, and then passed through the \texttt{ProFound} package \citep{2011MNRAS.416.2640R} for source detection and characterization. 
A `defragmenting' process is then conducted mostly automatically (with manual inspection and intervention when required) to piece segments back together for sources that have been fragmented. \texttt{ProFound} is then run in a `measure' mode to extract photometry for each source in all 9 bands separately. Artefacts in the image are screened out based on a series of logical arguments based on non-physical colors and sizes. More details are presented in Bellstedt et al. (submitted). 

The magnitudes (fluxes) in WAVES photometric catalogs come in two flavors: `total' and `color', the latter derived using the detection segment, whereas the total photometry is generated using a dilated segment to capture the full flux of the galaxy. Total photometry is more appropriate for e.g., source selection, whereas the color fluxes have slightly lower noise and hence are usually better for tasks like photometric redshifting. Similarly, as was the case for the other WW target selection approaches mentioned in Sec.~\ref{intro}, we tested both photometry versions and found that the `color' option gives consistently improved results for our classification, by about 1 percent point in all the metrics; we will therefore be using it henceforth. Avoiding `total' photometry makes our approach also immune to possible issues with flux extrapolation that could be present for such measurements.

The WD fields consist of four individual pointings, each spanning 4 square degrees within the VRO LSST DDF Regions and the G23 field. These four fields -- WD01, 02, 03, and 10 -- correspond to ELAIS-S1, XMMLSS, CDFS, and COSMOS, respectively. The data in these fields are more complex than in WW, and sourced from a diverse range of public and private survey programs conducted with VST and VISTA, including KiDS, KIDZ, WAVES-VST, VOICE \citep{2016heas.confE..26V}, VIDEO \citep{2013MNRAS.428.1281J}, ultraVISTA \citep{2012A&A...544A.156M}, and WAVES-VISTA (Bellstedt et al. submitted).

Our ML model requires all the $u$ to $K_s$ bands to be measured; as we use both magnitudes and colors, one missing band affects 8 further features. Therefore, in the photometric data, we keep only those objects that have all nine fluxes above 0. Missing data affects only a small fraction of the sample; in the entire input catalog limited to $Z<21.1$ mag (9,439,027  galaxies), about $1.5\%$ of the sources do not have flux reported in at least one of the 9 bands.  
We note that we do not differentiate between non-detections and non-observations, i.e.\ we ignore any galaxy that does not have positive flux in all 9 bands, whatever the cause. The sources with missing bands are, however, handled by other WAVES target selection approaches; see Suelves et al. (in prep.).

We also note that, although we use the `color' magnitudes, the flux limits themselves are imposed according to the `total' photometry. 
Moreover, we use only objects flagged as galaxies according to the approach described in Bellstedt et al. (submitted).  As a result, we are left with a dataset containing $8,600,759$ galaxies with 9-band detections, for the $Z<21.1$ mag limit. This will be our inference sample to which we will apply the trained ML classifier. The inference set is referred to as the \textit{photo-cat} hereafter.

\subsection{Spectroscopic data with true labels}
\label{sec:spectro}

Our ML model detailed below in Sec.~\ref{sec:method} is a supervised one, which means that it requires labels with ground truth to be trained and validated on. These labels -- class “0" for galaxies below the desired redshift threshold, and class “1" otherwise, both for the same fixed magnitude limit -- are derived from spectroscopic samples cross-matched with WAVES imaging. These samples cover both the main KiDS/VIKING/WW area and the WAVES-DDFs, where appropriate optical and NIR photometry is available. The input spectroscopic surveys and datasets are the same as those discussed in \cite{Jalan2024} and \cite{2024A&A...686A.170W}, and we join them into one parent catalog in the same way as described in \cite{Jalan2024}. The procedure includes, in particular, appropriate handling of objects present in more than one source spectroscopic dataset as well as redshift quality assignment. In what follows, we use only galaxies with quality flag $\mathrm{NQ}\geq 3$ to ensure reliable redshifts, where NQ is a redshift quality flag similar to that employed by GAMA, homogenized based on information from the input surveys (see \citealt{2024A&A...686A.170W} for details). The resulting imaging dataset cross-matched with spectroscopic labels is hereafter referred to as \textit{spec-z-cat}. At the fiducial $Z$-band magnitude limit of 21.1, the median redshift for the spec-z-cat 
is 0.24, while the median $Z$-band magnitude is 18.6. Key information on the constituent spectroscopic surveys for this magnitude limit is presented in Table \ref{tab:survey}, while the details of the selection criteria in the particular datasets and how they were combined are provided in sec.~5 of \cite{2024A&A...686A.170W} and Table~1 of \cite{Jalan2024}.

\begin{figure}
    \centering
    \includegraphics[width=\columnwidth]{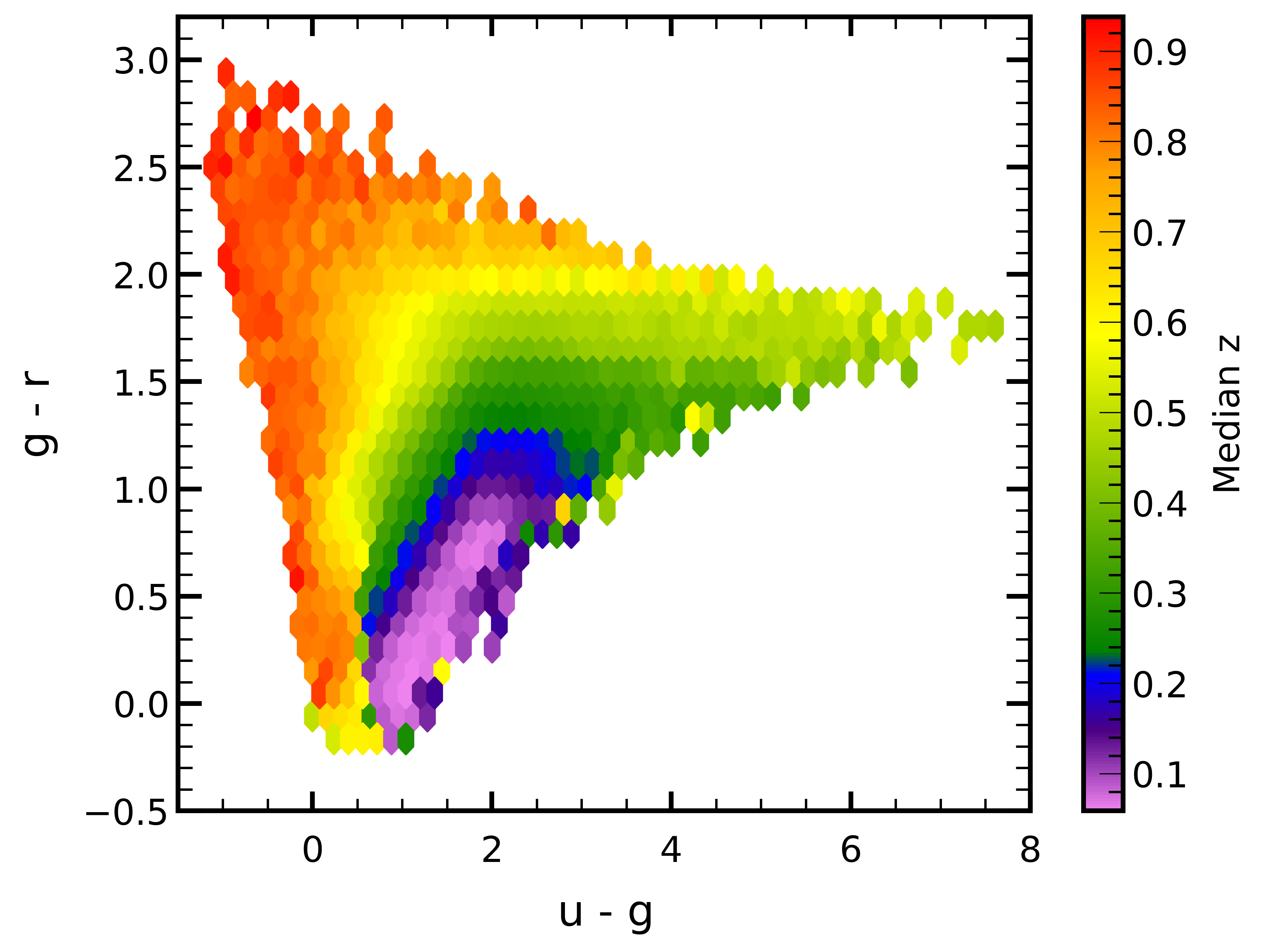}
    \caption{Distribution of galaxies from the spectroscopic catalog that we use to calibrate our classifier, on the $\text{g - r}$ vs. $\text{u - g}$ color plane. Each pixel is colored according to its median redshift. We show only the pixels with a minimum of 10 galaxies each.}
    \label{fig:color_color_training}
\end{figure}

\begin{table*}
\centering
    \caption{Input surveys included in the spectroscopic compilation that we use as the training and test sets.}
    \begin{tabular}{lrrrl}
    \hline\hline
    Survey    & Median redshift  & Median $Z$-mag  & Count  & References \\ 
\hline
    
    2dFGRS     &   0.113 &    17.26 &   82552  & \cite{Colless2001}\\
    GAMA       &   0.217 &    18.47 &  228163  & \cite{Baldry2010,Liske2015}  \\
    2dFLenS    &   0.297 &    18.30 &   35704  & \cite{Blake2016}\\
    HCOSMOS    &   0.305 &    19.45 &    1008  & \cite{Damjanov2018} \\
    G15-DEEP   &   0.325 &    20.26 &    1479  & \cite{gamaDriver2022} \\
    C3R2       &   0.363 &    20.62 &     165  & \cite{Masters2017,c3trMasters2019} \\
    DEVILS     &   0.372 &    20.07 &    9212  & \cite{devilsDavies2018} \\
    DESI EDR      &   0.382 &    19.71 &   84915  & \cite{desi2024} \\
    G10-COSMOS &   0.391 &    20.42 &    7971  &  \cite{g10cosmosDavies2015}\\
   VVDS       &   0.440 &    20.51 &    1213  & \cite{LeFevre2005,vvdsLeFevre2013} \\
    SDSS DR14       &   0.441 &    18.65 &   85631  & \cite{sdssabolfathi2018} \\
    OzDES      &   0.467 &    19.78 &     5460 & \cite{ozdesLidman2020} \\
    DEIMOS      &   0.517 &    20.37 &     642 & \cite{Hasinger_2018} \\
    GOODS      &   0.521 &    20.41 &     178  & \cite{goodsPopesso2009,Balestra2010} \\
    ACES       &   0.521 &    20.39 &     1384 & \cite{acesCooper2012} \\
    WIGGLEZ    &   0.583 &    20.49 &    21027 & \cite{wigglezDrinkwater2010} \\
    zCOSMOS    &   0.610 &    20.13 &     553  & priv. comm. (M. Salvato) \\
    VIPERS    &   0.629 &    20.62 &     4907  &  \cite{vipersScodeggio2018}\\
    LEGA-C    &   0.755 &    20.60 &     148   & \cite{lagcVanderwel2016} \\
    \hline
 full spec-z-cat & 0.242 & 18.60 & 572325 &  \\
 \hline
     full DESI DR1 (test only) & 0.296 & 19.07 & 868481 & \cite{DESI_DR1_2025arXiv250314745D}\\
     blind DESI DR1 (no spec-z-cat overlap) & 0.322 & 19.28 & 560984 & \\
    \hline\hline
\end{tabular}
\begin{tablenotes} 
\item The table is sorted in ascending median redshift of sources in the spec-z-cat. All the numbers apply after cross-matching the input spec-$z$ samples with the photometric data limited at $Z<21.1$ total mag and removing duplicates between surveys. We list only those surveys that have at least 100 cross-matched objects each. 
\end{tablenotes}
\label{tab:survey}

\end{table*}

We note that our spec-z-cat includes also objects at fainter magnitudes, and we use these to study our classifier's performance at various flux limits in Sec. \ref{change_cuts}. Figure \ref{fig:color_color_training} shows an example color-color distribution ($\text{g-r}$ vs. $\text{u-g}$) of the spec-z-cat mapped by the median spectroscopic redshift.  We see that low-redshift galaxies occupy a specific area on this diagram, although with some overlap with the high-$z$ ones, which is due to the complicated color-redshift relation. We choose to display this particular plane as these two colors are among the most important features for our classification (Sec.~\ref{sec:performance}).
We note, however, that the ML model uses the entire magnitude-color space and similar separation could be found also in other cross-sections. One further point to note regarding such diagrams is that their appearance is also driven by selection effects in the spectroscopic data that we have available; see below for more discussion.

The spectroscopic compilation described above, that we use to train our ML model, includes Dark Energy Spectroscopic Instrument Early Data Release \citep[DESI EDR,][]{desi2024} from the areas overlapping with WAVES, but not DESI Data Release 1 \citep[DR1,][]{DESI_DR1_2025arXiv250314745D}, which appeared in March 2025. This is because the final XGB predictions, which are included in the WW target selections as used by 4MOST, were made based on training without the latter dataset. Nevertheless, in this work we include DESI DR1 as an additional test set, used in two ways. First, we validate the results of our training on the full cross-match between DESI DR1 and the WAVES input catalog (`full DESI test'), which includes about 868k galaxies. However, due to considerable overlap between DESI and our spec-z-cat discussed above, we also design a fully `blind' DESI DR1 test set by removing from the DESI $\times$ WAVES photo-cat cross-match those galaxies which had already been included in the spec-z-cat. This gives almost 561k galaxies in such a ‘blind DESI test'. We note that this latter sample has very specific selections, as it does not include the bright-end galaxies from previous surveys such as 2dF, SDSS and GAMA. The details of both the full DESI DR1 test set and the blind one are provided in the bottom rows of Table~\ref{tab:survey}. For the purposes of this work, we only use DESI galaxies selected from DR1 with the following criteria: $\mathtt{ZCAT\_PRIMARY}==\mathrm{TRUE}$; $\mathtt{ZWARN}==0$; $\mathtt{SPECTYPE}==\mathrm{galaxy}$; and redshift $0.002<z<1.6$, overlapping with WAVES Wide and Deep fields.

To summarize, we will be evaluating the performance of our classifier on three test sets, of which two are fully independent of each other: one randomly selected from the spec-z-cat and unseen by the model in the training (`random test set'); another, including all the DESI DR1 galaxies overlapping with WAVES photo-cat which are not present in the spec-z-cat (`blind DESI test set'); and finally a `full DESI test set' which includes all DESI DR1 galaxies matched to the WAVES input photometric catalog in both Wide and Deep fields. The random test will be giving upper limits on the performance, as it is statistically identical to the training sample, but not so well matched to the inference sample, especially at the faint end. The DESI blind test, on the other hand, will likely be underestimating the metrics due to its very specific selections, which are different from a $Z$-band flux-limited selection.

\section{Challenges in selecting WAVES-Wide targets}
\label{Sec:problem}

One of the main challenges of supervised classification is the appropriate selection of training data. Typically, when working with real observational samples, the possibilities to obtain training sets that are sufficiently representative for the target (inference) catalog are limited. More common are situations when the former is a biased, often brighter and color-selected, subsample of the latter. This is indeed what we are facing here as well for the target selection in WW. Namely, at the planned WW photometric depth of $Z\sim21.1$ mag, galaxies with $z\lesssim0.2$ constitute a minority, which is not correctly reflected in the labeled data we have available.

\begin{figure}
    \centering
    \includegraphics[width=\columnwidth]{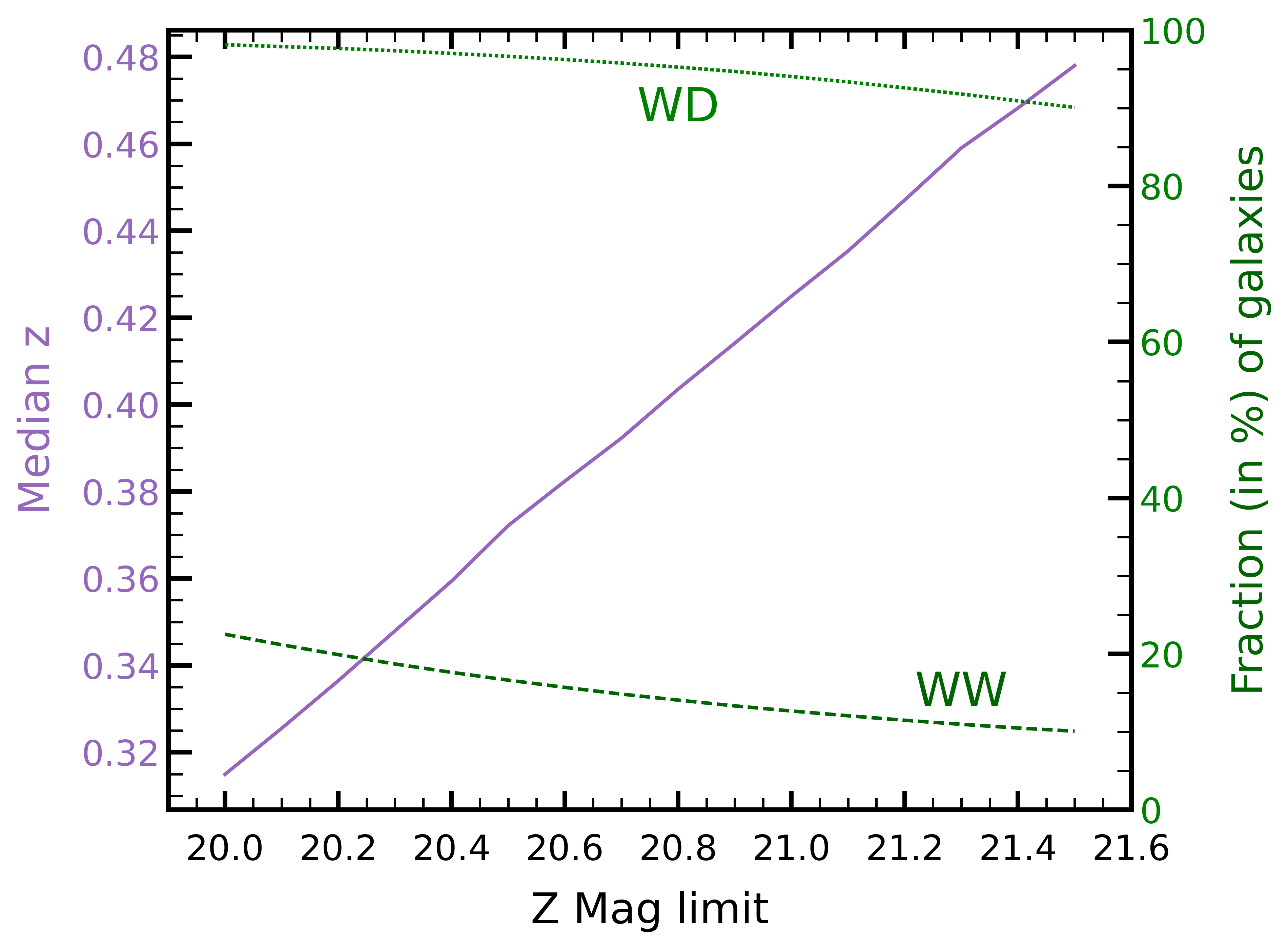}
    \caption{Dependence of the median redshift (left-hand axis, violet diagonal line) of galaxies selected as flux-limited up to the $Z$-band magnitude indicated on the x-axis, as derived from a SHARK/SURFS mock galaxy catalog. The right-hand axis and the green lines show what percentage of galaxies would constitute the WAVES targets for the $Z$-band flux limit as on the x-axis and the fiducial redshift limits: $z<0.2$ for WW (bottom green line) and $z<0.8$ for WD (top green line).}
    \label{fig:shark_selection}
\end{figure}

To quantify this effect, we start with a simulation dataset where we know both magnitudes and redshifts. For that, we employ a mock galaxy catalog built with the Shark semi-analytic model \citep{Shark} applied to the SURFS N-body simulations \citep{SURFS}, which was tailored to match the WAVES input data. The Shark semi-analytic model of galaxy formation generates self-consistent predictions from the far-ultraviolet to the far-infrared. Shark covers a wide range of baryon physics effects that are believed to play an important role in the formation and evolution of galaxies. For this paper, we use a lightcone generated from the default model presented in \cite{Shark} using the method described in \cite{Chauhan2019}.

\begin{figure*}
    \centering
    \includegraphics[width=0.7\textwidth]{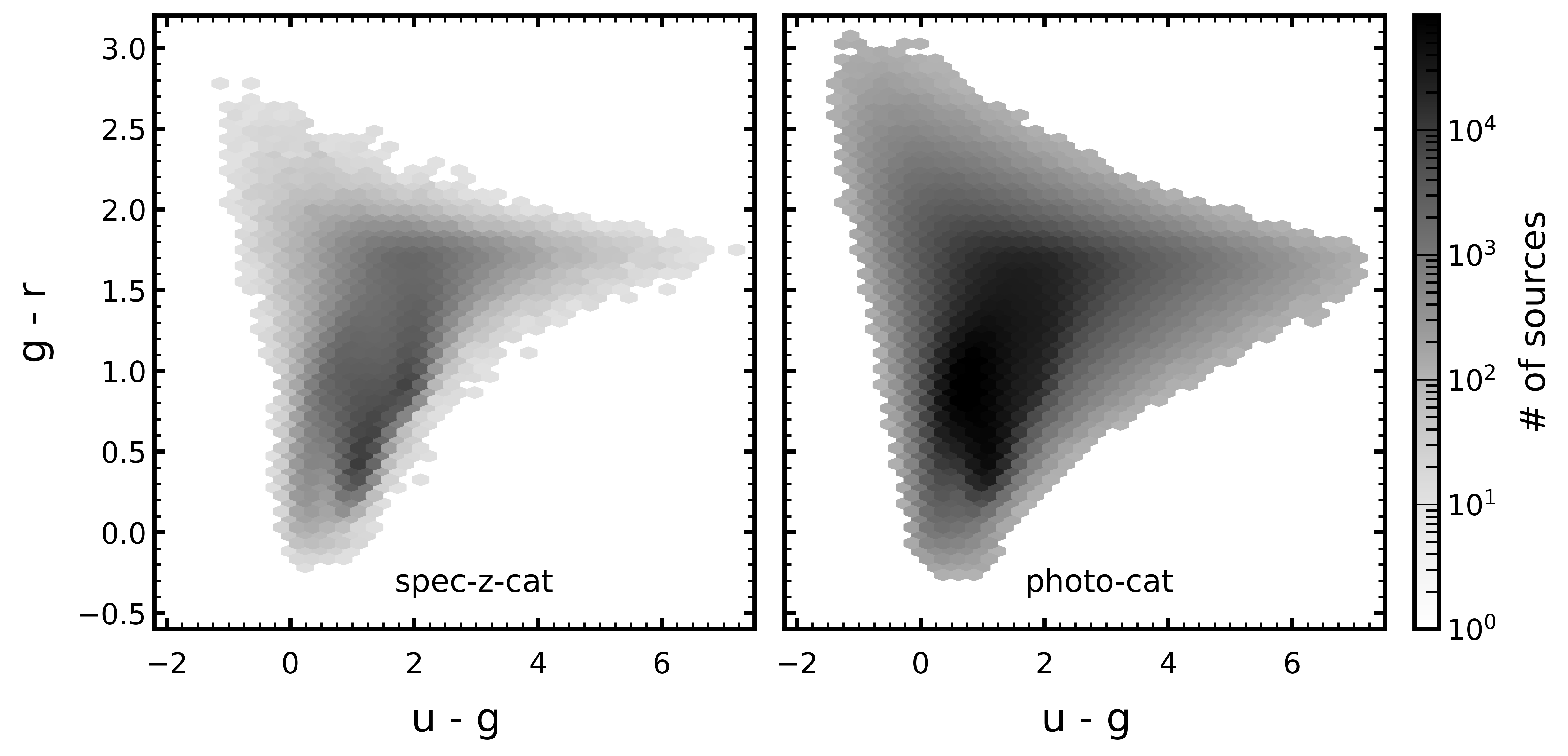}
    \caption{Number density of galaxies projected on the $\text{g - r}$ vs. $\text{u - g}$ color plane, for the spec-z-cat (left) that we use to calibrate our ML model, and for the entire WAVES photo-cat (right), from which WW targets are selected.  We show only the pixels with a minimum of 10 galaxies each for the spec-z-cat (left) and a minimum of 100 sources each for the photo-cat (right).}
    \label{fig:color_color_target_spec_numberdensity}
\end{figure*}

\cite{Lagos2019,Lagos2020} presented the FUV-to-FIR multi-wavelength predictions of Shark by combining the model with the generative SED model ProSpect \citep{prospect}, and using a novel method to compute dust attenuation parameters based on the radiative transfer analysis of the EAGLE hydrodynamical simulations of \cite{Trayford2020}. The model was shown to produce excellent agreement with the observed number counts from the FUV to the FIR \citep{Lagos2019}, and the redshift distribution of FIR-selected sources \citep{Lagos2020}. Here, we use a lightcone produced for WAVES following the same pipeline described in \cite{Lagos2019}.

In particular, our mock sample includes the 9-band flux information mimicking the KV measurements, as well as the “true" galaxy redshifts, although with a caveat that it is limited to $Z=21.5$ mag. We use these to generate Fig.~\ref{fig:shark_selection}, where we show two pieces of information. The violet diagonal line, corresponding to the left-hand y-axis, illustrates the median redshift of a galaxy sample selected as flux-limited to the $Z$-band magnitude provided on the x-axis. We can see that at $Z\sim21$ mag, this median $z$ is well above the WW target limit of $z\lesssim0.2$, which means that the majority of galaxies at this magnitude cut are at redshifts larger than our targets should have. Indeed, this is quantified with the lower green line, corresponding to the right-hand axis, and showing the percentage of $z<0.2$ galaxies brighter than the $Z$-band limit of the x-axis. Even for the brightest considered threshold of $Z<20$ mag, this ratio is close to 20\% and decreases for fainter flux limits. In contrast, as shown with the top green line, for WD with $z<0.8$ selection, more than 90\% of galaxies at the considered flux limits would be appropriate targets. Therefore, the task of target selection for WW is much more challenging than for WD. 

One additional caveat we need to highlight is that our mocks reach only to $z=1$, so in fact the violet diagonal one will be slightly shifted upwards in real data, while the green lines are overestimates. These should not be big effects, though, as the fraction of $z>1$ galaxies at the flux limits of interest is small.

The second challenge we are facing when designing the WW target selection is related to the limited completeness of currently available calibration data with true spectroscopic labels. This will be relevant for any methods of WW selection; even if ‘training' data are not needed (e.g. in template-fitting approaches), one still needs to verify if a given target assignment indeed yields correct results. In the best-case scenario, one would use a complete flux-limited galaxy subsample with true (spectroscopic) redshift labels reaching at least to the WAVES photometric limit on a sufficiently large sky area, to provide good statistics and minimize cosmic variance. As discussed in Sec.~\ref{sec:spectro}, at present such samples are not available. We have wide-angle datasets that are relatively deep but color-selected, or flux-limited but shallow; or very deep but small-area spectroscopic samples. Figure \ref{fig:color_color_target_spec_numberdensity} illustrates the resulting difference in coverage of the $\text{g-r}$ vs.\ $\text{u-g}$ color-color plane by the full photometric catalog and the spec-z-cat.

\begin{figure}
    \centering
    \includegraphics[width=0.95\columnwidth]{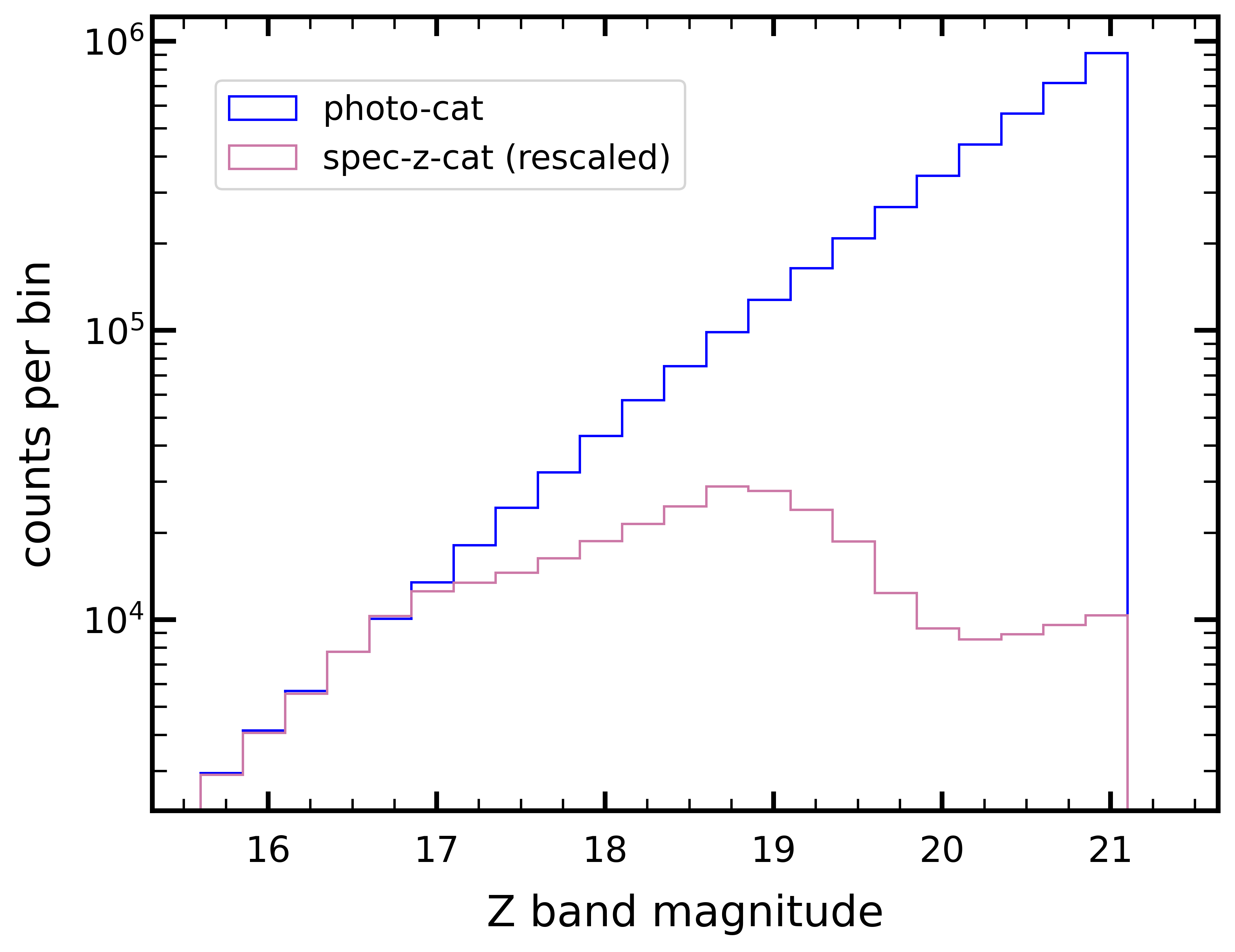}
    \caption{Comparison of $Z$-band number counts in the full WAVES photometric catalog (blue line) and the training set with spectroscopic redshift labels (purple line).  The scale on the y-axis gives actual counts per bin for the photo-cat, while for the spec-z-cat the counts have been rescaled to give a good match at the bright end.}
       \label{fig:zmag_speccat}
\end{figure}

Figure \ref{fig:zmag_speccat} shows the $Z$-magnitude distribution (number counts) of our spectroscopic calibration set (purple line), compared with that of the full parent WAVES photometric catalog (blue line). As the spec-z-cat has complicated geometry and we do not know its exact mask, we do not normalize the number counts by sky area; instead, we roughly match them at the bright end where both spec- and photo-data follow a power law. We see that, with such scaling, the spec-z-cat starts to depart from the photo-cat slope already at $Z\sim17$ mag, and for such normalization the amplitude of the former at the limit of WW is many times smaller than that of the photo-cat. This is driven by the fact that over the full WAVES-North area, flux-limited spectroscopic selection within our spec-z-cat was only available from the 2dFGRS and SDSS Main datasets. This has now changed with the DESI DR1 Bright Galaxy Sample of similar depth as GAMA, but here we describe the dataset that was used to calibrate WW target selection as submitted for 4MOST observations.

\begin{figure}
    \centering
    \includegraphics[width=\columnwidth]{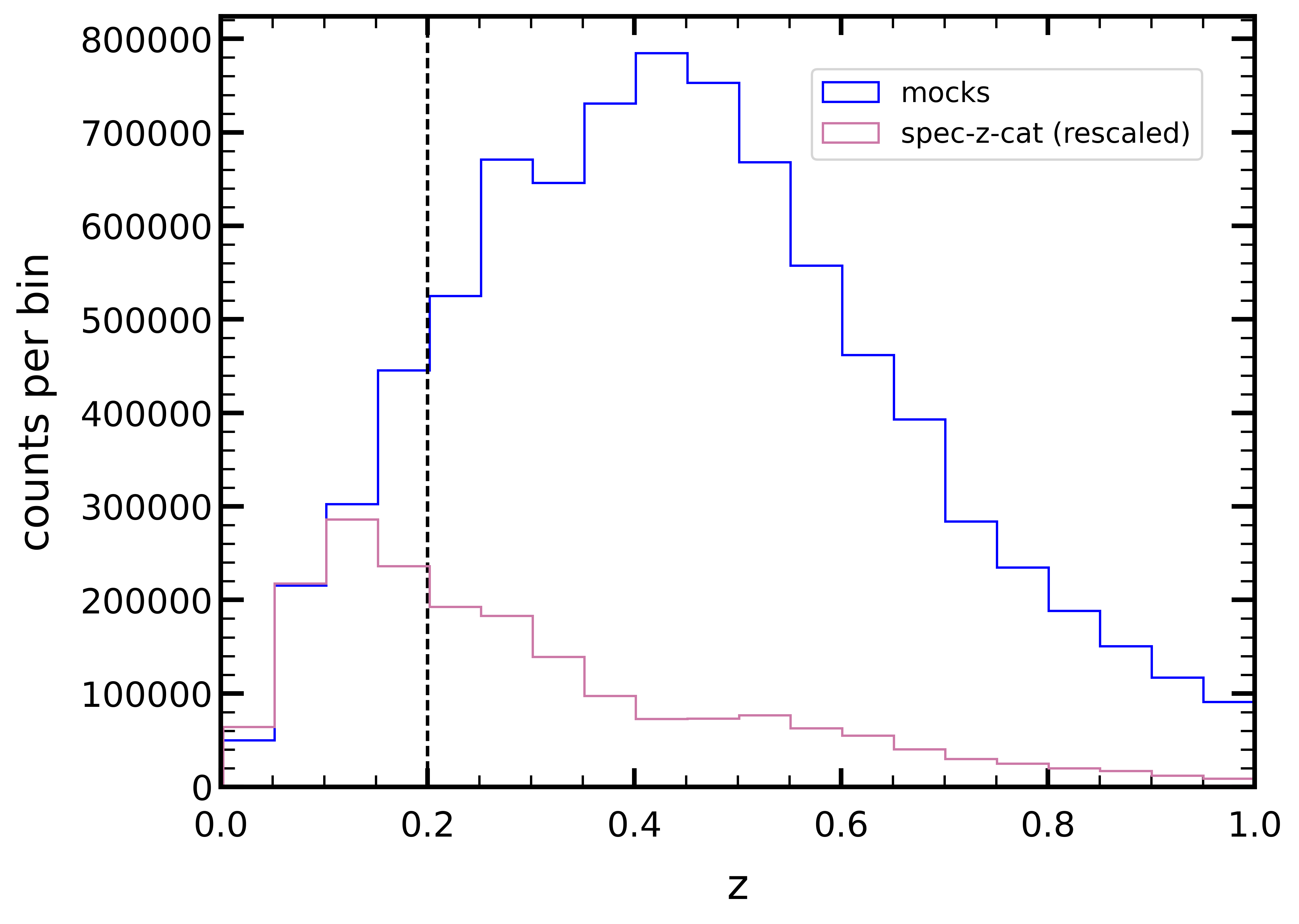}
    \caption{Comparison between redshift distributions of our spectroscopic training set (purple) and of the SHARK mock catalog (blue) for galaxies at the flux limit of $Z<21.1$ mag. The vertical line indicates the fiducial WW redshift limit for the targets. The redshift distribution of the spec-z-cat is rescaled to match the counts of the mock catalog at low redshift.}
    \label{fig:shark_z}
\end{figure}

\begin{figure}
    \centering
    \includegraphics[width=0.95\columnwidth]{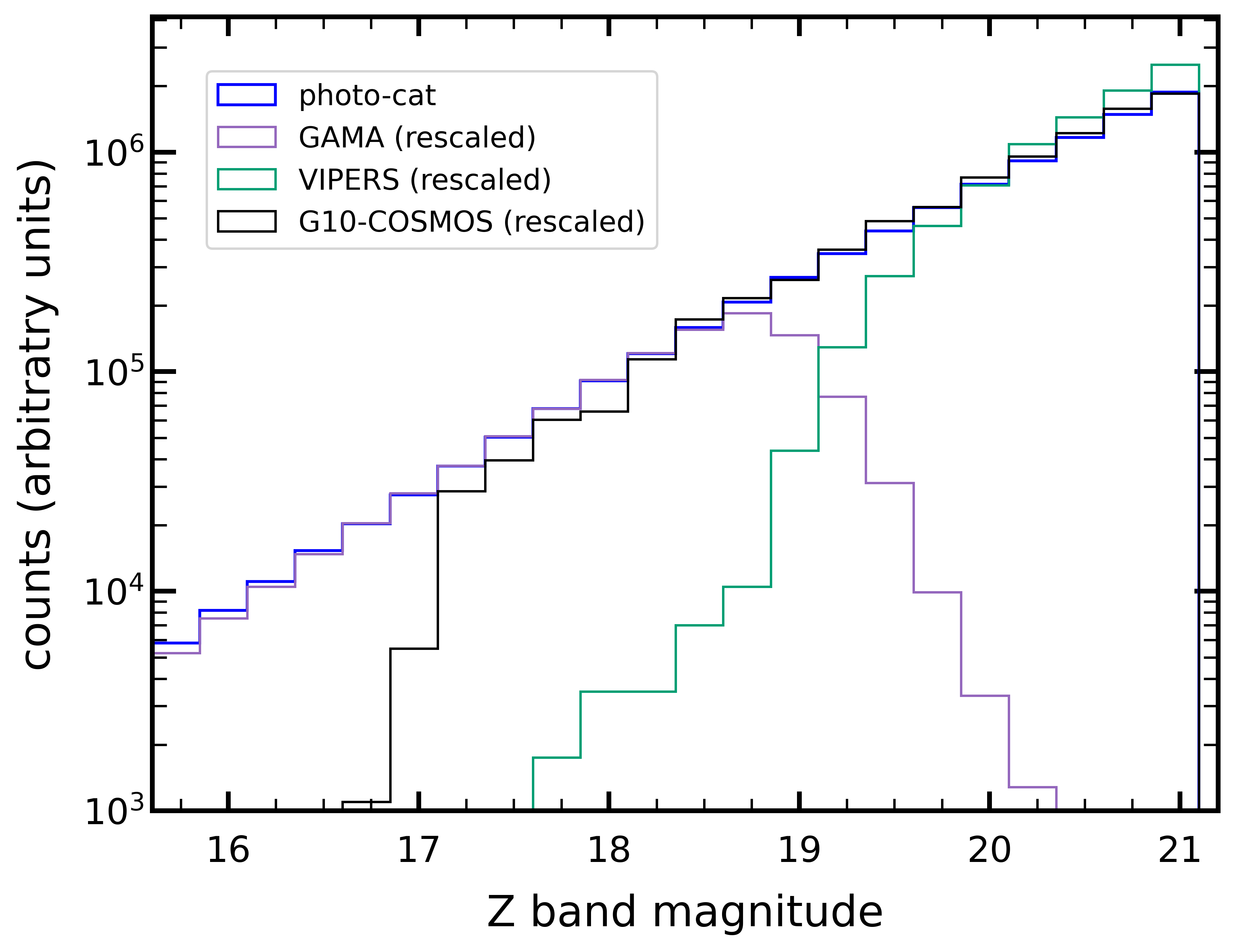}
    \caption{Comparison of $Z$-band number counts in the full WAVES photometric catalogue (blue line) with some of the spectroscopic datasets included in the spec-z-cat. The scale on the y-axis gives actual counts per bin for the photo-cat, while for the spec-z subsamples the counts have been rescaled to give a good match where the slope is the same as in the photo-cat.}
       \label{fig:zmag_surveys}
\end{figure}

In Figure \ref{fig:shark_z} we compare the redshift distribution of the spec-z-cat with that of the mock, as (obviously) we do not have true redshifts for the entire WAVES input catalog. Both the datasets are flux-limited to $Z<21.1$ mag. The normalization is again not by the sky area but roughly matched at low redshift, where both for mocks and the spec-z-cat we have 
$dN/dz \propto z^2$ approximately. Similarly to the magnitudes, also the redshift distribution of the spec-z-cat follows the `true' (mock) one only at the low-$z$ end and starts departing already at $z\sim0.15$. As mentioned in Sec.~\ref{sec:spectro}, the median redshift of the spec-z-cat for $Z<21.1$ mag is equal to 0.24, while, as shown in Fig.~\ref{fig:shark_selection}, a complete galaxy sample at this flux limit would have $\langle z \rangle \simeq 0.43$. As Figs.~\ref{fig:zmag_speccat} and \ref{fig:shark_z} indicate, this mismatch is driven by an underabundance of faint, higher-redshift sources in the spec-z-cat.

The fact that the spec-z-cat averaged over its full area does not constitute a flux-limited sample beyond $Z\sim 17$ mag does not mean that we do not have spectroscopic subsamples allowing us to train our ML classifier for fainter galaxies. On the contrary, several included surveys and datasets from the deep fields provide redshifts up to the WW flux limit, although some of them are over small sky areas and hence not very numerous. To illustrate that coverage,  in Fig.~\ref{fig:zmag_surveys} we show the $Z$-band number counts for four subsets, as compared to the photo-cat. As in Fig.~\ref{fig:zmag_speccat}, we again rescale the counts to match the photo-cat where the slopes are consistent. At the brightest end, we show GAMA (violet), which is highly complete to $Z=18.8$ mag (95\% completeness in the equatorial fields, \citealt{gamaDriver2022}).

At the faint end, we show in Fig.~\ref{fig:zmag_surveys} two example datasets: VIPERS in green and G10-COSMOS in black. The former, with which our photo-cat has overlap in the WD02 deep field, was color-selected to target $z>0.5$ galaxies \citep{VIPERS}, which in the $Z$-band counts translates roughly to selecting $Z>18$ mag with very good coverage at $Z\gtrsim 19.5$. G10-COSMOS \citep{g10cosmosDavies2015}, on the other hand, is one of the spectroscopic subsets in our compilation that lie in the WD10 (COSMOS) field. The counts shown in Fig.~\ref{fig:zmag_surveys} cover the range $Z\gtrsim 17$ mag, lacking the bright end only because it had already been included in the wide-angle surveys that also map WD10 (we show the counts after removing duplicates from, e.g., SDSS). Generally, COSMOS is the most comprehensively covered by spectroscopy well beyond the WAVES depth, and together with the other deep fields allows us to map the color-redshift relation beyond the flux limits of the wide-angle samples such as GAMA.

A more comprehensive analysis of how the spec-z-cat covers the color space of the WAVES photo-cat could be performed with tools such as self-organizing maps (SOM), in a similar way as in past works \citep[e.g.][]{Masters2015,Jalan2024}. In particular, \cite{Jalan2024} presented a way to quantify the (mis-)match between spectroscopic training and photometric inference sets by mapping them on the SOM, for the particular case of the KiDS-Bright and GAMA galaxy samples. Such a study would be, however, very challenging to undertake in our case, as our spec-z-cat is a conglomerate of numerous surveys and datasets of very different sky coverages and depths, which does not allow us to construct the sky masks to properly rescale the count for SOM cells per effective area. Nevertheless, in forthcoming work we will make an attempt to compare the color coverage of the photo-cat with current and future spectroscopic samples in particular sky areas, including the 4MOST Survey Programme Validation (SPV) data that were gathered in March 2026 and include many galaxies with $Z<21.1$ in the WAVES Wide and Deep areas.

The lower completeness of the spectroscopic training data at fainter, higher-redshift range, as compared to the photometric catalog, will have implications for our supervised learning approach. One particular effect is that the classifier would perform better if a brighter flux cut was adopted instead of the WW fiducial one. Nevertheless, our target selection performs satisfactorily overall, as also quantified in the companion paper by Suelves et al. (in prep.) where it is compared with the other WAVES methods. In the longer term, a possible mitigation is by providing more training examples populating different regions of the color space than what our current spec-z-cat offers, in a similar way as done in the Complete Calibration of the Color-Redshift Relation \citep{Masters2017,c3trMasters2019,4C3R2_2023,DC3R2_2024}.

\section{Methodology}
\label{sec:method}

In this Section, we describe how we selected the ML algorithm for the classification pipeline -- XGBoost -- as well as the algorithm and the pipeline itself.

\subsection{Decision Trees for classification}
\label{sub:decision_tree}

Decision trees are a class of supervised ML algorithms that have a hierarchical structure. Standard decision trees consider all input features and split the data into subsets based on these features. In the most commonly used CART \citep[Classification And Regression Trees,][]{breiman1984classification}, the split is decided by the Gini impurity, which calculates the probability of misclassifying a random data point into the wrong subset or class label. The Gini impurity is given by \citep{breiman1984classification}:
\begin{equation}
    \text{Gini Impurity} = 1 - \sum_{i}P_{i}^{2} \; .
\end{equation}
Here, $P_{i}$ is the probability of the data point belonging to class “$i$", the classes in our case being galaxies with $z \le z_\mathrm{lim}$ for class “0", or $z > z_\mathrm{lim}$ for class “1". The probability $P_{i}$ (and $1 - P_{i}$) is calculated as the fraction of galaxies with $z_\mathrm{true}\le z_\mathrm{lim}$ (and $z_\mathrm{true} > z_\mathrm{lim}$ respectively) within the subset generated after the split. For the true redshift label, we use the spectroscopic measurement from the calibration sample.

The data is then divided into two subsets (in the case of binary classification) based on this best split, and each subset forms a child node. This process of evaluating and splitting continues recursively, creating a tree-like structure with branches and nodes. The recursion terminates when a stopping criterion is met, such as reaching a predefined maximum depth, having fewer than a predefined minimum number of samples per node, achieving a decrease in impurity below a certain threshold, or when all the samples in a node belong to the same class for classification. 
In such a situation, the node is designated as a leaf node. In a classification tree, the leaf node assigns a class label, typically the majority class of the samples in that node. The probability of the classification for galaxies to be in a particular class in the leaf node is assigned as the fraction of galaxies falling within the class out of the total number of galaxies in the node. 

When looking for the most suitable approach for our binary classification, we tested several methods. Following early tests, we focused on two popular tree-based ML algorithms: Random Forest \citep[RF,][]{breiman2001random} and XGBoost \citep[XGB,][]{chen2016xgboost}. After a number of experiments, we decided to adopt the latter for several reasons. First, XGB did not require much hyperparameter tuning for our data, and it worked well already with default parameters. Second, the training time for XGB classifiers was significantly shorter than for the RF ones in our setup. Apart from supervised ML methods, we also tried the unsupervised approach of $k$-nearest neighbors \citep{fix1951discriminatory}. However, the results were not satisfactory. The reason could be that a classification like this is not “inherent" for the photometric feature space, unlike the star-galaxy separation.

The models were implemented using the \texttt{scikit-learn} library of Python \citep{2011JMLR...12.2825P}. For XGB, the \texttt{xgboost} library was applied through the scikit-learn API.

\subsection{XGBoost}

Like other supervised learning algorithms, XGB attempts to learn a function, $f(x)$, that maps input features $x$ to output labels $y$. The function $f$ is discrete for classification, and output labels are just $(0,1)$ for our binary case. XGB creates an ensemble of CART to minimize a loss function. The default loss function for XGB binary classification, which is also used in this work, is Log-Loss or binary cross-entropy. In the XGB library, this is specified as `binary:logistic' and is given by:

\begin{equation}
L(y, \hat{y}) = -\frac{1}{N}\sum_{i=1}^{N}\left[\,y_{i}\ln\hat{y}_{i} + (1-y_{i})\ln\left(1-\hat{y}_{i}\right)\right]
\end{equation}

where $y$ is the class label 0 or 1, and $\hat{y}$ is the probability of the output label being 0 or 1 assigned by the leaf node of the tree. XGB minimizes the loss function combined with regularization penalties  by constructing decision trees sequentially to minimize the gradient of the loss function. This process of adding new CART, or models in general, one after another, is termed as boosting. The regularization terms help prevent overfitting of the model to training data.  

In the RF algorithm, all the trees are created simultaneously instead of iteratively, and the majority vote of the 0 and 1 or yes and no decisions of all trees is the final output.  In  XGBoost, the leaf node holds a real number -- class score -- used to evaluate potential splits. The sum of scores of all trees for a galaxy having $z \leq z_\mathrm{lim}$, $S$, is then mapped into the probability of the classification at the leaf node via the sigmoid function:
\begin{equation}
P(z \leq z_\mathrm{lim}) = \frac{1}{1+\text{exp}(-S)} \; .
\end{equation}

If the classification probability is higher than a threshold probability, then the galaxy is assigned class 0, i.e.\ $z \leq z_\mathrm{lim}$. The default probability limit is 0.5, but in Sec.~\ref{sec:inference} we will also explore higher thresholds which reduce the number of galaxies assigned as targets. We will in particular find that relatively strict cuts are needed ($P>0.71$) if the size of the targeted sample in WW should not exceed the expected $\sim1$ million.

\subsection{Classification and metrics}
\label{Sec:metrics}

The model's performance in classifying galaxies according to their redshifts is evaluated using a two-class accuracy metric. However, since the survey specifically targets galaxies with redshifts less than $z_\mathrm{lim}$ (with the fiducial value $z_\mathrm{lim} = 0.2$), we mainly utilize purity and completeness metrics for class 0, corresponding to galaxies with redshifts below this threshold, as well as their harmonic mean, the F1 score, to evaluate the performance of the classifier. 

The two-class accuracy is defined as:
\begin{equation}
\label{eq:acc}
    \text{Accuracy} = \frac{(0,0)+(1,1)}{(0,0)+(0,1)+(1,0)+(1,1)}
\end{equation}
Here, $(a,b)$ refers to ground truth $a$, for which class $b$ is  
ascribed by the ML classifier.
Purity (also known as \textit{precision}), completeness (also \textit{recall}), and F1 score for class 0 are respectively given as:
\begin{equation}
\label{eq:pur}
\text{Purity, P}= \frac{(0,0)}{(0,0)+(1,0)} \, ,
\end{equation}
\begin{equation}
\label{eq:com}
    \text{Completeness, C}= \frac{(0,0)}{(0,0)+(0,1)} \, ,
\end{equation}
\begin{equation}
\label{eq:f1}
    \text{F1 score}= 2 \times \frac{\text{Purity} \times \text{Completeness}}{\text{Purity} +\text{Completeness}} \, .
\end{equation}
Our goal is to maximize these metrics, i.e.\ bring them as close to unity as possible. For the WAVES science goals, it is especially desirable to have Completeness above 0.95 \citep{WAVES}, but for the efficiency of the survey, also Purity should be kept at a high level.

\subsection{Training with the spec-z compilation}
\label{sec:training}

The XGB classification model is trained on the spectroscopically labeled dataset detailed in Sec.~\ref{sec:spectro}. As discussed in Sec.~\ref{sec:photo}, our input photometric catalog has 
fluxes available in 9 bands (4 from KiDS and 5 from VIKING). 
For the ML model, the input photometric information is provided in terms of features. For that, we used 9-band magnitudes corresponding to `color' fluxes as available in the database. 

In Sec.~\ref{sec:photo}, we also mentioned that for our analysis we only kept galaxies with the full 9-band information by eliminating objects with any missing bands. This is because,  although the XGBoost classifier has an implicit way of dealing with missing values, this built-in approach is 
not the best solution for imputing the missing fluxes in our case. XGBoost assigns missing values to a default direction (left or right) in each decision tree split.  The optimal direction is chosen by evaluating the loss function for each potential split. The reason that this sparsity-aware split-finding method is not suited for the missing flux values is that these values are not missing at random. What is more, in our work we also use all 36 colors (magnitude differences) constructed 
from the 9 magnitudes. If we wanted to impute those where at least 1 band is missing, this would affect several of our features and could introduce biases. Therefore, in total, we have 45 input features, and each has a numerical value associated. 

To select the best model and test its performance, a random subset of the dataset with true labels is held out, which is referred to as the random test set. In this work, we use randomly sampled 80\% of the spec-z-cat as the training set and the rest as the test set. For a given set of experiments, when both the magnitude and redshift cut are at their fiducial WW values, the test set is fixed. We, however, draw it anew when we change any of these thresholds. 

From the ML point of view, our labeled data (and hence a random test set extracted from them) are relatively balanced, as the galaxies with spec-z label $z \leq 0.2$ (Class 0) make up $\sim 45\%$ of it. We would like to reiterate, however, that this does not reflect the actual expected proportion between the true classes in the final data, which should be closer to 15\% at the fiducial flux limit (Sec.~\ref{Sec:problem}).

To check for overfitting and ensure the robustness of the model, we employ the popular $k$-fold cross-validation approach. In this method, the training data is divided into $k$ equally sized subsets (or “folds"). The model is trained $k$ times, each time using a different subset as a validation set and the remaining $k - 1$ subsets as part of the training set. This process is repeated $k$ times, and the performance metrics are averaged across the $k$ iterations to obtain a more reliable estimate of the model’s performance. For our implementation, we use $k=5$, meaning the procedure is repeated 5 times with each fold being used exactly once as the validation data. As mentioned earlier, in Sec. ~\ref{sub:decision_tree}, we found that the default hyperparameters provided satisfactory performance, and additional tuning did not lead to significant improvements.

The full photometric catalog defined in  Sec.~\ref{sec:photo} is the target set. For the final inference, to find the probability of the galaxies being desired targets from Class 0, i.e.\ having $z \leq 0.2$, we train the final XGB classifier on the entire spec-z-cat. Then, the trained classifier gives class labels and probabilities of classification for the target set galaxies.

\section{Classification Performance}
\label{sec:performance}
In this Section, we present the classification metrics for different test sets: a random test set selected from spec-z-cat and not used in training, and test sets consisting of DESI DR1. This will allow us to quantify the model performance on data with true labels. 

\subsection{Feature Importance}
\label{subsec:featureimportance}
We have kept the feature engineering aspect of the model at a minimal level, as we found that the classifier works well with the combination of the 9 KiDS+VIKING magnitudes plus all the possible 36 colors as the input features, as mentioned in Sec.~\ref{sec:training}. Even if this means repetitive information (as most combinations are then non-unique), such redundancy is useful and often desired for ML, as the models, especially tree-based, usually are ignorant of the fact that some features are combinations of others. In fact, in addition to colors, i.e.\ magnitude differences, extra information could be provided, such as, for instance, magnitude ratios, as shown in e.g. \citet{2019A&A...624A..13N}, \citet{2021A&A...649A..81N}, even if there is no direct physical meaning of such features. In this work, we do not employ such feature combinations since the basic setup of magnitudes and colors works sufficiently well. We also do not use other features such as flux errors or signal-to-noise, having tested that they do not improve the performance of our classifier.
 
For better understanding which bands and colors matter the most for WW target selection, we perform feature importance analysis. We do it in two different ways. In the first approach, we rank the features 
based on their effect on the F1 score (Eq.~\ref{eq:f1}) for the class of galaxies with redshift below 0.2 for the fiducial case. We run the classifier 30 times with shuffled values of one of the features and check the decrease in the F1 score. The relative mean of this decrease for each feature gives the importance of the particular feature for the classification, which we express in per cent. 
Table  \ref{tab:feature_imp_table} (middle column) 
shows that out of the total 45 input features, the $g-r$ and $u-g$ colors are the most important for our classification, followed by the $g$-band magnitude and some near-IR colors. The fact that the $g$ band appears in the 3 top features most relevant for our classifier is likely related to the fact that the redshifted 4000~\AA\ break passes through the VST $g$ filter close to $z=0.2$, which is the WW selection threshold.

In order to better understand the impact of each photometric feature on the XGB classification, we perform an additional feature importance analysis called SHAP \citep[SHapley Additive exPlanations,][]{Lundberg2020}. This approach takes into account the contribution of every instance of the feature to the model's final prediction of the probability, $P(z \leq z_\mathrm{lim})$ in our case. The Shapley value of a feature is the average of the individual contributions in different feature combinations, considering every possible order in which features could be added. Such analysis is useful because, unlike the permutation-based method above, SHAP takes into consideration the correlations between the features. 
The SHAP values for the ten most important features for classification are given in the right-hand-side column of Table \ref{tab:feature_imp_table}. 
It is interesting to note that, although their order and relative importance differ from what was given in the middle column of Table~\ref{tab:feature_imp_table},
all but one of the most important features are the same between the two approaches. Also, the top four, although shuffled, are the same, only that now the $J-K_s$ color is the most important in terms of its SHAP value. This color is sensitive to the shift of the ‘stellar bump', which is likely the reason why it is at least as important as the features including the  $g$-band. 

Based on the above analysis, we could in principle trim the least important features and build a new model with fewer magnitudes/colors. This would, however, be relevant if it allowed us to reduce the number of required bands and/or considerably reduced the computational time. However, as visible in Table~\ref{tab:feature_imp_table},  the top 10 features include combinations of all 9 KiDS+VIKING magnitudes. What is more,  even with the full set of 45 features, the XGB model runs very fast, and computations can be done on a laptop. We therefore chose not to re-run the whole framework and keep all the magnitudes and colors throughout.

\begin{table}[!t]
\caption{Feature importance ranking for our XGBoost classification model from the permutation feature importance and the Shapley analysis.}\label{tab:feature_imp_table}
\centering
\begin{threeparttable}
\begin{tabular}{|c|c|c|}
\hline
& & \\
\textbf{Feature name}   &  \textbf{Standard importance} & \textbf{SHAP score}\\       
&                                &  \\ 
\hline
$\text{g-r}$    &  7.26\% & 1.51\\
$\text{u-g}$    &  5.66\% & 0.94\\
mag\_gc &  2.01\% & 0.94\\
$\text{J-K}$    &  1.83\%  & 1.59\\
$\text{Z-Y}$    &  0.83\% & 0.29\\
$\text{H-K}$    & 0.63\% & 0.56\\
$\text{g-i}$    & 0.47\% & 0.54\\
$\text{r-i}$    & 0.38\% & 0.20 \\
$\text{Y-K}$    & 0.28\% & 0.27\\
$\text{r-Y}$   & 0.23\% & 0.14\\
\hline
\end{tabular}
 \begin{tablenotes}
           \footnotesize
            \item
           For the permutation feature importance, the scores are obtained from the relative decrease in the F1 score when the values of the feature are shuffled. Feature importance scores from the SHAP analysis are based on the mean of the absolute SHAP values for the features. The features are ordered in the importance ranking from the permutation feature importance. We show the 10 most important features from each method.
        \end{tablenotes}
\end{threeparttable}
\end{table}

\subsection{Results for the default WAVES-Wide thresholds}
\label{results_ww}

\begin{table*}
\centering
\begin{threeparttable}
    \caption[]{Classification performance for the test sets.}
    \label{metrics}
    
    \begin{tabular}{lccc} 
        \hline
        \noalign{\smallskip}
        \textbf{Metric} & \textbf{Random test} & \textbf{Full DESI test} & \textbf{Blind DESI test}\\
        \noalign{\smallskip}
        \hline
        \noalign{\smallskip}
        Purity (Eq.~\ref{eq:pur}) & $94.71 \%$ $\pm 0.11\%$ & $93.59 \%$ $\pm 0.04\%$  & $91.15 \%$ $\pm 0.06\%$\\
        Completeness (Eq.~\ref{eq:com}) & $94.68 \%$  $\pm 0.12\%$ & $91.87 \%$ $\pm 0.07\%$ & $88.72 \%$ $\pm 0.11\%$\\
        F1 score (Eq.~\ref{eq:f1}) & $94.70 \%$ $\pm 0.10\%$ & $92.72 \%$ $\pm 0.03\%$ & $89.92 \%$ $\pm 0.04 \%$ \\
        Accuracy (Eq.~\ref{eq:acc}) & $95.64 \%$ $\pm 0.08\%$ & $95.87 \%$ $\pm 0.02\%$ & $95.33 \%$ $\pm 0.01 \%$\\
        \noalign{\smallskip}
        \hline
    \end{tabular}
 \begin{tablenotes}
           \footnotesize
            \item
           The first three metrics are computed for class ‘0', i.e., WW targets, while the accuracy is a two-class metric including both positives and negatives (targets and non-targets). See Sec.~\ref{Sec:metrics} for metric definitions and Sec.~\ref{results_ww} and Appendix \ref{appendix_err} for details on error calculation. The values are for the random test set selected from the spec-z-cat, an independent test set consisting of DESI Data Release 1 sources, and a blind test set constructed of DESI DR1 after removing sources overlapping with spec-z-cat.
        \end{tablenotes}
\end{threeparttable}
\end{table*}

Here we will quantify the performance of our classifier for the default WAVES-Wide selection thresholds, which are $Z<21.1$ mag and $z<0.2$.
The classification metrics for the random test set and the two DESI-based ones, computed according to the definitions from Sec.~\ref{Sec:metrics} (Eqs. \ref{eq:acc}-\ref{eq:f1}), are presented in Table \ref{metrics}. Here and below, the errors on the metrics are calculated by propagating the Poissonian errors on the number of (0,0), (0,1), (1,0), and (0,1) objects, where 0 is the target class and 1 are non-targets, and first position in parentheses is the true class while the second is the class assigned by the classifier. 

It is important to note that the number of objects identified as a certain class varies with 5-fold training and testing. Hence, the errors in these numbers are correlated. We calculate this correlation via the Pearson coefficient from the 5-fold testing. For more details, see Appendix \ref{appendix_err}.
As can be seen, the values of the one-class metrics (purity, completeness and F1) depend on the test set. For the random subset of the spec-z-cat, they are consistently close to 95\%, while the two-class accuracy is even higher. Except for the accuracy, the other metrics are, however, at lower levels if DESI is used for testing, with the completeness and F1 score lying even below 90\% for the blind DESI test set. We would like to emphasize, however, that the numbers for that test set should be regarded as `pessimistic' lower limits of the performance of our classifier. As already mentioned, such a DESI selection is very different from a $Z$-band flux-limited sample that would be ideally required to robustly assess the classification.

\begin{figure}
    \includegraphics[width=\columnwidth]{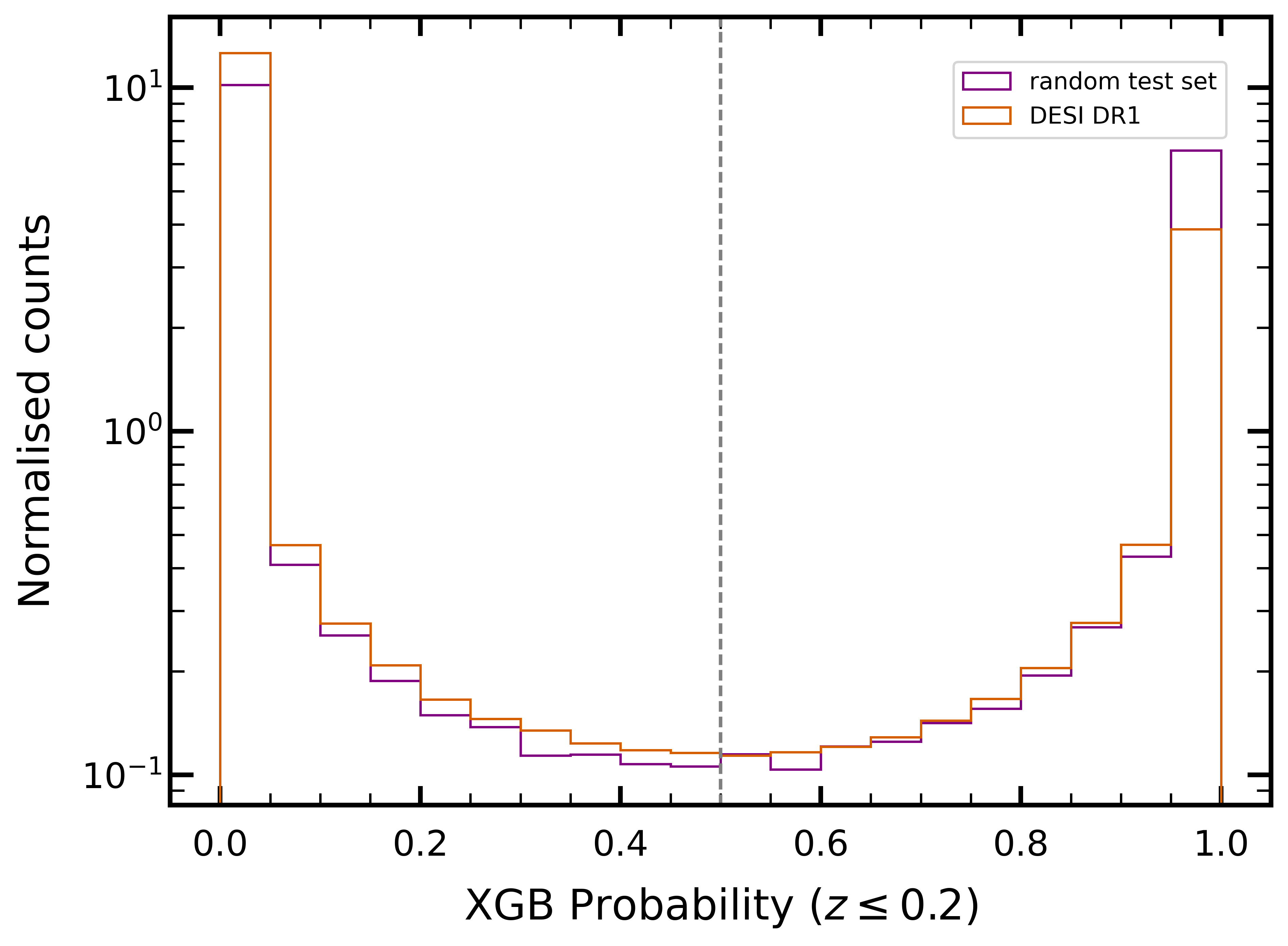}
    \caption{Distribution of XGB-assigned probability that test sample sources are the desired WAVES-Wide targets, i.e., have $z \leq 0.2$. The default classification probability threshold, shown with the vertical line, is 0.5. The purple histogram refers to the test set randomly selected from the spec-z-cat, and the rust-orange line refers to DESI DR1 being used as an independent test set.}
    \label{fig:prob_hist}
\end{figure}

The values of the metrics discussed above were provided for a cut in classifier probability at 0.5, i.e.\ objects with $P_\mathrm{XGB}>0.5$ were allocated to class 0 and the others to class 1. This probability threshold could, however, be varied (in particular, increased) for more reliable classification. It is then informative to look at the distribution of probabilities assigned to the full test sample (with both the positive and negative true class, i.e.\ galaxies at any true redshift); this is presented in Fig.~\ref{fig:prob_hist}. A reliable classifier should have two sharp peaks at $P=0$ and $P=1$, which correspond to class 1 and 0, respectively. This is what we observe indeed; in addition, relatively few sources are assigned probabilities close to 0.5, which would indicate uncertain predictions. From the plot, we see that varying the probability threshold, based on which the targets are assigned, between even $P\sim 0.3 \div 0.7$ would hardly affect the classification.

\begin{figure}
    \includegraphics[width=\columnwidth]{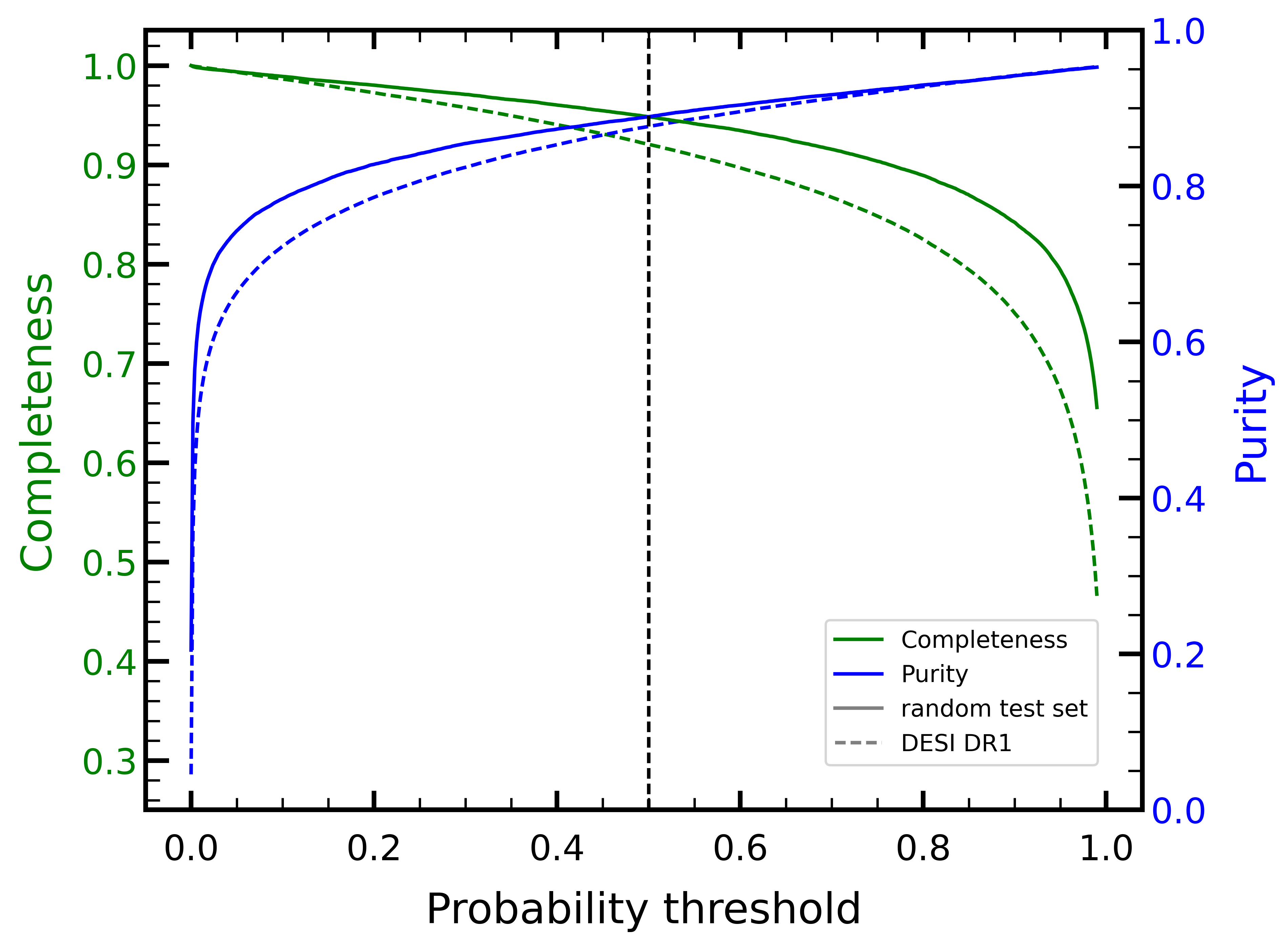}
    \caption{Classification metrics as a function of the probability threshold of the XGB classifier to select WAVES-Wide targets. The solid lines represent the metrics for the test set selected randomly from spec-z-cat; the dashed lines represent the metrics for DESI DR1 used as an independent test set.}
    \label{fig:compurity}
\end{figure}

To further verify the appropriate threshold for classification probability, in Fig.~\ref{fig:compurity} we display the performance metrics as a function of $P_\mathrm{XGB}$. We observe that completeness and purity change almost linearly except for a quick steepening at the extreme end of $ P\sim1$ and $ P\sim0$, respectively. Overall, the four presented curves suggest that our originally adopted cut of $P=0.5$ to differentiate between WW targets and non-targets is a good compromise. At the same time, Fig.~\ref{fig:compurity} could be further used to decide how to move the probability threshold in case a higher value of completeness or purity is required.

\begin{figure*}
\centering
    \includegraphics[width=0.75\textwidth]{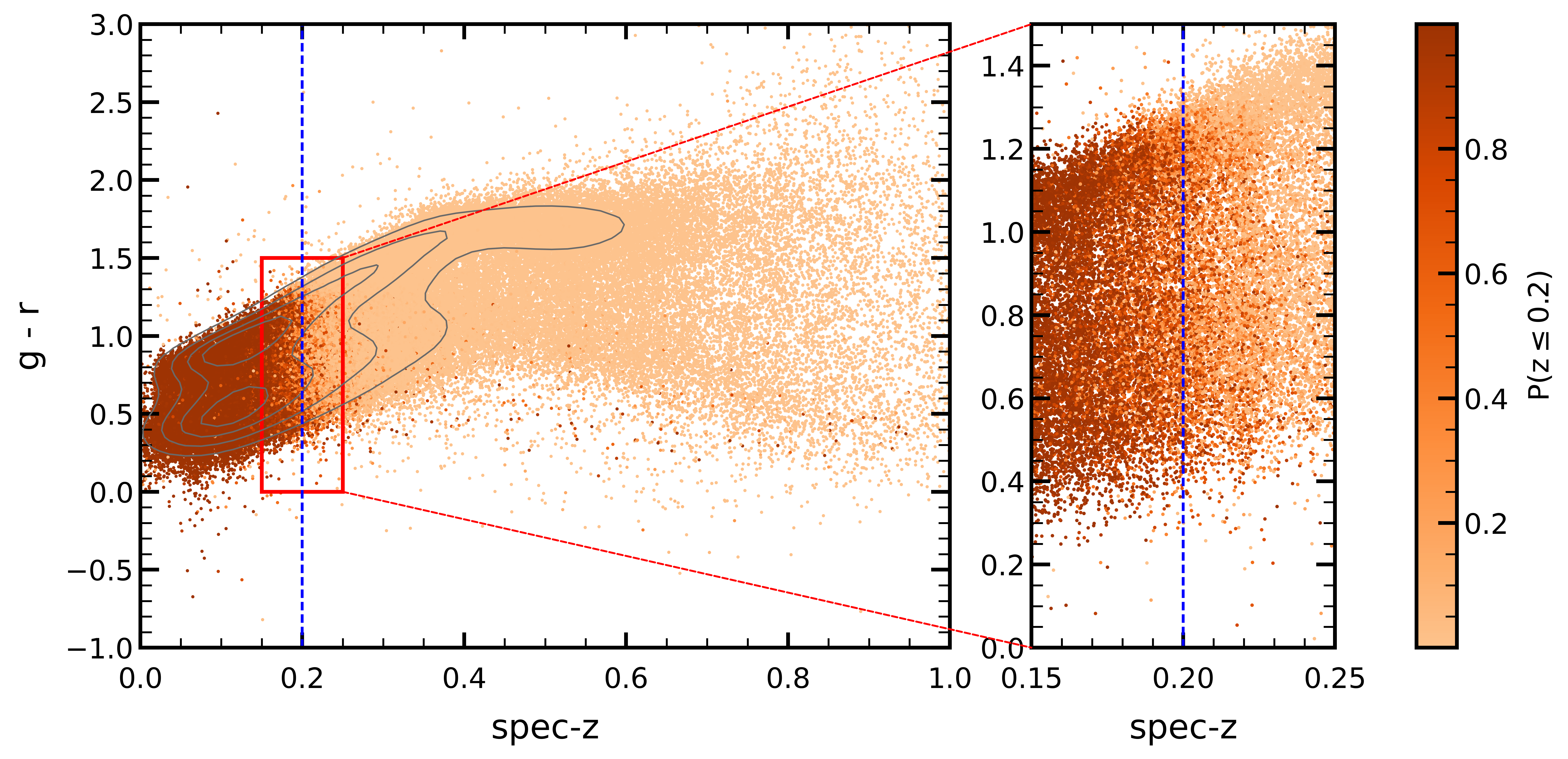}
    \caption{Distribution of the galaxies from the test set on the observed $\text{g-r}$ color -- true redshift plane, color-coded by the probability of being a WAVES-Wide target as assigned by our classifier. Dark indicates a high probability of having $z<0.2$, while light means assignment as a non-target. The vertical dashed line shows the WW division threshold. Grey contours indicate the region of the largest density of galaxies in the diagram. On the right-hand side, we enlarge the area close to the threshold to highlight the region where most of the misclassification happens.}
    \label{fig:gr_z_prob}
\end{figure*}

We further visualize the performance of our classifier by mapping the XGB-assigned probabilities on the color-redshift plane. For the latter, we choose $\text{g-r}$, which was identified as the most important feature for our classifier. This is illustrated in Fig.~\ref{fig:gr_z_prob}. Each dot in this plot is one galaxy from the test set, and galaxies are color-coded by the assigned probability of being a target (dark being high and light being low $P$). The grey contours illustrate the distribution of the sources where their density is the largest. The bimodality in the contours is due to both red (larger $\text{g-r}$) and blue galaxies being present in our test set. The vertical dashed line indicates the $z=0.2$ threshold for the WW targets. An ideal classifier would label all the galaxies to the left of this line as targets (giving them $P>0.5$, i.e.\ dark color) while those to the right would all have low $P$ and light color. Although our classifier is not perfect, we can see that, consistent with previously quoted values of completeness and purity metrics, a very small fraction of objects end up as false negatives (5.3\%) or false positives (3.7\%). In addition, these incorrect predictions are mostly located close to the threshold line. Some outliers exist, but they are not numerous. For instance, only about 100 galaxies at $z>0.4$ from the test set have $P(z<0.2)>0.7$. A similar picture emerges if we use DESI DR1 as the test set instead. The relevant plot is shown as Fig.~\ref{fig:gr_z_prob_desi} in the Appendix \ref{app:DESI_performance}. 

\begin{figure}
    \includegraphics[width=\columnwidth]{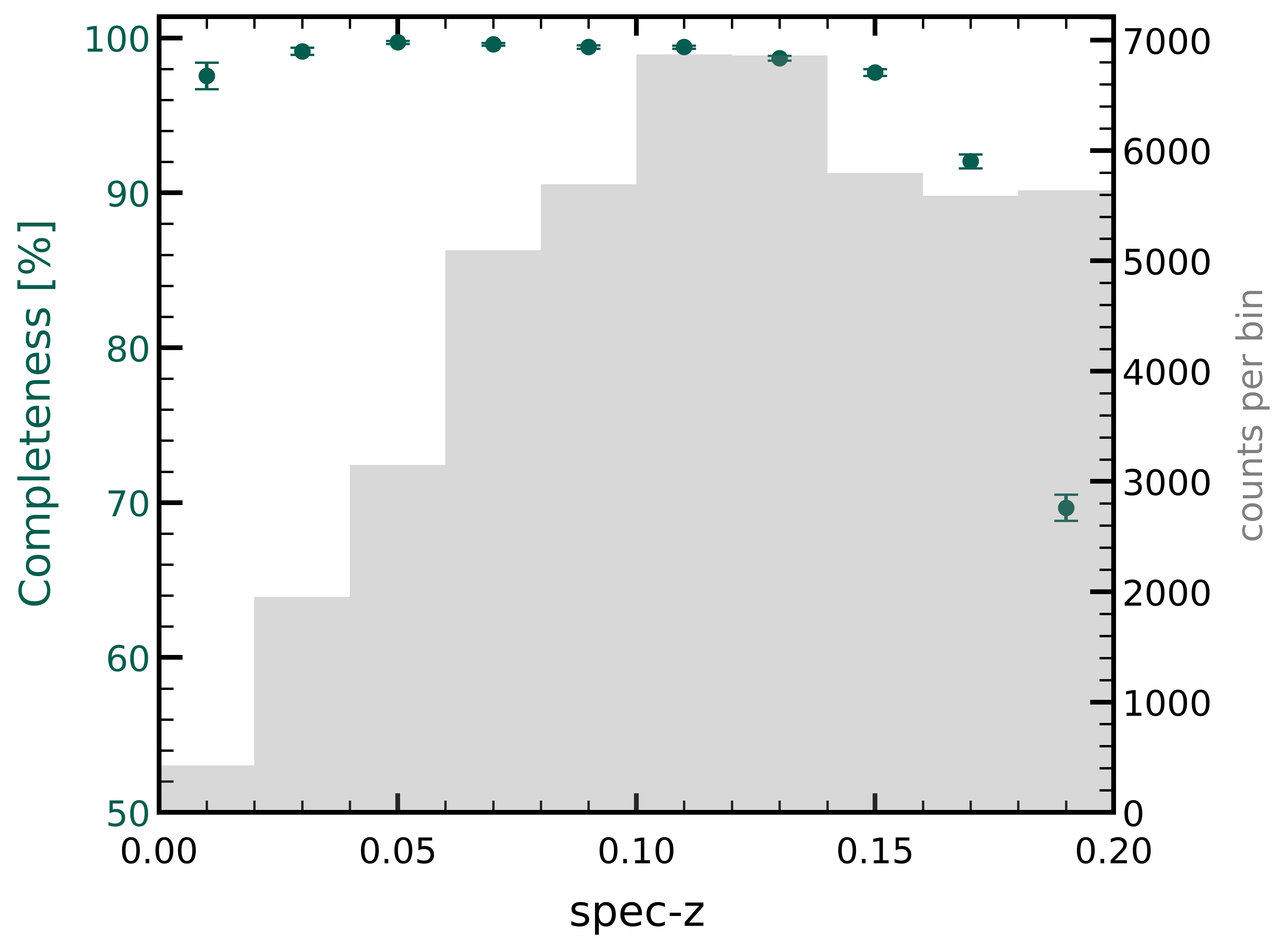}
    \caption{Completeness as a function of true redshift for the desired target (Class '0') galaxies, computed per redshift bin of width 0.02. The scatter on the metric is from the Poissonian error on the number of correctly identified (0,0) and missed (0,1) galaxies. The histogram in grey shows the distribution of redshifts in the random test sample.}
    \label{fig:dcompdz}
\end{figure}

\begin{figure}
    \includegraphics[width=\columnwidth]{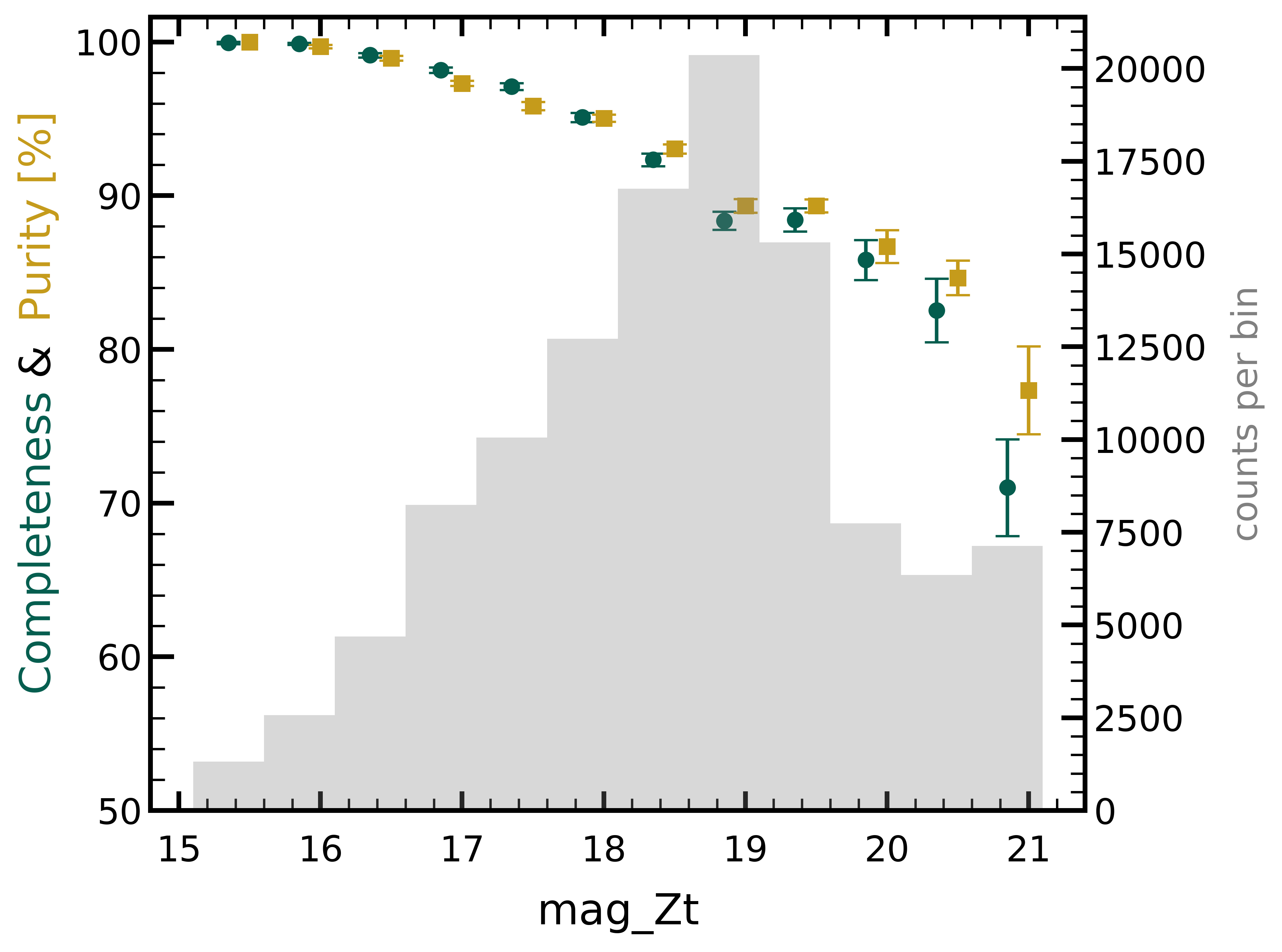}
    \caption{Completeness (dark green) and purity (olive) as a function of $Z$-band magnitude for sources in the test set, computed per magnitude bin of width 0.5. For better visualization, olive dots representing purity are shifted by 0.15 along the x-axis. The scatter in the metrics arises from Poissonian errors associated with the counts of correctly identified (0,0), contaminant (1,0), and missed (0,1) galaxies. The histogram in grey shows the Z band magnitude distribution in the random test set.}
    \label{fig:dcompdmagz}
\end{figure}

The final test we provide in this section is to quantify the performance of our classifier as a function of the (true) redshift, $Z$-band apparent magnitude, and the $\text{g-r}$ color, for the fiducial WW redshift and magnitude thresholds and using the $P=0.5$ cut between the classes. For that, we partition the test sample into narrow bins of the respective quantities and compute the statistics per bin. Figures \ref{fig:dcompdz} and \ref{fig:dcompdmagz} illustrate, respectively, the redshift and magnitude dependence of completeness (dark green points) and purity (olive points, only for magnitudes). 
Clear trends emerge: the XGB classification performs best at the bright, low-redshift end, deteriorating for galaxies close to the WW redshift threshold and magnitude limit. This fall-off is more abrupt for redshifts: up to $z \lesssim 0.15$, completeness stays at a level of $\sim99\%$ and quickly plummets to $\sim 70\%$ at $z\sim 0.2$. As a function of magnitude, the deterioration is more gradual, although the completeness reaches the value of $\sim 70\%$ close to the magnitude limit of the sample. It is also worth noting that, as the median $Z$-band magnitude of the test set is 18.6 mag, the averaged metrics are biased towards the brighter end. 
The trends as a function of redshift and magnitude are likely driven by two aspects: one is the limitations of the training set, whose very complete flux-limited component from GAMA reaches to $z<0.2$ at $r<20$ mag (equivalent to roughly $Z< 19.3$ mag). Secondly, our classifier seems to be most uncertain in its predictions close to the redshift threshold (cf.\ Fig.~\ref{fig:gr_z_prob}).

\begin{figure}
    \includegraphics[width=\columnwidth]{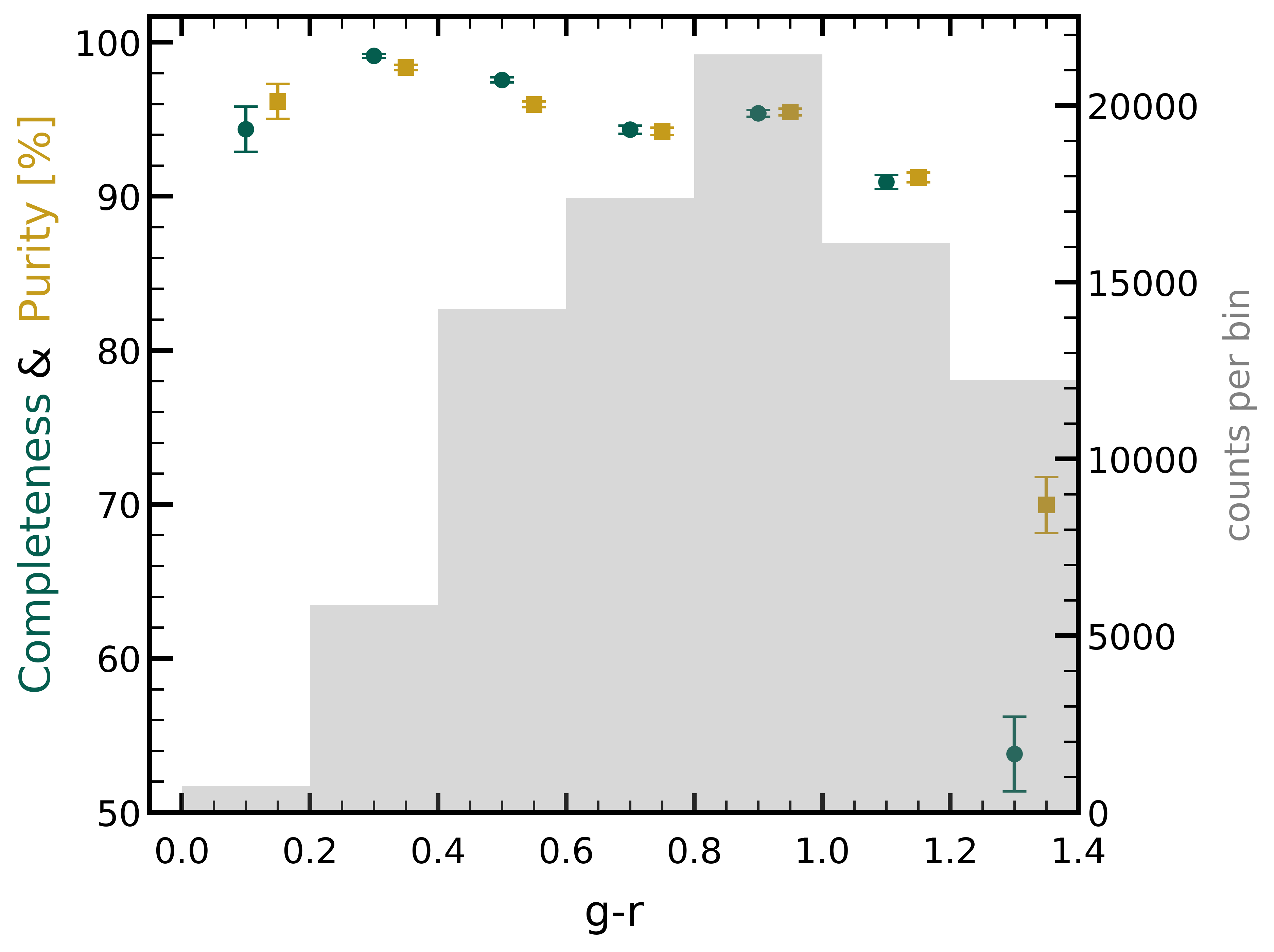}
    \caption{Completeness (dark green) and purity (olive) as a function of the $\text{g-r}$ color for sources in the test set, computed per bin width of 0.2 mag. For better visualization, olive dots representing purity are shifted by 0.05 mag along the x-axis. The histogram in grey shows the color distribution of galaxies in the random test set.}
    \label{fig:dcompcolor}
\end{figure}

As far as the color dependence is concerned, no strong trends are visible, as illustrated in Fig.~\ref{fig:dcompcolor}. For $g-r \lesssim1.2$, both completeness and purity stay at similarly high levels above 90\%. Only for the reddest galaxies with $\text{g-r} \sim 1.3$ mag, we observe considerable deterioration in the two metrics. This is likely driven by the fact that such reddened galaxies are mostly above the WW threshold redshift of $z=0.2$ (cf. Fig.~\ref{fig:gr_z_prob}); there is then a large imbalance between the true and false classes in the training for this color range. However, as very few galaxies at $z<0.2$ have $g-r>1.3$, the classification results could be improved by simply removing a posteriori all such red objects from the sample.

Another factor that will contribute to the classifier's performance will be the signal-to-noise ratio (S/N) of the photometry. We have checked the trends in that respect and found that, as expected, both completeness and purity deteriorate with lowered S/N. For instance, in the $g$-band they go down to levels of $\sim85\%$ at S/N of roughly 50, and worsen further with lowered S/N. As magnitudes (fluxes) and their errors are highly correlated, these trends are qualitatively very similar to what is observed as a function of magnitude itself.

We have also looked into any possible trends in magnitude and color distributions for the misclassified sources (i.e., those missed out -- false negatives, FN, and contaminants -- false positives, FP). While such objects are typically fainter and bluer than the correctly assigned WW targets (true positives), we do not see significant differences between FP and FN that would point to a particular sensitivity of our classifier. Similar conclusions are reached if we use DESI DR1 as the test set instead of the random subsample of the spec-z-cat. The relevant plots are presented in Appendix \ref{app:DESI_performance}.

\subsection{Going beyond the default WAVES-Wide selections}\label{change_cuts}

\begin{figure}
    \includegraphics[width=\columnwidth]{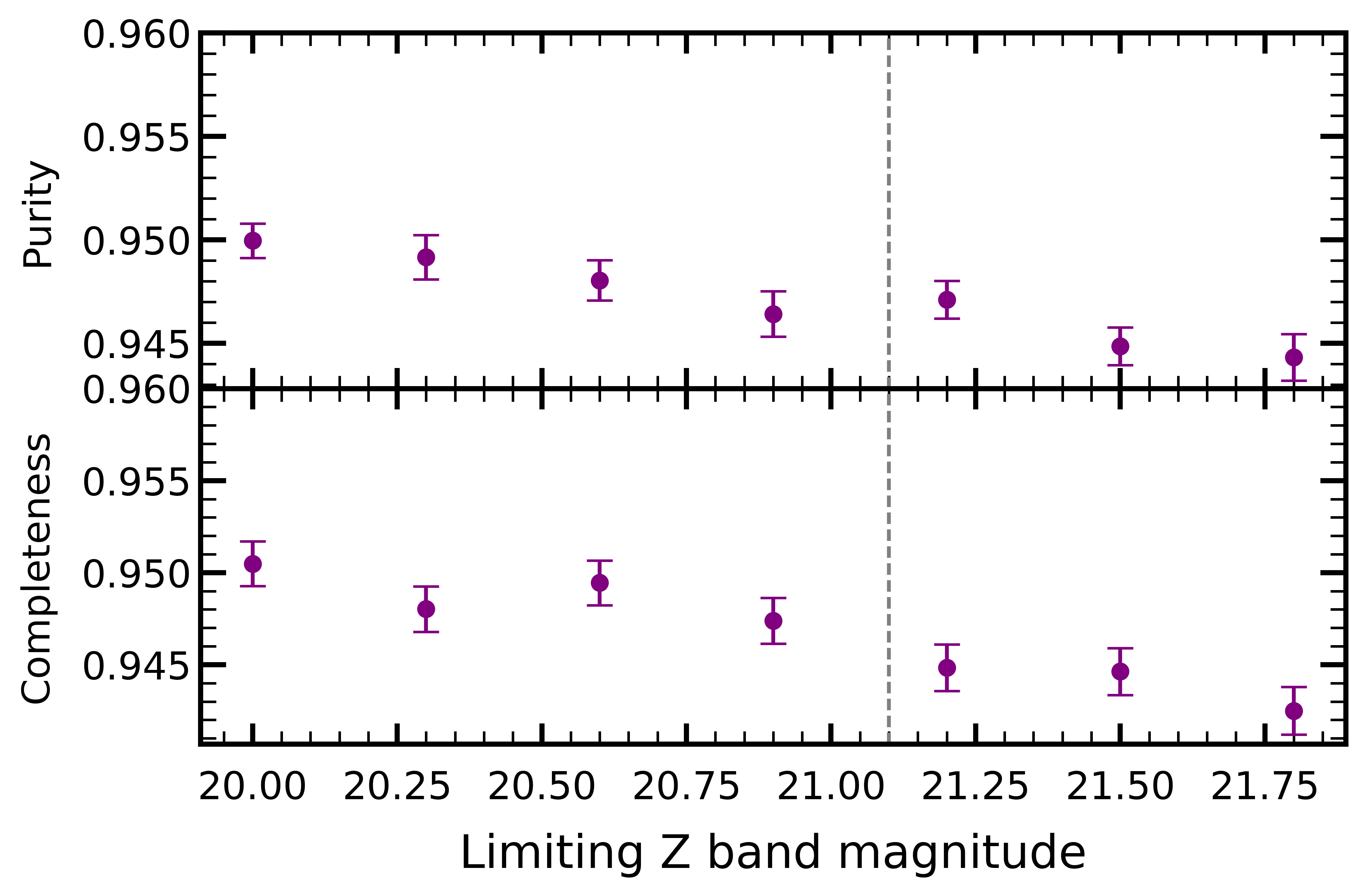}
    \caption{Performance of XGB-based selection of WAVES-Wide targets for different thresholds of the $Z$-mag cut (x-axis) with the redshift limit fixed at the fiducial $z<0.2$ value. The scatter on the metrics is from the Poissonian error on the number of True Positives, False Positives, and False Negatives. The vertical dashed line shows the WW fiducial $Z$-band cut.}
    \label{fig:mag_change}
\end{figure}

As we specified previously, the currently planned WAVES-Wide selections are $Z<21.1$ mag and $z<0.2$. However, these thresholds could still be revised by the survey depending on the efficiency and time availability of the observations. It is then interesting to explore how our approach will perform if we vary one or both of the limits, still keeping them close to the fiducial values. In particular, we would like to find out if, for our binary classifier, there exists a more optimal combination of the redshift-magnitude thresholds that would still provide a similar number of WW targets as for the default cuts. For this purpose, we ran a set of new XGB classifications where each time the samples with true labels (train, validation, and test) were changed after imposing new $Z$-band magnitude and/or redshift thresholds. For these tests, only samples drawn from the spec-z-cat were used (i.e.\ no DESI DR1 data were employed).

We start by varying the flux limit in the $Z$-band for selection, while the redshift cut is the fiducial one, i.e.\ $z<0.2$. For that purpose, from the spec-z-cat described in Sec. \ref{sec:spectro} we select subsamples at different $Z$-mag limits ranging from 20.0 to 21.8 in steps of 0.3 mag (i.e.\ 7 possibilities). The magnitude range we chose here starts from roughly 1 mag brighter than the fiducial cut, while the upper limit is driven by the input catalog going only to $Z=22$ mag\footnote{The KiDS+VIKING source data reach fainter than this, but for practical reasons we work with a catalog limited to $Z=22$ mag.}. We then train and test separate XGB classifiers for these flux-limited catalogs. The results are illustrated in Fig.~\ref{fig:mag_change} for the statistics introduced in Sec.~\ref{Sec:metrics}. In the plot, each data point with error bars is derived from the training and testing at a particular magnitude limit. We notice that all the 1-class statistics consistently deteriorate when the flux limit is decreased, although not by a large margin: from roughly 0.95 at $Z<20$ limit to about 0.945 at $Z<21.8$ mag. This deterioration is likely driven by two factors. First, as illustrated earlier in Fig.~\ref{fig:shark_selection}, increasing the magnitude threshold for the targets increases the mean redshift of the entire input sample to select from, and makes the percentage of the required $z<0.2$ targets smaller. This means that it becomes more difficult to properly select the targets as the flux limit is lowered.
Second, our spectroscopic training set is most complete at the bright end, 
therefore, a supervised classifier like ours will perform worse the more faint objects we add.

\begin{figure}
    \centering
    \includegraphics[width=\columnwidth]{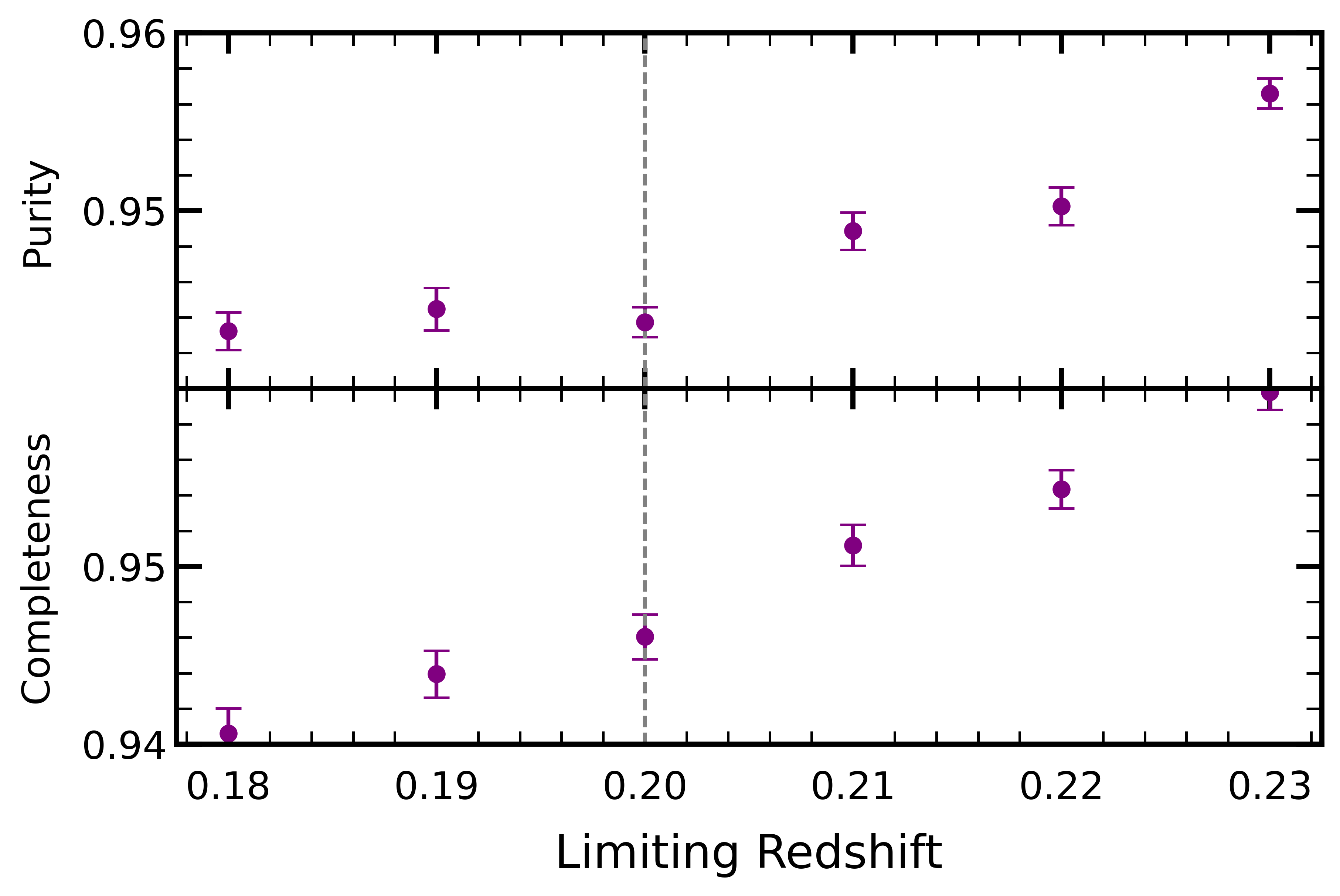}
    \caption{Performance of XGB-based selection of WAVES-Wide targets for different thresholds of the redshift limit (x-axis) with the $Z$-mag cut fixed at the fiducial $Z=21.1$ mag value. The scatter on the metrics is from the Poissonian error on the number of True Positives, False Positives, and False Negatives. The vertical dashed line shows the WW fiducial redshift cut.}
    \label{fig:z_change}
\end{figure}

We now turn to the second set of experiments, where we fix the $Z$-mag cut at the fiducial value of 21.1 but move the redshift threshold for the targets. This means that we now have one selection of galaxies with spec-z labels for training and testing -- the same as in the fiducial case --, but we train 4 additional classifiers in addition to the one analyzed in the previous section; namely, we look at target selection for cases with $z_\mathrm{lim}=0.18,0.19,0.21,0.22$. As previously for varied magnitude limits, also here we compute the same metrics and show them in Fig.~\ref{fig:z_change}, where this time the changing redshift threshold for WW targets is shown on the x-axis. We find that, for the fiducial magnitude cut, increasing the redshift limit between $z=0.18$ and $z=0.22$ improves all the analyzed statistics, although again the differences are not substantial, roughly between 0.945 and 0.955. A likely explanation for this improvement is that, as can be appreciated from the violet line in Fig.~\ref{fig:shark_selection}, the median redshift of galaxies flux-limited to $Z=21.1$ is about 0.43. Therefore, increasing the redshift selection threshold in our analyzed range reduces the imbalance between targets and non-targets, which improves the performance of the binary classifier. The closer we are to the median redshift with the threshold, the easier it should be for the classifier.

As the final test, we vary the magnitude and redshift threshold for the WW target selection jointly. Here the number of possibilities is very large if we kept the same ranges and steps as in the two previous paragraphs; therefore we discuss only a sub-selection for 3 different redshift limits ($z<0.18, z<0.20, z<0.22$) and 4 magnitude thresholds ($Z \leq 20.0, Z\leq 20.6, Z\leq 21.2, Z\leq21.8$), which gives 12 combinations in total. We report the metrics for the XGB classifiers in Table \ref{grid}, in which the first column is the redshift limit for the target selection, column two provides particular $Z$-mag cuts, and the remaining columns include the metrics computed for the particular redshift-magnitude threshold combination. The results confirm our earlier findings: the metrics are maximized if the redshift limit is increased and the magnitude limit decreased. In other words, our supervised classifier would work best, for the currently available training data, if WW target selection went deeper in terms of redshift, but shallower in terms of flux, than the fiducial adopted values. We note, however, that the differences in the metrics between the most optimal and the fiducial case are not very big, roughly at the 1\% level, even if they are statistically significant, taking into account the typical scatter of $\sim0.1\%$.

\begin{table}[!t]
\caption{Performance of XGB-based selection of WAVES-Wide targets for different combinations of the redshift and $Z$-band magnitude limits}  
\label{grid}
\begin{threeparttable}
\centering
\begin{tabular}{|c|c|c|c|c|}
\hline
& & & &\\
\textbf{Redshift}   &  \textbf{$Z$-mag} & \textbf{Purity}   & \textbf{Completeness} & \textbf{F1 Score} \\
\textbf{cut}        &  \textbf{cut}     &    &  &  \\
\hline
        & 20.0 & 94.5 \% & 94.5 \% & 94.5  \%\\
0.18    & 20.6 & 94.4 \% & 94.5 \% & 94.5  \%\\
        & 21.2 & 94.1 \% & 94.2 \% & 94.2  \%\\
        & 21.8 & 94.0 \% & 93.7 \% & 93.9   \%\\
\hline
        & 20.0 & 94.9 \% & 94.9 \% & 94.9 \%\\
0.20    & 20.6 & 94.8 \% & 94.8 \% & 94.8 \% \\
        & 21.2 & 94.7 \% & 94.8 \% & 94.8 \% \\
        & 21.8 & 94.6 \% & 94.1 \% & 94.3 \% \\
\hline
        & 20.0 & \textbf{95.4} \% & 95.7 \% & \textbf{95.6} \% \\
0.22    & 20.6 & 95.1 \% & \textbf{95.8} \% & 95.4 \% \\
        & 21.2 & 95.1 \% & 95.6 \% & 95.3 \% \\
        & 21.8 & 94.8 \% & 95.1 \% & 95.0 \% \\
\hline

\end{tabular}
 \begin{tablenotes}
           \footnotesize
            \item
            The first and second columns represent different combinations of the redshift limit and $Z$-band magnitude limit. Columns 3-5 present, respectively, purity, completeness, and F1 score for the given combination of cuts. We highlight the instances when the given metric is maximized on this particular grid. The scatter of all the metrics takes the value of about 0.1\%, so we do not show it here.
        \end{tablenotes}
\end{threeparttable}
\end{table}

\section{Selection of WAVES-Wide targets from the entire input catalog}
\label{sec:inference}

We will now employ our approach to select WAVES-Wide targets from the entire input photometric catalog. For that purpose, we train the XGB classifier on the full spec-z set of  572,325 galaxies (at the fiducial $Z<21.1$ mag limit).
As previously, our true class ‘0' are galaxies at $z<0.2$. For the probability threshold of $P>0.5$, our model allocates about 1.19 million target candidates out of over 8.6 million of all objects with 9-band detections and pre-assigned to be galaxies according to the WAVES selection procedure. This gives about 13.8\% of input galaxies being labeled as targets with our approach, as compared to 12.1\% predicted from the SHARK mocks (Sec.~\ref{Sec:problem}). Taking into account that the threshold of 0.5 in the probability is rather (too) permissive, this percentage of WW targets could be reduced by increasing the probability cut of the classifier. The ratio predicted by the mocks would be matched if we adapted the threshold in such a way as to obtain about 1.04 million targets.

\begin{figure}
    \includegraphics[width = \columnwidth]{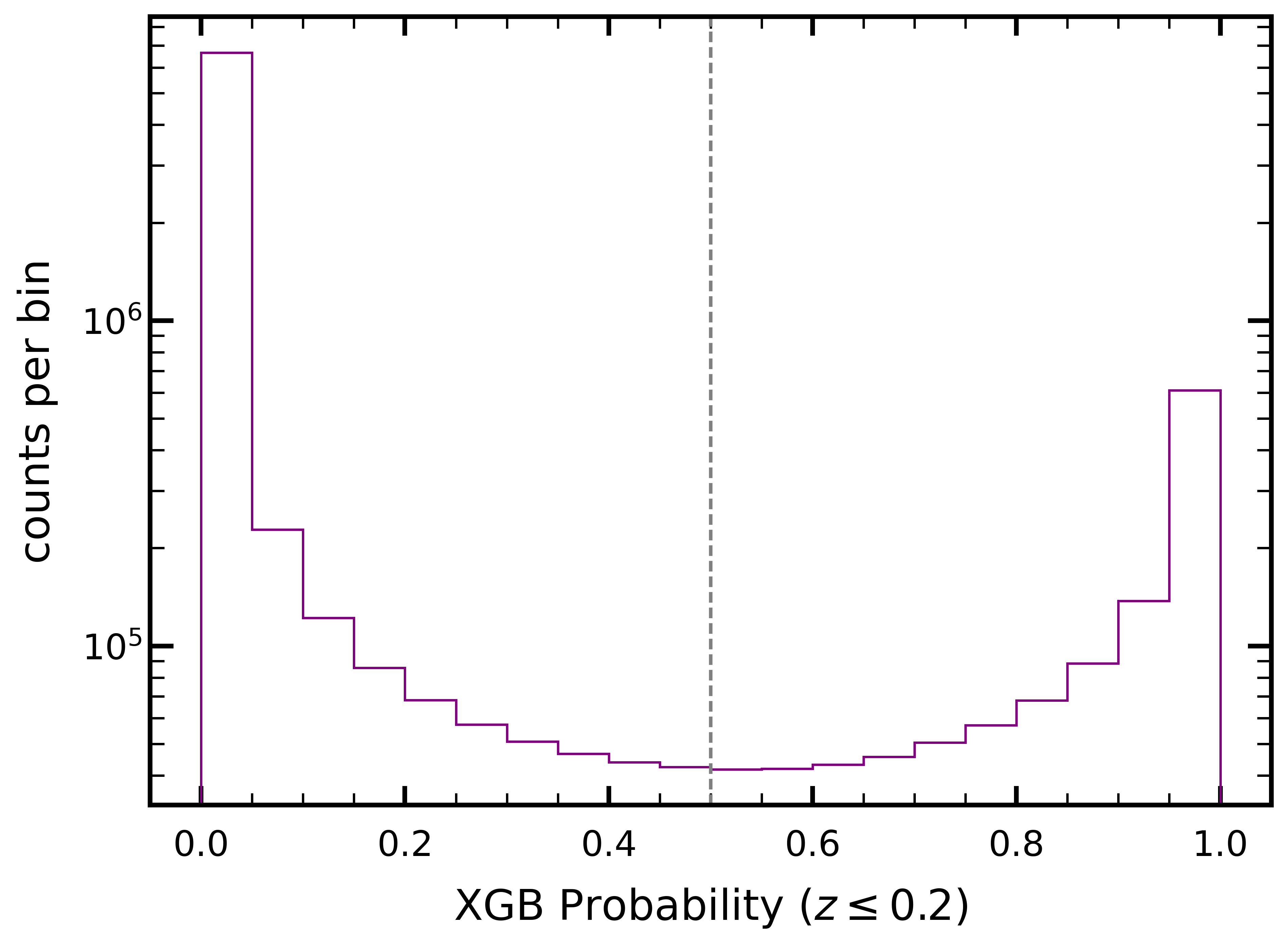}
    \caption{Distribution of XGB-assigned probability of inference set galaxies being the desired WAVES-Wide targets, i.e., having $z \leq 0.2$. The default classification probability threshold, shown with the vertical line, is 0.5.}
      \label{fig:prob_hist_inf}
\end{figure}

For the resulting catalog of WW target candidates, we cannot compute such performance statistics as in the previous sections, as these require true labels (redshifts) to be known. We can, however, still analyze what the classifier provides by looking into the properties of the resulting catalog. The first relevant quantity to check is the probability of being a target as assigned by XGB. This is illustrated in Fig.~\ref{fig:prob_hist_inf}, which shows the histogram of the XGB probabilities for the entire photometric catalog in the same way as was done previously in Fig.~\ref{fig:prob_hist} for the spectroscopic test sets. The allocated targets are to the right of the vertical $P=0.5$ line. The first observation to make is the very pronounced peak at $P\sim0$, which is consistent with the fact that most of the $Z<21.1$ mag galaxies are above the $z=0.2$ redshift threshold. 
The probability distribution at the intermediate ranges of $0.3\lesssim P \lesssim 0.7$ flattens out, although there still is a clear peak at $P\sim1$. This suggests that the probability cut at 0.5 may not be optimal in the final photometric set, and a more stringent threshold could be adopted to improve the [robustness/reliability/confidence] of the selection. 

\begin{figure}
    \includegraphics[width = \columnwidth]{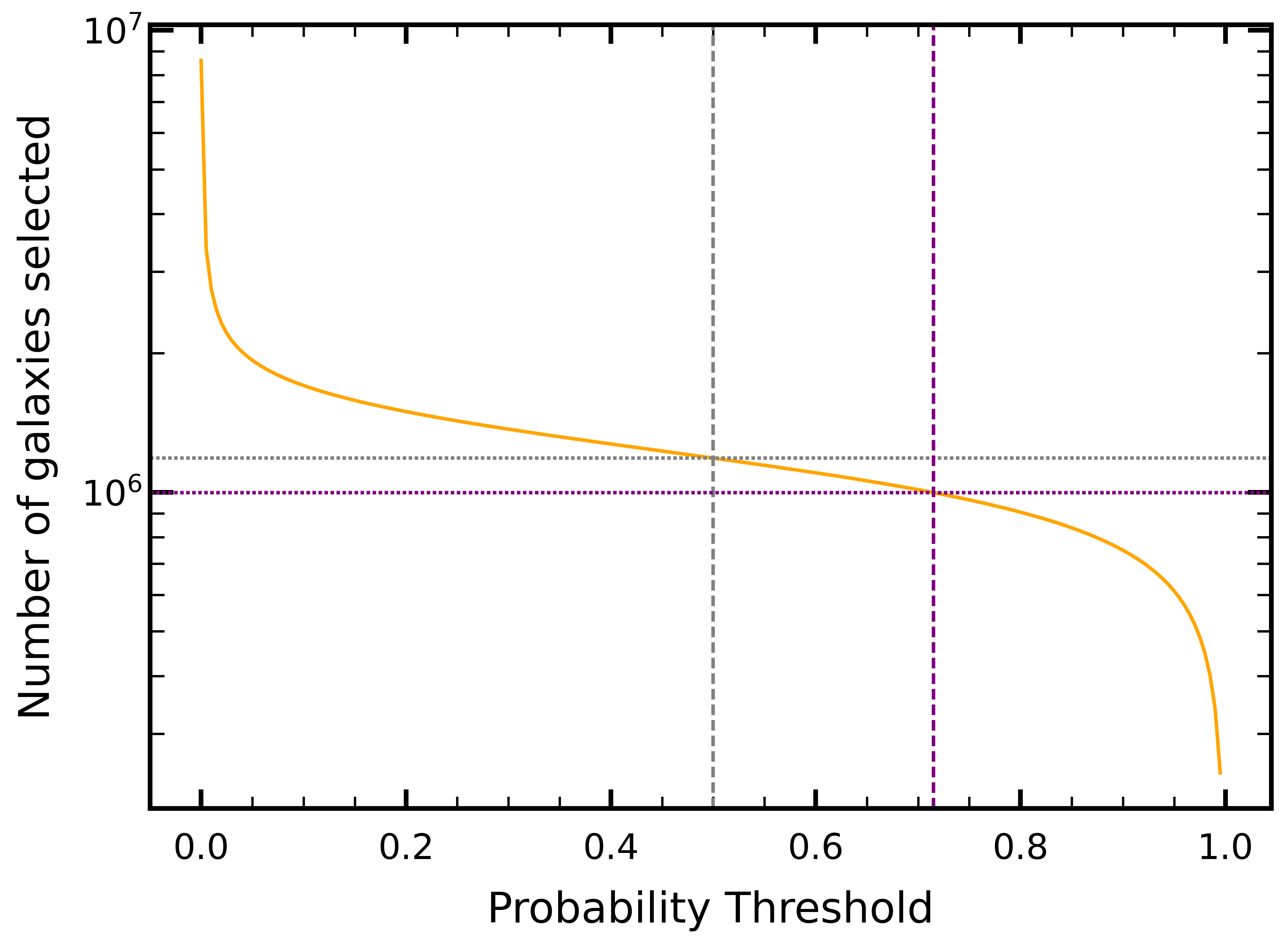}
    \caption{The number of galaxies selected as targets for WAVES-wide as a function of the $P(z \leq 0.2)$ classification threshold. The grey lines show the number of targets selected when the classification probability threshold is the default one, i.e., $P>0.5$. The purple line shows the probability criteria needed to select 1 million targets ($P>0.715$).}
    \label{fig:num_thres}
\end{figure}

A way to decrease the number of possible targets -- which might be preferable from the point of view of time allocation on the 4MOST instrument -- is by raising the probability cut over the fiducial $P>0.5$. In Fig.~\ref{fig:num_thres} we show how the number of targets would change when sliding through the $P$ threshold all the way from 0 to 1. Obviously, at the extremes, we have either the entire photometric dataset assigned as `targets' ($P_\mathrm{cut}=0$) or none ($P_\mathrm{cut}=1$). We also see that the target number changes very rapidly close to the edges, while it varies much more slowly over the range of $0.1\lesssim P_\mathrm{cut} \lesssim 0.8$. This diagram can be used to decide what probability cut to take in order to obtain a fixed number of target candidates. As already mentioned, for the fiducial cut of $P>0.5$, we have about 1.19 million targets; this would be reduced to 1 million if we take $P>0.715$.

\begin{figure*}
\centering
    \includegraphics[width = 0.9\textwidth]{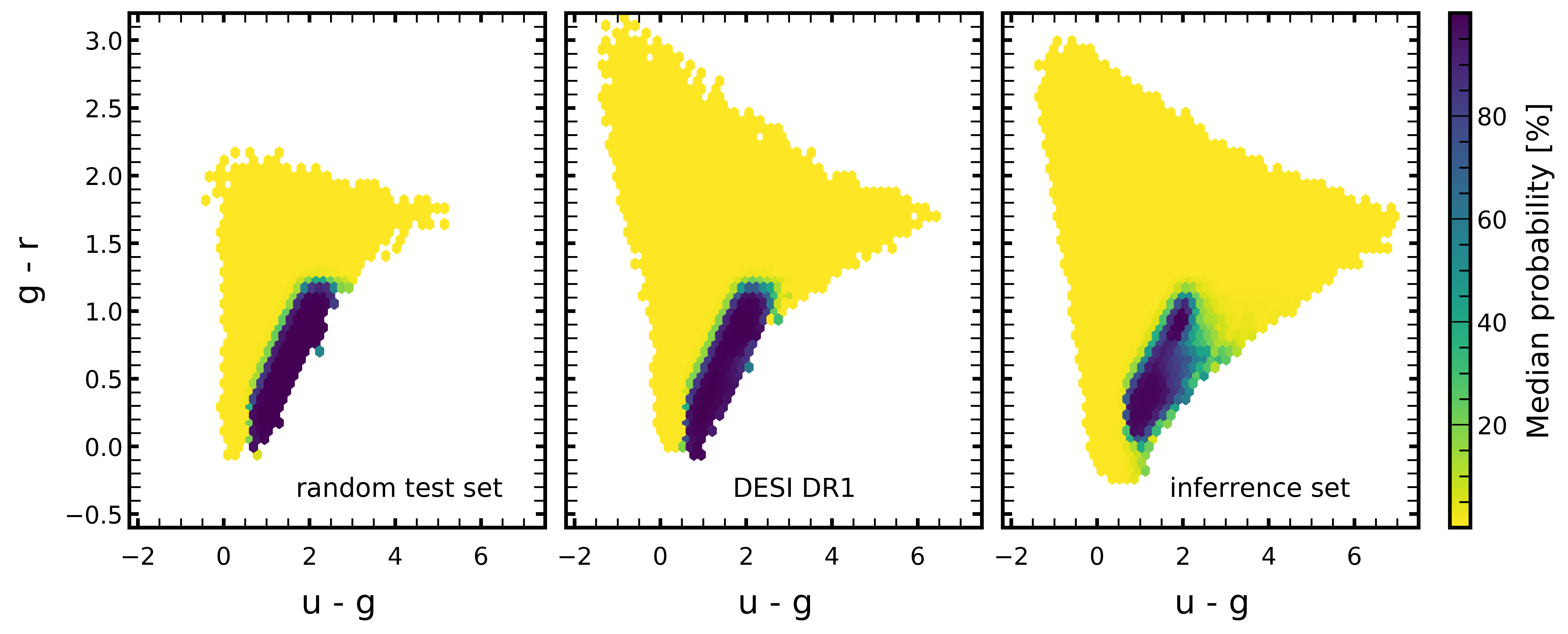}
    \caption{Distribution of WAVES galaxies on the observed $\text{g - r}$ vs. $\text{u - g}$ color-color plane. Each pixel is color-coded with the median probability of the source being assigned as a WW target by XGBoost, i.e., having $z<0.2$. Left panel: random test set; middle panel: DESI DR1 test set; right panel: entire photometric sample at the WAVES-Wide fiducial flux limit of $Z<21.1$ mag. For the random test set, we only show pixels with at least 10 galaxies each; for the DESI DR1 test set, the minimum count per hexbin is 20, while for the inference sample, this threshold is 100.}
    \label{fig:colcol}
\end{figure*}

In Figure \ref{fig:colcol}, we take a closer look at how our classifier maps the probabilities onto the color space. For this, we show the observed $\text{g-r}$ vs. $\text{u-g}$ color-color planes for the random test set (left panel), DESI test set (middle panel) and the entire photometric inference set (right panel). Pixels are color-coded by the median XGB probability of having $z<0.2$. We notice that the classifier seems more confident for the test sets, where there is a clearer separation between dark (targets) and light (non-targets) pixels; this is more diluted for the entire inference sample. Secondly, in the photometric dataset from which the observational targets will be chosen, as well as in the DESI test set, there are galaxies having colors not well-represented in the random test set, the latter being statistically consistent with the training. We note, however, that these `extrapolation' objects are mostly redder than galaxies at $z<0.2$; for instance, many have $\text{g-r}\gtrsim1.3$, which is the limit for WW targets (cf.\ Fig.~\ref{fig:gr_z_prob}). They are then `true negatives' and are properly labeled by the model as such ($P\sim0$). Some lie at the outskirts of the color distribution, which either indicates very rare sources or suggests considerable scatter in the photometry (especially in the $u$ band). Similar observations can be made for the general feature space, i.e.\ when other colors are also taken into account. 

As we already discussed in Sec.~\ref{sec:spectro}, at the faint end of the WAVES selection we only had color- or galaxy-type-selected datasets with true redshift labels for the training. This complication would affect any empirical methodology that aims to map the color space to redshift for the purpose of WW target assignment. In our case, we can conclude that our classifier assigns relatively low probabilities of being a target to the objects whose colors are not well represented in the current spectroscopic training. A cut on probability above the fiducial $P=0.5$ threshold should mitigate the contamination by false positives (galaxies chosen as targets which in fact have $z>0.2$).

Due to the high dimensionality of the problem, a quantitative comparison of training and inference set coverage would require using an appropriate tool, such as self-organizing maps \citep[SOM, e.g.][]{Jalan2024}. We defer this to future work, where we will be able to also use 4MOST SPV data and locate them on the SOM together with the spectroscopic training sets.

The eventual WAVES target selection combines our approach with others developed by the team. Suelves et al. (in prep.) demonstrate that the XGB-based target selection presented here performs at least equally well as the other methods employed in WAVES, including the training-free spectral energy distribution fitting approach `TopZ' \citep{TOPz_GAMA}. We note that independently of the method, color cuts on the inference catalog could be applied a priori to remove highly unlikely targets. This would reduce the size of the photometric catalog as well as possibly the number of misclassified targets.

\section{Summary and conclusions}
\label{sec:conclusions}
In this paper, we presented a supervised machine-learning (ML) approach to select observational targets for the forthcoming WAVES-Wide galaxy survey on 4MOST. The targets will be both flux- and redshift-limited, while their redshifts are not known a priori. In our method, rather than estimating individual redshifts from multi-band photometry, we performed binary classification to decide if a galaxy should be a target or not, based on object locations in magnitude-color space. In particular, we used the fast and efficient XGBoost classifier, trained on a labeled sample derived from a compilation of relevant spectroscopic surveys overlapping with WAVES photometry, from which we use 9 broad-band magnitudes and 36 associated colors. XGB, an ensemble tree-based ML model, is not very demanding computationally and also provides output probabilities of objects belonging to a given class, rather than just discrete labels.

Using the labeled sample separated into train and test sets, we quantified the performance of our XGB classifier. This was done for the default WW limits of $Z<21.1$ mag \& $z<0.2$, but also when one or both of these thresholds were varied. If a random test set, of the same properties as the training one, is used, the classification metrics such as completeness, purity or accuracy are at the level of $\sim95\%$. Small improvement in such statistics is visible if the magnitude limit of the selection is lowered and/or the redshift cut is increased, but the differences are within 1\%. We also examined the probabilities of objects belonging to the desired class (i.e.\ being a target) that XGB provides. For the test set, we found that the default limit of $P>0.5$ is optimal as far as the various metrics are concerned.

If instead the model performance is quantified from a cross-match of DESI DR1 (not used in training) with WAVES photometry, the metrics have slightly lower values than for the random test set, with completeness the most affected, deteriorating to $\sim92\%$. Last but not least, if we use a DESI DR1 subset totally disjoint from the entire original spec-z-cat (used for classifier training and validation), the completeness falls below 89\%, with purity at the level of $\sim91\%$. These latter numbers should, however, be considered lower limits for our classifier's performance, as such a ‘blind DESI test set' does not include any bright-end galaxies nor those of the fainter ones from the previous wide-angle surveys, such as 2dF, SDSS or GAMA. It is then very different from the actual WAVES flux-limited selection and therefore unlikely to be representative.

As with any supervised ML model, our classification method has its limitations related to the composition of the training set. Here in particular, our approach requires spectroscopic redshifts for the class labels in the calibration set, overlapping on the sky with input photometry for the color-based features we use. At present, within the general WAVES area, surveys with true redshifts do not provide complete coverage of the color-redshift space to fully represent WW targets on sufficiently wide areas. As a result, our training set is biased towards brighter, redder, and lower-redshift galaxies as compared to a general galaxy sample magnitude-limited to $Z<21.1$, from which targets will be selected. This leads to deterioration of the classifier's performance close to the flux and redshift cuts (i.e.\ at $z \gtrsim 0.18$ and $Z \gtrsim 19$). On the other hand, we do not see strong trends in classification metrics as a function of color, except for worsening at the reddest end, where most of the galaxies are anyway redshifted beyond the WW limit. Biases in the training set may also affect classification results once the trained model is applied to the overall inference sample. Predictions may need to be extrapolated in the regions of the color space that were not well covered by training. 

The results of our XGBoost classification are used together with other approaches for WAVES-Wide target selection to prepare the final list of galaxies to be observed by the 4MOST instrument. 

In a companion paper by Seulves et al. (in prep.), our approach is extended to deliver photo-zs and their probability density functions. Those predictions are then compared with what the three other photo-z methods deliver, and all four approaches are also combined via logarithmic pooling. As shown there, the photo-zs derived via our classifier perform similarly well as the other methods, and in particular give the smallest scatter with respect to the true values.

The probability of redshift being less than the desired redshift derived from the logarithmic pooling of photo-zs is used for the target selection. The performance of our supervised classification approach could be further improved if the training set with true redshift labels is extended by adding more of fainter galaxies within WAVES flux limits, to better map the color-redshift relation. Some such galaxies are available from DESI DR1, which had not been used to train our model but was employed to test its performance in this work.
In the long term, WAVES itself, and 4MOST more generally, will deliver spectroscopic redshifts for such objects, and in future work we will use the already available 4MOST Survey Programme Validation data for early tests. However, as the target selection for 4MOST observations is already fixed, these new redshifts will not serve to revise the target selection but rather to verify a posteriori the proposed approaches such as ours.

\section*{Acknowledgments}
{\footnotesize
We would like to thank Matthijs Mars and Julia Krajecka for their contributions at the early stages of this project.

GK \& MB are supported by the Polish National Science Center through grant no.\ 2020/38/E/ST9/00395 and by the Polish Ministry of Science and Higher Education under the agreement no.\ 2026/WK/06.
SB acknowledges funding by the Australian Research Council (ARC) Laureate Fellowship scheme (FL220100191).
ET acknowledges funding from the HTM (grant TK202), ETAg (grant PRG1006, PRG3034) and the EU Horizon Europe (EXCOSM, grant No.\ 101159513).

WAVES is a joint European-Australian project based around a spectroscopic campaign using the 4-metre Multi-Object Spectroscopic Telescope. The WAVES input catalogue is based on data taken from the European Southern Observatory’s VST and VISTA telescopes (ID 179.A-2004 and ID 177.A-3016). Complementary imaging of the WAVES regions is being obtained by a number of independent survey programmes, including GALEX MIS, VST, WISE, Herschel-ATLAS, and ASKAP, providing UV to radio coverage. WAVES is funded by the ARC (Australia) and the participating institutions. The WAVES website is \url{https://wavesurvey.org}.}

\bibliographystyle{mnras}
\bibliography{refs}

@ARTICLE{Tempel2020a,
       author = {{Tempel}, E. and {Norberg}, P. and {Tuvikene}, T. and {Bensby}, T. and {Chiappini}, C. and {Christlieb}, N. and {Cioni}, M. -R.~L. and {Comparat}, J. and {Davies}, L.~J.~M. and {Guiglion}, G. and {Koch}, A. and {Kordopatis}, G. and {Krumpe}, M. and {Loveday}, J. and {Merloni}, A. and {Micheva}, G. and {Minchev}, I. and {Roukema}, B.~F. and {Sorce}, J.~G. and {Starkenburg}, E. and {Storm}, J. and {Swann}, E. and {Thi}, W.~F. and {Traven}, G. and {de Jong}, R.~S.},
        title = "{Probabilistic fibre-to-target assignment algorithm for multi-object spectroscopic surveys}",
      journal = {\aap},
         year = 2020,
        month = mar,
       volume = {635},
          eid = {A101},
        pages = {A101},
          doi = {10.1051/0004-6361/201937228},
archivePrefix = {arXiv},
       eprint = {2001.09348},
 primaryClass = {astro-ph.IM},
       adsurl = {https://ui.adsabs.harvard.edu/abs/2020A&A...635A.101T}
}

@ARTICLE{WAVES,
       author = {{Driver}, S.~P. and {Liske}, J. and {Davies}, L.~J.~M. and {Robotham}, A.~S.~G. and {Baldry}, I.~K. and {Brown}, M.~J.~I. and {Cluver}, M. and {Kuijken}, K. and {Loveday}, J. and {McMahon}, R. and {Meyer}, M.~J. and {Norberg}, P. and {Owers}, M. and {Power}, C. and {Taylor}, E.~N. and {WAVES Team}},
        title = "{4MOST Consortium Survey 7: Wide-Area VISTA Extragalactic Survey (WAVES)}",
      journal = {The Messenger},
         year = 2019,
        month = mar,
       volume = {175},
        pages = {46-49},
          doi = {10.18727/0722-6691/5126},
archivePrefix = {arXiv},
       eprint = {1903.02473},
 primaryClass = {astro-ph.GA},
       adsurl = {https://ui.adsabs.harvard.edu/abs/2019Msngr.175...46D}
}

@ARTICLE{4MOST,
       author = {{de Jong}, R.~S. and {Agertz}, O. and {Berbel}, A.~A. and {Aird}, J. and {Alexander}, D.~A. and {Amarsi}, A. and {Anders}, F. and {Andrae}, R. and {Ansarinejad}, B. and {Ansorge}, W. and {Antilogus}, P. and {Anwand-Heerwart}, H. and {Arentsen}, A. and {Arnadottir}, A. and {Asplund}, M. and {Auger}, M. and {Azais}, N. and {Baade}, D. and {Baker}, G. and {Baker}, S. and {Balbinot}, E. and {Baldry}, I.~K. and {Banerji}, M. and {Barden}, S. and {Barklem}, P. and {Barth{\'e}l{\'e}my-Mazot}, E. and {Battistini}, C. and {Bauer}, S. and {Bell}, C.~P.~M. and {Bellido-Tirado}, O. and {Bellstedt}, S. and {Belokurov}, V. and {Bensby}, T. and {Bergemann}, M. and {Bestenlehner}, J.~M. and {Bielby}, R. and {Bilicki}, M. and {Blake}, C. and {Bland-Hawthorn}, J. and {Boeche}, C. and {Boland}, W. and {Boller}, T. and {Bongard}, S. and {Bongiorno}, A. and {Bonifacio}, P. and {Boudon}, D. and {Brooks}, D. and {Brown}, M.~J.~I. and {Brown}, R. and {Br{\"u}ggen}, M. and {Brynnel}, J. and {Brzeski}, J. and {Buchert}, T. and {Buschkamp}, P. and {Caffau}, E. and {Caillier}, P. and {Carrick}, J. and {Casagrande}, L. and {Case}, S. and {Casey}, A. and {Cesarini}, I. and {Cescutti}, G. and {Chapuis}, D. and {Chiappini}, C. and {Childress}, M. and {Christlieb}, N. and {Church}, R. and {Cioni}, M. -R.~L. and {Cluver}, M. and {Colless}, M. and {Collett}, T. and {Comparat}, J. and {Cooper}, A. and {Couch}, W. and {Courbin}, F. and {Croom}, S. and {Croton}, D. and {Daguis{\'e}}, E. and {Dalton}, G. and {Davies}, L.~J.~M. and {Davis}, T. and {de Laverny}, P. and {Deason}, A. and {Dionies}, F. and {Disseau}, K. and {Doel}, P. and {D{\"o}scher}, D. and {Driver}, S.~P. and {Dwelly}, T. and {Eckert}, D. and {Edge}, A. and {Edvardsson}, B. and {Youssoufi}, D.~E. and {Elhaddad}, A. and {Enke}, H. and {Erfanianfar}, G. and {Farrell}, T. and {Fechner}, T. and {Feiz}, C. and {Feltzing}, S. and {Ferreras}, I. and {Feuerstein}, D. and {Feuillet}, D. and {Finoguenov}, A. and {Ford}, D. and {Fotopoulou}, S. and {Fouesneau}, M. and {Frenk}, C. and {Frey}, S. and {Gaessler}, W. and {Geier}, S. and {Gentile Fusillo}, N. and {Gerhard}, O. and {Giannantonio}, T. and {Giannone}, D. and {Gibson}, B. and {Gillingham}, P. and {Gonz{\'a}lez-Fern{\'a}ndez}, C. and {Gonzalez-Solares}, E. and {Gottloeber}, S. and {Gould}, A. and {Grebel}, E.~K. and {Gueguen}, A. and {Guiglion}, G. and {Haehnelt}, M. and {Hahn}, T. and {Hansen}, C.~J. and {Hartman}, H. and {Hauptner}, K. and {Hawkins}, K. and {Haynes}, D. and {Haynes}, R. and {Heiter}, U. and {Helmi}, A. and {Aguayo}, C.~H. and {Hewett}, P. and {Hinton}, S. and {Hobbs}, D. and {Hoenig}, S. and {Hofman}, D. and {Hook}, I. and {Hopgood}, J. and {Hopkins}, A. and {Hourihane}, A. and {Howes}, L. and {Howlett}, C. and {Huet}, T. and {Irwin}, M. and {Iwert}, O. and {Jablonka}, P. and {Jahn}, T. and {Jahnke}, K. and {Jarno}, A. and {Jin}, S. and {Jofre}, P. and {Johl}, D. and {Jones}, D. and {J{\"o}nsson}, H. and {Jordan}, C. and {Karovicova}, I. and {Khalatyan}, A. and {Kelz}, A. and {Kennicutt}, R. and {King}, D. and {Kitaura}, F. and {Klar}, J. and {Klauser}, U. and {Kneib}, J. -P. and {Koch}, A. and {Koposov}, S. and {Kordopatis}, G. and {Korn}, A. and {Kosmalski}, J. and {Kotak}, R. and {Kovalev}, M. and {Kreckel}, K. and {Kripak}, Y. and {Krumpe}, M. and {Kuijken}, K. and {Kunder}, A. and {Kushniruk}, I. and {Lam}, M.~I. and {Lamer}, G. and {Laurent}, F. and {Lawrence}, J. and {Lehmitz}, M. and {Lemasle}, B. and {Lewis}, J. and {Li}, B. and {Lidman}, C. and {Lind}, K. and {Liske}, J. and {Lizon}, J. -L. and {Loveday}, J. and {Ludwig}, H. -G. and {McDermid}, R.~M. and {Maguire}, K. and {Mainieri}, V. and {Mali}, S. and {Mandel}, H. and {Mandel}, K. and {Mannering}, L. and {Martell}, S. and {Martinez Delgado}, D. and {Matijevic}, G. and {McGregor}, H. and {McMahon}, R. and {McMillan}, P. and {Mena}, O. and {Merloni}, A. and {Meyer}, M.~J. and {Michel}, C. and {Micheva}, G. and {Migniau}, J. -E. and {Minchev}, I. and {Monari}, G. and {Muller}, R. and {Murphy}, D. and {Muthukrishna}, D. and {Nandra}, K. and {Navarro}, R. and {Ness}, M. and {Nichani}, V. and {Nichol}, R. and {Nicklas}, H. and {Niederhofer}, F. and {Norberg}, P. and {Obreschkow}, D. and {Oliver}, S. and {Owers}, M. and {Pai}, N. and {Pankratow}, S. and {Parkinson}, D. and {Paschke}, J. and {Paterson}, R. and {Pecontal}, A. and {Parry}, I. and {Phillips}, D. and {Pillepich}, A. and {Pinard}, L. and {Pirard}, J. and {Piskunov}, N. and {Plank}, V. and {Pl{\"u}schke}, D. and {Pons}, E. and {Popesso}, P. and {Power}, C. and {Pragt}, J. and {Pramskiy}, A. and {Pryer}, D. and {Quattri}, M. and {Queiroz}, A.~B. d. A. and {Quirrenbach}, A. and {Rahurkar}, S. and {Raichoor}, A. and {Ramstedt}, S. and {Rau}, A. and {Recio-Blanco}, A. and {Reiss}, R. and {Renaud}, F. and {Revaz}, Y. and {Rhode}, P. and {Richard}, J. and {Richter}, A.~D. and {Rix}, H. -W. and {Robotham}, A.~S.~G. and {Roelfsema}, R. and {Romaniello}, M. and {Rosario}, D. and {Rothmaier}, F. and {Roukema}, B. and {Ruchti}, G. and {Rupprecht}, G. and {Rybizki}, J. and {Ryde}, N. and {Saar}, A. and {Sadler}, E. and {Sahl{\'e}n}, M. and {Salvato}, M. and {Sassolas}, B. and {Saunders}, W. and {Saviauk}, A. and {Sbordone}, L. and {Schmidt}, T. and {Schnurr}, O. and {Scholz}, R. -D. and {Schwope}, A. and {Seifert}, W. and {Shanks}, T. and {Sheinis}, A. and {Sivov}, T. and {Sk{\'u}lad{\'o}ttir}, {\'A}. and {Smartt}, S. and {Smedley}, S. and {Smith}, G. and {Smith}, R. and {Sorce}, J. and {Spitler}, L. and {Starkenburg}, E. and {Steinmetz}, M. and {Stilz}, I. and {Storm}, J. and {Sullivan}, M. and {Sutherland}, W. and {Swann}, E. and {Tamone}, A. and {Taylor}, E.~N. and {Teillon}, J. and {Tempel}, E. and {ter Horst}, R. and {Thi}, W. -F. and {Tolstoy}, E. and {Trager}, S. and {Traven}, G. and {Tremblay}, P. -E. and {Tresse}, L. and {Valentini}, M. and {van de Weygaert}, R. and {van den Ancker}, M. and {Veljanoski}, J. and {Venkatesan}, S. and {Wagner}, L. and {Wagner}, K. and {Walcher}, C.~J. and {Waller}, L. and {Walton}, N. and {Wang}, L. and {Winkler}, R. and {Wisotzki}, L. and {Worley}, C.~C. and {Worseck}, G. and {Xiang}, M. and {Xu}, W. and {Yong}, D. and {Zhao}, C. and {Zheng}, J. and {Zscheyge}, F. and {Zucker}, D.},
        title = "{4MOST: Project overview and information for the First Call for Proposals}",
      journal = {The Messenger},
         year = 2019,
        month = mar,
       volume = {175},
        pages = {3-11},
          doi = {10.18727/0722-6691/5117},
archivePrefix = {arXiv},
       eprint = {1903.02464},
 primaryClass = {astro-ph.IM},
       adsurl = {https://ui.adsabs.harvard.edu/abs/2019Msngr.175....3D}
}

@ARTICLE{KiDS,
       author = {{de Jong}, Jelte T.~A. and {Verdoes Kleijn}, Gijs A. and {Kuijken}, Konrad H. and {Valentijn}, Edwin A.},
        title = "{The Kilo-Degree Survey}",
      journal = {Experimental Astronomy},
         year = 2013,
        month = jan,
       volume = {35},
       number = {1-2},
        pages = {25-44},
          doi = {10.1007/s10686-012-9306-1},
archivePrefix = {arXiv},
       eprint = {1206.1254},
 primaryClass = {astro-ph.CO},
       adsurl = {https://ui.adsabs.harvard.edu/abs/2013ExA....35...25D}
}

@ARTICLE{VIKING,
       author = {{Edge}, A. and {Sutherland}, W. and {Kuijken}, K. and {Driver}, S. and {McMahon}, R. and {Eales}, S. and {Emerson}, J.~P.},
        title = "{The VISTA Kilo-degree Infrared Galaxy (VIKING) Survey: Bridging the Gap between Low and High Redshift}",
      journal = {The Messenger},
         year = 2013,
        month = dec,
       volume = {154},
        pages = {32-34},
       adsurl = {https://ui.adsabs.harvard.edu/abs/2013Msngr.154...32E}
}

@ARTICLE{Bellstedt20,
       author = {{Bellstedt}, Sabine and {Driver}, Simon P. and {Robotham}, Aaron S.~G. and {Davies}, Luke J.~M. and {Bogue}, Kamran R.~J. and {Cook}, Robin H.~W. and {Hashemizadeh}, Abdolhosein and {Koushan}, Soheil and {Taylor}, Edward N. and {Thorne}, Jessica E. and {Turner}, Ryan J. and {Wright}, Angus H.},
        title = "{Galaxy And Mass Assembly (GAMA): assimilation of KiDS into the GAMA database}",
      journal = {\mnras},
         year = 2020,
        month = aug,
       volume = {496},
       number = {3},
        pages = {3235-3256},
          doi = {10.1093/mnras/staa1466},
archivePrefix = {arXiv},
       eprint = {2005.11215},
 primaryClass = {astro-ph.GA},
       adsurl = {https://ui.adsabs.harvard.edu/abs/2020MNRAS.496.3235B}
}

@ARTICLE{Cook24,
       author = {{Cook}, Todd L. and {Bandi}, Behnood and {Philipsborn}, Sam and {Loveday}, Jon and {Bellstedt}, Sabine and {Driver}, Simon P. and {Robotham}, Aaron S.~G. and {Bilicki}, Maciej and {Kaur}, Gursharanjit and {Tempel}, Elmo and {Baldry}, Ivan and {Gruen}, Daniel and {Longhetti}, Marcella and {Iovino}, Angela and {Holwerda}, Benne W. and {Demarco}, Ricardo},
        title = "{Wide Area VISTA Extra-galactic Survey (WAVES): unsupervised star-galaxy separation on the WAVES-Wide photometric input catalogue using UMAP and HDBSCAN}",
      journal = {\mnras},
         year = 2024,
        month = dec,
       volume = {535},
       number = {3},
        pages = {2129-2148},
          doi = {10.1093/mnras/stae2389},
archivePrefix = {arXiv},
       eprint = {2406.11611},
 primaryClass = {astro-ph.GA},
       adsurl = {https://ui.adsabs.harvard.edu/abs/2024MNRAS.535.2129C}
}

@ARTICLE{VIPERS,
       author = {{Guzzo}, L. and {Scodeggio}, M. and {Garilli}, B. and {Granett}, B.~R. and {Fritz}, A. and {Abbas}, U. and {Adami}, C. and {Arnouts}, S. and {Bel}, J. and {Bolzonella}, M. and {Bottini}, D. and {Branchini}, E. and {Cappi}, A. and {Coupon}, J. and {Cucciati}, O. and {Davidzon}, I. and {De Lucia}, G. and {de la Torre}, S. and {Franzetti}, P. and {Fumana}, M. and {Hudelot}, P. and {Ilbert}, O. and {Iovino}, A. and {Krywult}, J. and {Le Brun}, V. and {Le F{\`e}vre}, O. and {Maccagni}, D. and {Ma{\l}ek}, K. and {Marulli}, F. and {McCracken}, H.~J. and {Paioro}, L. and {Peacock}, J.~A. and {Polletta}, M. and {Pollo}, A. and {Schlagenhaufer}, H. and {Tasca}, L.~A.~M. and {Tojeiro}, R. and {Vergani}, D. and {Zamorani}, G. and {Zanichelli}, A. and {Burden}, A. and {Di Porto}, C. and {Marchetti}, A. and {Marinoni}, C. and {Mellier}, Y. and {Moscardini}, L. and {Nichol}, R.~C. and {Percival}, W.~J. and {Phleps}, S. and {Wolk}, M.},
        title = "{The VIMOS Public Extragalactic Redshift Survey (VIPERS). An unprecedented view of galaxies and large-scale structure at 0.5 < z < 1.2}",
      journal = {\aap},
         year = 2014,
        month = jun,
       volume = {566},
          eid = {A108},
        pages = {A108},
          doi = {10.1051/0004-6361/201321489},
archivePrefix = {arXiv},
       eprint = {1303.2623},
 primaryClass = {astro-ph.CO},
       adsurl = {https://ui.adsabs.harvard.edu/abs/2014A&A...566A.108G}
}

@ARTICLE{DESI,
       author = {{DESI Collaboration} and {Aghamousa}, Amir and {Aguilar}, Jessica and {Ahlen}, Steve and {Alam}, Shadab and {Allen}, Lori E. and {Allende Prieto}, Carlos and {Annis}, James and {Bailey}, Stephen and {Balland}, Christophe and {Ballester}, Otger and {Baltay}, Charles and {Beaufore}, Lucas and {Bebek}, Chris and {Beers}, Timothy C. and {Bell}, Eric F. and {Bernal}, Jos{\'e} Luis and {Besuner}, Robert and {Beutler}, Florian and {Blake}, Chris and {Bleuler}, Hannes and {Blomqvist}, Michael and {Blum}, Robert and {Bolton}, Adam S. and {Briceno}, Cesar and {Brooks}, David and {Brownstein}, Joel R. and {Buckley-Geer}, Elizabeth and {Burden}, Angela and {Burtin}, Etienne and {Busca}, Nicolas G. and {Cahn}, Robert N. and {Cai}, Yan-Chuan and {Cardiel-Sas}, Laia and {Carlberg}, Raymond G. and {Carton}, Pierre-Henri and {Casas}, Ricard and {Castander}, Francisco J. and {Cervantes-Cota}, Jorge L. and {Claybaugh}, Todd M. and {Close}, Madeline and {Coker}, Carl T. and {Cole}, Shaun and {Comparat}, Johan and {Cooper}, Andrew P. and {Cousinou}, M. -C. and {Crocce}, Martin and {Cuby}, Jean-Gabriel and {Cunningham}, Daniel P. and {Davis}, Tamara M. and {Dawson}, Kyle S. and {de la Macorra}, Axel and {De Vicente}, Juan and {Delubac}, Timoth{\'e}e and {Derwent}, Mark and {Dey}, Arjun and {Dhungana}, Govinda and {Ding}, Zhejie and {Doel}, Peter and {Duan}, Yutong T. and {Ealet}, Anne and {Edelstein}, Jerry and {Eftekharzadeh}, Sarah and {Eisenstein}, Daniel J. and {Elliott}, Ann and {Escoffier}, St{\'e}phanie and {Evatt}, Matthew and {Fagrelius}, Parker and {Fan}, Xiaohui and {Fanning}, Kevin and {Farahi}, Arya and {Farihi}, Jay and {Favole}, Ginevra and {Feng}, Yu and {Fernandez}, Enrique and {Findlay}, Joseph R. and {Finkbeiner}, Douglas P. and {Fitzpatrick}, Michael J. and {Flaugher}, Brenna and {Flender}, Samuel and {Font-Ribera}, Andreu and {Forero-Romero}, Jaime E. and {Fosalba}, Pablo and {Frenk}, Carlos S. and {Fumagalli}, Michele and {Gaensicke}, Boris T. and {Gallo}, Giuseppe and {Garcia-Bellido}, Juan and {Gaztanaga}, Enrique and {Pietro Gentile Fusillo}, Nicola and {Gerard}, Terry and {Gershkovich}, Irena and {Giannantonio}, Tommaso and {Gillet}, Denis and {Gonzalez-de-Rivera}, Guillermo and {Gonzalez-Perez}, Violeta and {Gott}, Shelby and {Graur}, Or and {Gutierrez}, Gaston and {Guy}, Julien and {Habib}, Salman and {Heetderks}, Henry and {Heetderks}, Ian and {Heitmann}, Katrin and {Hellwing}, Wojciech A. and {Herrera}, David A. and {Ho}, Shirley and {Holland}, Stephen and {Honscheid}, Klaus and {Huff}, Eric and {Hutchinson}, Timothy A. and {Huterer}, Dragan and {Hwang}, Ho Seong and {Illa Laguna}, Joseph Maria and {Ishikawa}, Yuzo and {Jacobs}, Dianna and {Jeffrey}, Niall and {Jelinsky}, Patrick and {Jennings}, Elise and {Jiang}, Linhua and {Jimenez}, Jorge and {Johnson}, Jennifer and {Joyce}, Richard and {Jullo}, Eric and {Juneau}, St{\'e}phanie and {Kama}, Sami and {Karcher}, Armin and {Karkar}, Sonia and {Kehoe}, Robert and {Kennamer}, Noble and {Kent}, Stephen and {Kilbinger}, Martin and {Kim}, Alex G. and {Kirkby}, David and {Kisner}, Theodore and {Kitanidis}, Ellie and {Kneib}, Jean-Paul and {Koposov}, Sergey and {Kovacs}, Eve and {Koyama}, Kazuya and {Kremin}, Anthony and {Kron}, Richard and {Kronig}, Luzius and {Kueter-Young}, Andrea and {Lacey}, Cedric G. and {Lafever}, Robin and {Lahav}, Ofer and {Lambert}, Andrew and {Lampton}, Michael and {Landriau}, Martin and {Lang}, Dustin and {Lauer}, Tod R. and {Le Goff}, Jean-Marc and {Le Guillou}, Laurent and {Le Van Suu}, Auguste and {Lee}, Jae Hyeon and {Lee}, Su-Jeong and {Leitner}, Daniela and {Lesser}, Michael and {Levi}, Michael E. and {L'Huillier}, Benjamin and {Li}, Baojiu and {Liang}, Ming and {Lin}, Huan and {Linder}, Eric and {Loebman}, Sarah R. and {Luki{\'c}}, Zarija and {Ma}, Jun and {MacCrann}, Niall and {Magneville}, Christophe and {Makarem}, Laleh and {Manera}, Marc and {Manser}, Christopher J. and {Marshall}, Robert and {Martini}, Paul and {Massey}, Richard and {Matheson}, Thomas and {McCauley}, Jeremy and {McDonald}, Patrick and {McGreer}, Ian D. and {Meisner}, Aaron and {Metcalfe}, Nigel and {Miller}, Timothy N. and {Miquel}, Ramon and {Moustakas}, John and {Myers}, Adam and {Naik}, Milind and {Newman}, Jeffrey A. and {Nichol}, Robert C. and {Nicola}, Andrina and {Nicolati da Costa}, Luiz and {Nie}, Jundan and {Niz}, Gustavo and {Norberg}, Peder and {Nord}, Brian and {Norman}, Dara and {Nugent}, Peter and {O'Brien}, Thomas and {Oh}, Minji and {Olsen}, Knut A.~G. and {Padilla}, Cristobal and {Padmanabhan}, Hamsa and {Padmanabhan}, Nikhil and {Palanque-Delabrouille}, Nathalie and {Palmese}, Antonella and {Pappalardo}, Daniel and {P{\^a}ris}, Isabelle and {Park}, Changbom and {Patej}, Anna and {Peacock}, John A. and {Peiris}, Hiranya V. and {Peng}, Xiyan and {Percival}, Will J. and {Perruchot}, Sandrine and {Pieri}, Matthew M. and {Pogge}, Richard and {Pollack}, Jennifer E. and {Poppett}, Claire and {Prada}, Francisco and {Prakash}, Abhishek and {Probst}, Ronald G. and {Rabinowitz}, David and {Raichoor}, Anand and {Ree}, Chang Hee and {Refregier}, Alexandre and {Regal}, Xavier and {Reid}, Beth and {Reil}, Kevin and {Rezaie}, Mehdi and {Rockosi}, Constance M. and {Roe}, Natalie and {Ronayette}, Samuel and {Roodman}, Aaron and {Ross}, Ashley J. and {Ross}, Nicholas P. and {Rossi}, Graziano and {Rozo}, Eduardo and {Ruhlmann-Kleider}, Vanina and {Rykoff}, Eli S. and {Sabiu}, Cristiano and {Samushia}, Lado and {Sanchez}, Eusebio and {Sanchez}, Javier and {Schlegel}, David J. and {Schneider}, Michael and {Schubnell}, Michael and {Secroun}, Aur{\'e}lia and {Seljak}, Uros and {Seo}, Hee-Jong and {Serrano}, Santiago and {Shafieloo}, Arman and {Shan}, Huanyuan and {Sharples}, Ray and {Sholl}, Michael J. and {Shourt}, William V. and {Silber}, Joseph H. and {Silva}, David R. and {Sirk}, Martin M. and {Slosar}, Anze and {Smith}, Alex and {Smoot}, George F. and {Som}, Debopam and {Song}, Yong-Seon and {Sprayberry}, David and {Staten}, Ryan and {Stefanik}, Andy and {Tarle}, Gregory and {Sien Tie}, Suk and {Tinker}, Jeremy L. and {Tojeiro}, Rita and {Valdes}, Francisco and {Valenzuela}, Octavio and {Valluri}, Monica and {Vargas-Magana}, Mariana and {Verde}, Licia and {Walker}, Alistair R. and {Wang}, Jiali and {Wang}, Yuting and {Weaver}, Benjamin A. and {Weaverdyck}, Curtis and {Wechsler}, Risa H. and {Weinberg}, David H. and {White}, Martin and {Yang}, Qian and {Yeche}, Christophe and {Zhang}, Tianmeng and {Zhao}, Gong-Bo and {Zheng}, Yi and {Zhou}, Xu and {Zhou}, Zhimin and {Zhu}, Yaling and {Zou}, Hu and {Zu}, Ying},
        title = "{The DESI Experiment Part I: Science,Targeting, and Survey Design}",
      journal = {arXiv e-prints},
         year = 2016,
        month = oct,
          eid = {arXiv:1611.00036},
        pages = {arXiv:1611.00036},
          doi = {10.48550/arXiv.1611.00036},
archivePrefix = {arXiv},
       eprint = {1611.00036},
 primaryClass = {astro-ph.IM},
       adsurl = {https://ui.adsabs.harvard.edu/abs/2016arXiv161100036D}
}

@ARTICLE{SDSS-III,
       author = {{Eisenstein}, Daniel J. and {Weinberg}, David H. and {Agol}, Eric and {Aihara}, Hiroaki and {Allende Prieto}, Carlos and {Anderson}, Scott F. and {Arns}, James A. and {Aubourg}, {\'E}ric and {Bailey}, Stephen and {Balbinot}, Eduardo and {Barkhouser}, Robert and {Beers}, Timothy C. and {Berlind}, Andreas A. and {Bickerton}, Steven J. and {Bizyaev}, Dmitry and {Blanton}, Michael R. and {Bochanski}, John J. and {Bolton}, Adam S. and {Bosman}, Casey T. and {Bovy}, Jo and {Brandt}, W.~N. and {Breslauer}, Ben and {Brewington}, Howard J. and {Brinkmann}, J. and {Brown}, Peter J. and {Brownstein}, Joel R. and {Burger}, Dan and {Busca}, Nicolas G. and {Campbell}, Heather and {Cargile}, Phillip A. and {Carithers}, William C. and {Carlberg}, Joleen K. and {Carr}, Michael A. and {Chang}, Liang and {Chen}, Yanmei and {Chiappini}, Cristina and {Comparat}, Johan and {Connolly}, Natalia and {Cortes}, Marina and {Croft}, Rupert A.~C. and {Cunha}, Katia and {da Costa}, Luiz N. and {Davenport}, James R.~A. and {Dawson}, Kyle and {De Lee}, Nathan and {Porto de Mello}, Gustavo F. and {de Simoni}, Fernando and {Dean}, Janice and {Dhital}, Saurav and {Ealet}, Anne and {Ebelke}, Garrett L. and {Edmondson}, Edward M. and {Eiting}, Jacob M. and {Escoffier}, Stephanie and {Esposito}, Massimiliano and {Evans}, Michael L. and {Fan}, Xiaohui and {Femen{\'\i}a Castell{\'a}}, Bruno and {Dutra Ferreira}, Leticia and {Fitzgerald}, Greg and {Fleming}, Scott W. and {Font-Ribera}, Andreu and {Ford}, Eric B. and {Frinchaboy}, Peter M. and {Garc{\'\i}a P{\'e}rez}, Ana Elia and {Gaudi}, B. Scott and {Ge}, Jian and {Ghezzi}, Luan and {Gillespie}, Bruce A. and {Gilmore}, G. and {Girardi}, L{\'e}o and {Gott}, J. Richard and {Gould}, Andrew and {Grebel}, Eva K. and {Gunn}, James E. and {Hamilton}, Jean-Christophe and {Harding}, Paul and {Harris}, David W. and {Hawley}, Suzanne L. and {Hearty}, Frederick R. and {Hennawi}, Joseph F. and {Gonz{\'a}lez Hern{\'a}ndez}, Jonay I. and {Ho}, Shirley and {Hogg}, David W. and {Holtzman}, Jon A. and {Honscheid}, Klaus and {Inada}, Naohisa and {Ivans}, Inese I. and {Jiang}, Linhua and {Jiang}, Peng and {Johnson}, Jennifer A. and {Jordan}, Cathy and {Jordan}, Wendell P. and {Kauffmann}, Guinevere and {Kazin}, Eyal and {Kirkby}, David and {Klaene}, Mark A. and {Knapp}, G.~R. and {Kneib}, Jean-Paul and {Kochanek}, C.~S. and {Koesterke}, Lars and {Kollmeier}, Juna A. and {Kron}, Richard G. and {Lampeitl}, Hubert and {Lang}, Dustin and {Lawler}, James E. and {Le Goff}, Jean-Marc and {Lee}, Brian L. and {Lee}, Young Sun and {Leisenring}, Jarron M. and {Lin}, Yen-Ting and {Liu}, Jian and {Long}, Daniel C. and {Loomis}, Craig P. and {Lucatello}, Sara and {Lundgren}, Britt and {Lupton}, Robert H. and {Ma}, Bo and {Ma}, Zhibo and {MacDonald}, Nicholas and {Mack}, Claude and {Mahadevan}, Suvrath and {Maia}, Marcio A.~G. and {Majewski}, Steven R. and {Makler}, Martin and {Malanushenko}, Elena and {Malanushenko}, Viktor and {Mandelbaum}, Rachel and {Maraston}, Claudia and {Margala}, Daniel and {Maseman}, Paul and {Masters}, Karen L. and {McBride}, Cameron K. and {McDonald}, Patrick and {McGreer}, Ian D. and {McMahon}, Richard G. and {Mena Requejo}, Olga and {M{\'e}nard}, Brice and {Miralda-Escud{\'e}}, Jordi and {Morrison}, Heather L. and {Mullally}, Fergal and {Muna}, Demitri and {Murayama}, Hitoshi and {Myers}, Adam D. and {Naugle}, Tracy and {Neto}, Angelo Fausti and {Nguyen}, Duy Cuong and {Nichol}, Robert C. and {Nidever}, David L. and {O'Connell}, Robert W. and {Ogando}, Ricardo L.~C. and {Olmstead}, Matthew D. and {Oravetz}, Daniel J. and {Padmanabhan}, Nikhil and {Paegert}, Martin and {Palanque-Delabrouille}, Nathalie and {Pan}, Kaike and {Pandey}, Parul and {Parejko}, John K. and {P{\^a}ris}, Isabelle and {Pellegrini}, Paulo and {Pepper}, Joshua and {Percival}, Will J. and {Petitjean}, Patrick and {Pfaffenberger}, Robert and {Pforr}, Janine and {Phleps}, Stefanie and {Pichon}, Christophe and {Pieri}, Matthew M. and {Prada}, Francisco and {Price-Whelan}, Adrian M. and {Raddick}, M. Jordan and {Ramos}, Beatriz H.~F. and {Reid}, I. Neill and {Reyle}, Celine and {Rich}, James and {Richards}, Gordon T. and {Rieke}, George H. and {Rieke}, Marcia J. and {Rix}, Hans-Walter and {Robin}, Annie C. and {Rocha-Pinto}, Helio J. and {Rockosi}, Constance M. and {Roe}, Natalie A. and {Rollinde}, Emmanuel and {Ross}, Ashley J. and {Ross}, Nicholas P. and {Rossetto}, Bruno and {S{\'a}nchez}, Ariel G. and {Santiago}, Basilio and {Sayres}, Conor and {Schiavon}, Ricardo and {Schlegel}, David J. and {Schlesinger}, Katharine J. and {Schmidt}, Sarah J. and {Schneider}, Donald P. and {Sellgren}, Kris and {Shelden}, Alaina and {Sheldon}, Erin and {Shetrone}, Matthew and {Shu}, Yiping and {Silverman}, John D. and {Simmerer}, Jennifer and {Simmons}, Audrey E. and {Sivarani}, Thirupathi and {Skrutskie}, M.~F. and {Slosar}, An{\v{z}}e and {Smee}, Stephen and {Smith}, Verne V. and {Snedden}, Stephanie A. and {Stassun}, Keivan G. and {Steele}, Oliver and {Steinmetz}, Matthias and {Stockett}, Mark H. and {Stollberg}, Todd and {Strauss}, Michael A. and {Szalay}, Alexander S. and {Tanaka}, Masayuki and {Thakar}, Aniruddha R. and {Thomas}, Daniel and {Tinker}, Jeremy L. and {Tofflemire}, Benjamin M. and {Tojeiro}, Rita and {Tremonti}, Christy A. and {Vargas Maga{\~n}a}, Mariana and {Verde}, Licia and {Vogt}, Nicole P. and {Wake}, David A. and {Wan}, Xiaoke and {Wang}, Ji and {Weaver}, Benjamin A. and {White}, Martin and {White}, Simon D.~M. and {Wilson}, John C. and {Wisniewski}, John P. and {Wood-Vasey}, W. Michael and {Yanny}, Brian and {Yasuda}, Naoki and {Y{\`e}che}, Christophe and {York}, Donald G. and {Young}, Erick and {Zasowski}, Gail and {Zehavi}, Idit and {Zhao}, Bo},
        title = "{SDSS-III: Massive Spectroscopic Surveys of the Distant Universe, the Milky Way, and Extra-Solar Planetary Systems}",
      journal = {\aj},
         year = 2011,
        month = sep,
       volume = {142},
       number = {3},
          eid = {72},
        pages = {72},
          doi = {10.1088/0004-6256/142/3/72},
archivePrefix = {arXiv},
       eprint = {1101.1529},
 primaryClass = {astro-ph.IM},
       adsurl = {https://ui.adsabs.harvard.edu/abs/2011AJ....142...72E}
}

@ARTICLE{SDSS-IV-LRG,
       author = {{Prakash}, Abhishek and {Licquia}, Timothy C. and {Newman}, Jeffrey A. and {Ross}, Ashley J. and {Myers}, Adam D. and {Dawson}, Kyle S. and {Kneib}, Jean-Paul and {Percival}, Will J. and {Bautista}, Julian E. and {Comparat}, Johan and {Tinker}, Jeremy L. and {Schlegel}, David J. and {Tojeiro}, Rita and {Ho}, Shirley and {Lang}, Dustin and {Rao}, Sandhya M. and {McBride}, Cameron K. and {Ben Zhu}, Guangtun and {Brownstein}, Joel R. and {Bailey}, Stephen and {Bolton}, Adam S. and {Delubac}, Timoth{\'e}e and {Mariappan}, Vivek and {Blanton}, Michael R. and {Reid}, Beth and {Schneider}, Donald P. and {Seo}, Hee-Jong and {Carnero Rosell}, Aurelio and {Prada}, Francisco},
        title = "{The SDSS-IV Extended Baryon Oscillation Spectroscopic Survey: Luminous Red Galaxy Target Selection}",
      journal = {\apjs},
         year = 2016,
        month = jun,
       volume = {224},
       number = {2},
          eid = {34},
        pages = {34},
          doi = {10.3847/0067-0049/224/2/34},
archivePrefix = {arXiv},
       eprint = {1508.04478},
 primaryClass = {astro-ph.CO},
       adsurl = {https://ui.adsabs.harvard.edu/abs/2016ApJS..224...34P}
}

@inproceedings{chen2016xgboost,
author = {Chen, Tianqi and Guestrin, Carlos},
title = {XGBoost: A Scalable Tree Boosting System},
year = {2016},
isbn = {9781450342322},
publisher = {Association for Computing Machinery},
address = {New York, NY, USA},
url = {https://doi.org/10.1145/2939672.2939785},
doi = {10.1145/2939672.2939785},
booktitle = {Proceedings of the 22nd ACM SIGKDD International Conference on Knowledge Discovery and Data Mining},
pages = {785–794},
numpages = {10},
location = {San Francisco, California, USA},
series = {KDD '16}
}

@ARTICLE{2011JMLR...12.2825P,
       author = {{Pedregosa}, Fabian and {Varoquaux}, Ga{\"e}l and {Gramfort}, Alexandre and {Michel}, Vincent and {Thirion}, Bertrand and {Grisel}, Olivier and {Blondel}, Mathieu and {M{\"u}ller}, Andreas and {Nothman}, Joel and {Louppe}, Gilles and {Prettenhofer}, Peter and {Weiss}, Ron and {Dubourg}, Vincent and {Vanderplas}, Jake and {Passos}, Alexandre and {Cournapeau}, David and {Brucher}, Matthieu and {Perrot}, Matthieu and {Duchesnay}, {\'E}douard},
        title = "{Scikit-learn: Machine Learning in Python}",
      journal = {Journal of Machine Learning Research},
         year = 2011,
        month = oct,
       volume = {12},
        pages = {2825-2830},
          doi = {10.48550/arXiv.1201.0490},
archivePrefix = {arXiv},
       eprint = {1201.0490},
 primaryClass = {cs.LG},
       adsurl = {https://ui.adsabs.harvard.edu/abs/2011JMLR...12.2825P}
}

@ARTICLE{NG22,
       author = {{Newman}, Jeffrey A. and {Gruen}, Daniel},
        title = "{Photometric Redshifts for Next-Generation Surveys}",
      journal = {\araa},
         year = 2022,
        month = aug,
       volume = {60},
        pages = {363-414},
          doi = {10.1146/annurev-astro-032122-014611},
archivePrefix = {arXiv},
       eprint = {2206.13633},
 primaryClass = {astro-ph.CO},
       adsurl = {https://ui.adsabs.harvard.edu/abs/2022ARA&A..60..363N}
}

@ARTICLE{Bilicki21,
       author = {{Bilicki}, M. and {Dvornik}, A. and {Hoekstra}, H. and {Wright}, A.~H. and {Chisari}, N.~E. and {Vakili}, M. and {Asgari}, M. and {Giblin}, B. and {Heymans}, C. and {Hildebrandt}, H. and {Holwerda}, B.~W. and {Hopkins}, A. and {Johnston}, H. and {Kannawadi}, A. and {Kuijken}, K. and {Nakoneczny}, S.~J. and {Shan}, H.~Y. and {Sonnenfeld}, A. and {Valentijn}, E.},
        title = "{Bright galaxy sample in the Kilo-Degree Survey Data Release 4. Selection, photometric redshifts, and physical properties}",
      journal = {\aap},
         year = 2021,
        month = sep,
       volume = {653},
          eid = {A82},
        pages = {A82},
          doi = {10.1051/0004-6361/202140352},
archivePrefix = {arXiv},
       eprint = {2101.06010},
 primaryClass = {astro-ph.GA},
       adsurl = {https://ui.adsabs.harvard.edu/abs/2021A&A...653A..82B}
}

@ARTICLE{Porredon21,
       author = {{Porredon}, A. and {Crocce}, M. and {Fosalba}, P. and {Elvin-Poole}, J. and {Carnero Rosell}, A. and {Cawthon}, R. and {Eifler}, T.~F. and {Fang}, X. and {Ferrero}, I. and {Krause}, E. and {MacCrann}, N. and {Weaverdyck}, N. and {Abbott}, T.~M.~C. and {Aguena}, M. and {Allam}, S. and {Amon}, A. and {Avila}, S. and {Bacon}, D. and {Bertin}, E. and {Bhargava}, S. and {Bridle}, S.~L. and {Brooks}, D. and {Carrasco Kind}, M. and {Carretero}, J. and {Castander}, F.~J. and {Choi}, A. and {Costanzi}, M. and {da Costa}, L.~N. and {Pereira}, M.~E.~S. and {De Vicente}, J. and {Desai}, S. and {Diehl}, H.~T. and {Doel}, P. and {Drlica-Wagner}, A. and {Eckert}, K. and {Fert{\'e}}, A. and {Flaugher}, B. and {Frieman}, J. and {Garc{\'\i}a-Bellido}, J. and {Gaztanaga}, E. and {Gerdes}, D.~W. and {Giannantonio}, T. and {Gruen}, D. and {Gruendl}, R.~A. and {Gschwend}, J. and {Gutierrez}, G. and {Hartley}, W.~G. and {Hinton}, S.~R. and {Hollowood}, D.~L. and {Honscheid}, K. and {Hoyle}, B. and {James}, D.~J. and {Jarvis}, M. and {Kuehn}, K. and {Kuropatkin}, N. and {Maia}, M.~A.~G. and {Marshall}, J.~L. and {Menanteau}, F. and {Miquel}, R. and {Morgan}, R. and {Palmese}, A. and {Pandey}, S. and {Paz-Chinch{\'o}n}, F. and {Plazas}, A.~A. and {Rodriguez-Monroy}, M. and {Roodman}, A. and {Samuroff}, S. and {Sanchez}, E. and {Scarpine}, V. and {Serrano}, S. and {Sevilla-Noarbe}, I. and {Smith}, M. and {Soares-Santos}, M. and {Suchyta}, E. and {Swanson}, M.~E.~C. and {Tarle}, G. and {To}, C. and {Varga}, T.~N. and {Weller}, J. and {Wilkinson}, R.~D. and {DES Collaboration}},
        title = "{Dark Energy Survey Year 3 results: Optimizing the lens sample in a combined galaxy clustering and galaxy-galaxy lensing analysis}",
      journal = {\prd},
         year = 2021,
        month = feb,
       volume = {103},
       number = {4},
          eid = {043503},
        pages = {043503},
          doi = {10.1103/PhysRevD.103.043503},
archivePrefix = {arXiv},
       eprint = {2011.03411},
 primaryClass = {astro-ph.CO},
       adsurl = {https://ui.adsabs.harvard.edu/abs/2021PhRvD.103d3503P}
}

@ARTICLE{Treyer24,
       author = {{Treyer}, M. and {Ait Ouahmed}, R. and {Pasquet}, J. and {Arnouts}, S. and {Bertin}, E. and {Fouchez}, D.},
        title = "{CNN photometric redshifts in the SDSS at r {\ensuremath{\leq}} 20}",
      journal = {\mnras},
         year = 2024,
        month = jan,
       volume = {527},
       number = {1},
        pages = {651-671},
          doi = {10.1093/mnras/stad3171},
archivePrefix = {arXiv},
       eprint = {2310.02173},
 primaryClass = {astro-ph.CO},
       adsurl = {https://ui.adsabs.harvard.edu/abs/2024MNRAS.527..651T}
}

@ARTICLE{TOPz,
       author = {{Laur}, J. and {Tempel}, E. and {Tamm}, A. and {Kipper}, R. and {Liivam{\"a}gi}, L.~J. and {Hern{\'a}n-Caballero}, A. and {Muru}, M.~M. and {Chaves-Montero}, J. and {D{\'\i}az-Garc{\'\i}a}, L.~A. and {Turner}, S. and {Tuvikene}, T. and {Queiroz}, C. and {Bom}, C.~R. and {Fern{\'a}ndez-Ontiveros}, J.~A. and {Gonz{\'a}lez Delgado}, R.~M. and {Civera}, T. and {Abramo}, R. and {Alcaniz}, J. and {Ben{\'\i}tez}, N. and {Bonoli}, S. and {Carneiro}, S. and {Cenarro}, J. and {Crist{\'o}bal-Hornillos}, D. and {Dupke}, R. and {Ederoclite}, A. and {L{\'o}pez-Sanjuan}, C. and {Mar{\'\i}n-Franch}, A. and {de Oliveira}, C.~M. and {Moles}, M. and {Sodr{\'e}}, L. and {Taylor}, K. and {Varela}, J. and {V{\'a}zquez Rami{\'o}}, H.},
        title = "{TOPz: Photometric redshifts for J-PAS}",
      journal = {\aap},
         year = 2022,
        month = dec,
       volume = {668},
          eid = {A8},
        pages = {A8},
          doi = {10.1051/0004-6361/202243881},
archivePrefix = {arXiv},
       eprint = {2209.01040},
 primaryClass = {astro-ph.IM},
       adsurl = {https://ui.adsabs.harvard.edu/abs/2022A&A...668A...8L}
}

@ARTICLE{2013MNRAS.432.1483C,
       author = {{Carrasco Kind}, Matias and {Brunner}, Robert J.},
        title = "{TPZ: photometric redshift PDFs and ancillary information by using prediction trees and random forests}",
      journal = {\mnras},
         year = 2013,
        month = jun,
       volume = {432},
       number = {2},
        pages = {1483-1501},
          doi = {10.1093/mnras/stt574},
archivePrefix = {arXiv},
       eprint = {1303.7269},
 primaryClass = {astro-ph.CO},
       adsurl = {https://ui.adsabs.harvard.edu/abs/2013MNRAS.432.1483C}
}

@ARTICLE{2022MNRAS.512.3662D,
       author = {{Duncan}, Kenneth J.},
        title = "{All-purpose, all-sky photometric redshifts for the Legacy Imaging Surveys Data Release 8}",
      journal = {\mnras},
         year = 2022,
        month = may,
       volume = {512},
       number = {3},
        pages = {3662-3683},
          doi = {10.1093/mnras/stac608},
archivePrefix = {arXiv},
       eprint = {2203.01949},
 primaryClass = {astro-ph.GA},
       adsurl = {https://ui.adsabs.harvard.edu/abs/2022MNRAS.512.3662D}
}

@ARTICLE{2021MNRAS.500.1557B,
       author = {{Baldry}, Ivan K. and {Sullivan}, Tricia and {Rani}, Raffaele and {Turner}, Sebastian},
        title = "{Compact galaxies and the size-mass galaxy distribution from a colour-selected sample at 0.04 < z < 0.15 supplemented by ugrizYJHK photometric redshifts}",
      journal = {\mnras},
         year = 2021,
        month = nov,
       volume = {500},
       number = {2},
        pages = {1557-1574},
          doi = {10.1093/mnras/staa3327},
archivePrefix = {arXiv},
       eprint = {2008.09625},
 primaryClass = {astro-ph.GA},
       adsurl = {https://ui.adsabs.harvard.edu/abs/2021MNRAS.500.1557B}
}

@ARTICLE{Payerne2024,
       author = {{Payerne}, Constantin and {d'Assignies Doumerg}, William and {Y{\`e}che}, Christophe and {Ruhlmann-Kleider}, Vanina and {Raichoor}, Anand and {Lang}, Dusting and {Aguilar}, Jessica Nicole and {Ahlen}, Steven and {Arnouts}, St{\'e}phane and {Bianchi}, Davide and {Brooks}, David and {Claybaugh}, Todd and {Cole}, Shaun and {de la Macorra}, Axel and {Dey}, Arjun and {Dey}, Biprateep and {Doel}, Peter and {Font-Ribera}, Andreu and {Forero-Romero}, Jaime E. and {Gontcho}, Satya Gontcho A. and {Gutierrez}, Gaston and {Gwyn}, Stephen and {Honscheid}, Klaus and {Juneau}, Stephanie and {Lambert}, Andrew and {Landriau}, Martin and {Guillou}, Laurent Le and {Levi}, Michael E. and {Magneville}, Christophe and {Manera}, Marc and {Meisner}, Aaron and {Miquel}, Ramon and {Moustakas}, John and {Newman}, Jeffrey A. and {Palanque-Delabrouille}, Nathalie and {Percival}, Will and {Picouet}, Vincent and {Prada}, Francisco and {P{\'e}rez-R{\`a}fols}, Ignasi and {Rossi}, Graziano and {Sanchez}, Eusebio and {Sawicki}, Marcin and {Schlegel}, David and {Schubnell}, Michael and {Sprayberry}, David and {Tarl{\'e}}, Gregory and {Weaver}, Benjamin A. and {Zou}, Hu},
        title = "{Selection of high-redshift Lyman-Break Galaxies from broadband and wide photometric surveys}",
      journal = {\jcap},
         year = 2025,
        month = may,
       volume = {2025},
       number = {5},
          eid = {031},
        pages = {031},
          doi = {10.1088/1475-7516/2025/05/031},
archivePrefix = {arXiv},
       eprint = {2410.08062},
 primaryClass = {astro-ph.CO},
       adsurl = {https://ui.adsabs.harvard.edu/abs/2025JCAP...05..031P}
}

@ARTICLE{Jalan2024,
       author = {{Jalan}, Priyanka and {Bilicki}, Maciej and {Hellwing}, Wojciech A. and {Wright}, Angus H. and {Dvornik}, Andrej and {Georgiou}, Christos and {Heymans}, Catherine and {Hildebrandt}, Hendrik and {Joudaki}, Shahab and {Kuijken}, Konrad and {Mahony}, Constance and {Nakoneczny}, Szymon Jan and {Radovich}, Mario and {van den Busch}, Jan Luca and {Yan}, Ziang and {Yoon}, Mijin},
        title = "{Enhancing photometric redshift catalogs through color-space analysis: Application to KiDS-bright galaxies}",
      journal = {\aap},
         year = 2024,
        month = dec,
       volume = {692},
          eid = {A177},
        pages = {A177},
          doi = {10.1051/0004-6361/202452424},
archivePrefix = {arXiv},
       eprint = {2409.14799},
 primaryClass = {astro-ph.GA},
       adsurl = {https://ui.adsabs.harvard.edu/abs/2024A&A...692A.177J}
}

@ARTICLE{2024A&A...686A.170W,
       author = {{Wright}, Angus H. and {Kuijken}, Konrad and {Hildebrandt}, Hendrik and {Radovich}, Mario and {Bilicki}, Maciej and {Dvornik}, Andrej and {Getman}, Fedor and {Heymans}, Catherine and {Hoekstra}, Henk and {Li}, Shun-Sheng and {Miller}, Lance and {Napolitano}, Nicola R. and {Xia}, Qianli and {Asgari}, Marika and {Brescia}, Massimo and {Buddelmeijer}, Hugo and {Burger}, Pierre and {Castignani}, Gianluca and {Cavuoti}, Stefano and {de Jong}, Jelte and {Edge}, Alastair and {Giblin}, Benjamin and {Giocoli}, Carlo and {Harnois-D{\'e}raps}, Joachim and {Jalan}, Priyanka and {Joachimi}, Benjamin and {John William}, Anjitha and {Joudaki}, Shahab and {Kannawadi}, Arun and {Kaur}, Gursharanjit and {La Barbera}, Francesco and {Linke}, Laila and {Mahony}, Constance and {Maturi}, Matteo and {Moscardini}, Lauro and {Nakoneczny}, Szymon J. and {Paolillo}, Maurizio and {Porth}, Lucas and {Puddu}, Emanuella and {Reischke}, Robert and {Schneider}, Peter and {Sereno}, Mauro and {Shan}, HuanYuan and {Sif{\'o}n}, Crist{\'o}bal and {St{\"o}lzner}, Benjamin and {Tr{\"o}ster}, Tilman and {Valentijn}, Edwin and {van den Busch}, Jan Luca and {Verdoes Kleijn}, Gijs and {Wittje}, Anna and {Yan}, Ziang and {Yao}, Ji and {Yoon}, Mijin and {Zhang}, Yun-Hao},
        title = "{The fifth data release of the Kilo Degree Survey: Multi-epoch optical/NIR imaging covering wide and legacy-calibration fields}",
      journal = {\aap},
         year = 2024,
        month = jun,
       volume = {686},
          eid = {A170},
        pages = {A170},
          doi = {10.1051/0004-6361/202346730},
       adsurl = {https://ui.adsabs.harvard.edu/abs/2024A&A...686A.170W}
}

@ARTICLE{Shark,
       author = {{Lagos}, Claudia del P. and {Tobar}, Rodrigo J. and {Robotham}, Aaron S.~G. and {Obreschkow}, Danail and {Mitchell}, Peter D. and {Power}, Chris and {Elahi}, Pascal J.},
        title = "{Shark: introducing an open source, free, and flexible semi-analytic model of galaxy formation}",
      journal = {\mnras},
         year = 2018,
        month = dec,
       volume = {481},
       number = {3},
        pages = {3573-3603},
          doi = {10.1093/mnras/sty2440},
archivePrefix = {arXiv},
       eprint = {1807.11180},
 primaryClass = {astro-ph.GA},
       adsurl = {https://ui.adsabs.harvard.edu/abs/2018MNRAS.481.3573L}
}

@ARTICLE{SURFS,
       author = {{Elahi}, Pascal J. and {Welker}, Charlotte and {Power}, Chris and {Lagos}, Claudia del P. and {Robotham}, Aaron S.~G. and {Ca{\~n}as}, Rodrigo and {Poulton}, Rhys},
        title = "{SURFS: Riding the waves with Synthetic UniveRses For Surveys}",
      journal = {\mnras},
         year = 2018,
        month = apr,
       volume = {475},
       number = {4},
        pages = {5338-5359},
          doi = {10.1093/mnras/sty061},
archivePrefix = {arXiv},
       eprint = {1712.01988},
 primaryClass = {astro-ph.GA},
       adsurl = {https://ui.adsabs.harvard.edu/abs/2018MNRAS.475.5338E}
}

@book{breiman1984classification,
  title={Classification and Regression Trees},
  author={Breiman, L. and Friedman, J. and Stone, C.J. and Olshen, R.A.},
  isbn={9780412048418},
  lccn={83019708},
  url={https://books.google.pl/books?id=JwQx-WOmSyQC},
  year={1984},
  publisher={Taylor \& Francis}
}

@article{breiman2001random,
  title={Random forests},
  author={Breiman, Leo},
  journal={Machine learning},
  volume={45},
  pages={5--32},
  year={2001},
  publisher={Springer}
}

@book{fix1951discriminatory,
  title={Discriminatory Analysis: Nonparametric Discrimination: Consistency Properties},
  author={Fix, E. and Hodges, J.L.},
  url={https://books.google.pl/books?id=4XwytAEACAAJ},
  year={1951},
  publisher={USAF School of Aviation Medicine}
}

@ARTICLE{Colless2001,
       author = {{Colless}, Matthew and {Dalton}, Gavin and {Maddox}, Steve and {Sutherland}, Will and {Norberg}, Peder and {Cole}, Shaun and {Bland-Hawthorn}, Joss and {Bridges}, Terry and {Cannon}, Russell and {Collins}, Chris and {Couch}, Warrick and {Cross}, Nicholas and {Deeley}, Kathryn and {De Propris}, Roberto and {Driver}, Simon P. and {Efstathiou}, George and {Ellis}, Richard S. and {Frenk}, Carlos S. and {Glazebrook}, Karl and {Jackson}, Carole and {Lahav}, Ofer and {Lewis}, Ian and {Lumsden}, Stuart and {Madgwick}, Darren and {Peacock}, John A. and {Peterson}, Bruce A. and {Price}, Ian and {Seaborne}, Mark and {Taylor}, Keith},
        title = "{The 2dF Galaxy Redshift Survey: spectra and redshifts}",
      journal = {\mnras},
         year = 2001,
        month = dec,
       volume = {328},
       number = {4},
        pages = {1039-1063},
          doi = {10.1046/j.1365-8711.2001.04902.x},
archivePrefix = {arXiv},
       eprint = {astro-ph/0106498},
 primaryClass = {astro-ph},
       adsurl = {https://ui.adsabs.harvard.edu/abs/2001MNRAS.328.1039C}
}

@ARTICLE{Baldry2010,
       author = {{Baldry}, I.~K. and {Robotham}, A.~S.~G. and {Hill}, D.~T. and {Driver}, S.~P. and {Liske}, J. and {Norberg}, P. and {Bamford}, S.~P. and {Hopkins}, A.~M. and {Loveday}, J. and {Peacock}, J.~A. and {Cameron}, E. and {Croom}, S.~M. and {Cross}, N.~J.~G. and {Doyle}, I.~F. and {Dye}, S. and {Frenk}, C.~S. and {Jones}, D.~H. and {van Kampen}, E. and {Kelvin}, L.~S. and {Nichol}, R.~C. and {Parkinson}, H.~R. and {Popescu}, C.~C. and {Prescott}, M. and {Sharp}, R.~G. and {Sutherland}, W.~J. and {Thomas}, D. and {Tuffs}, R.~J.},
        title = "{Galaxy And Mass Assembly (GAMA): the input catalogue and star-galaxy separation}",
      journal = {\mnras},
         year = 2010,
        month = may,
       volume = {404},
       number = {1},
        pages = {86-100},
          doi = {10.1111/j.1365-2966.2010.16282.x},
archivePrefix = {arXiv},
       eprint = {0910.5120},
 primaryClass = {astro-ph.CO},
       adsurl = {https://ui.adsabs.harvard.edu/abs/2010MNRAS.404...86B}
}

@ARTICLE{Blake2016,
       author = {{Blake}, Chris and {Amon}, Alexandra and {Childress}, Michael and {Erben}, Thomas and {Glazebrook}, Karl and {Harnois-Deraps}, Joachim and {Heymans}, Catherine and {Hildebrandt}, Hendrik and {Hinton}, Samuel R. and {Janssens}, Steven and {Johnson}, Andrew and {Joudaki}, Shahab and {Klaes}, Dominik and {Kuijken}, Konrad and {Lidman}, Chris and {Marin}, Felipe A. and {Parkinson}, David and {Poole}, Gregory B. and {Wolf}, Christian},
        title = "{The 2-degree Field Lensing Survey: design and clustering measurements}",
      journal = {\mnras},
         year = 2016,
        month = nov,
       volume = {462},
       number = {4},
        pages = {4240-4265},
          doi = {10.1093/mnras/stw1990},
archivePrefix = {arXiv},
       eprint = {1608.02668},
 primaryClass = {astro-ph.CO},
       adsurl = {https://ui.adsabs.harvard.edu/abs/2016MNRAS.462.4240B}
}

@ARTICLE{Damjanov2018,
       author = {{Damjanov}, Ivana and {Zahid}, H. Jabran and {Geller}, Margaret J. and {Fabricant}, Daniel G. and {Hwang}, Ho Seong},
        title = "{hCOSMOS: A Dense Spectroscopic Survey of r {\ensuremath{\leq}} 21.3 Galaxies in the COSMOS field}",
      journal = {\apjs},
         year = 2018,
        month = feb,
       volume = {234},
       number = {2},
          eid = {21},
        pages = {21},
          doi = {10.3847/1538-4365/aaa01c},
archivePrefix = {arXiv},
       eprint = {1710.01733},
 primaryClass = {astro-ph.GA},
       adsurl = {https://ui.adsabs.harvard.edu/abs/2018ApJS..234...21D}
}

@ARTICLE{c3trMasters2019,
       author = {{Masters}, Daniel C. and {Stern}, Daniel K. and {Cohen}, Judith G. and {Capak}, Peter L. and {Stanford}, S. Adam and {Hernitschek}, Nina and {Galametz}, Audrey and {Davidzon}, Iary and {Rhodes}, Jason D. and {Sanders}, Dave and {Mobasher}, Bahram and {Castander}, Francisco and {Pruett}, Kerianne and {Fotopoulou}, Sotiria},
        title = "{The Complete Calibration of the Color-Redshift Relation (C3R2) Survey: Analysis and Data Release 2}",
      journal = {\apj},
         year = 2019,
        month = jun,
       volume = {877},
       number = {2},
          eid = {81},
        pages = {81},
          doi = {10.3847/1538-4357/ab184d},
archivePrefix = {arXiv},
       eprint = {1904.06394},
 primaryClass = {astro-ph.GA},
       adsurl = {https://ui.adsabs.harvard.edu/abs/2019ApJ...877...81M}
}

@ARTICLE{devilsDavies2018,
       author = {{Davies}, L.~J.~M. and {Robotham}, A.~S.~G. and {Driver}, S.~P. and {Lagos}, C.~P. and {Cortese}, L. and {Mannering}, E. and {Foster}, C. and {Lidman}, C. and {Hashemizadeh}, A. and {Koushan}, S. and {O'Toole}, S. and {Baldry}, I.~K. and {Bilicki}, M. and {Bland-Hawthorn}, J. and {Bremer}, M.~N. and {Brown}, M.~J.~I. and {Bryant}, J.~J. and {Catinella}, B. and {Croom}, S.~M. and {Grootes}, M.~W. and {Holwerda}, B.~W. and {Jarvis}, M.~J. and {Maddox}, N. and {Meyer}, M. and {Moffett}, A.~J. and {Phillipps}, S. and {Taylor}, E.~N. and {Windhorst}, R.~A. and {Wolf}, C.},
        title = "{Deep Extragalactic VIsible Legacy Survey (DEVILS): motivation,design, and target catalogue}",
      journal = {\mnras},
         year = 2018,
        month = oct,
       volume = {480},
       number = {1},
        pages = {768-799},
          doi = {10.1093/mnras/sty1553},
archivePrefix = {arXiv},
       eprint = {1806.05808},
 primaryClass = {astro-ph.GA},
       adsurl = {https://ui.adsabs.harvard.edu/abs/2018MNRAS.480..768D}
}

@ARTICLE{desi2024,
       author = {{DESI Collaboration} and {Adame}, A.~G. and {Aguilar}, J. and {Ahlen}, S. and {Alam}, S. and {Aldering}, G. and {Alexander}, D.~M. and {Alfarsy}, R. and {Allende Prieto}, C. and {Alvarez}, M. and {Alves}, O. and {Anand}, A. and {Andrade-Oliveira}, F. and {Armengaud}, E. and {Asorey}, J. and {Avila}, S. and {Aviles}, A. and {Bailey}, S. and {Balaguera-Antol{\'\i}nez}, A. and {Ballester}, O. and {Baltay}, C. and {Bault}, A. and {Bautista}, J. and {Behera}, J. and {Beltran}, S.~F. and {BenZvi}, S. and {Beraldo e Silva}, L. and {Bermejo-Climent}, J.~R. and {Berti}, A. and {Besuner}, R. and {Beutler}, F. and {Bianchi}, D. and {Blake}, C. and {Blum}, R. and {Bolton}, A.~S. and {Brieden}, S. and {Brodzeller}, A. and {Brooks}, D. and {Brown}, Z. and {Buckley-Geer}, E. and {Burtin}, E. and {Cabayol-Garcia}, L. and {Cai}, Z. and {Canning}, R. and {Cardiel-Sas}, L. and {Carnero Rosell}, A. and {Castander}, F.~J. and {Cervantes-Cota}, J.~L. and {Chabanier}, S. and {Chaussidon}, E. and {Chaves-Montero}, J. and {Chen}, S. and {Chen}, X. and {Chuang}, C. and {Claybaugh}, T. and {Cole}, S. and {Cooper}, A.~P. and {Cuceu}, A. and {Davis}, T.~M. and {Dawson}, K. and {de Belsunce}, R. and {de la Cruz}, R. and {de la Macorra}, A. and {Della Costa}, J. and {de Mattia}, A. and {Demina}, R. and {Demirbozan}, U. and {DeRose}, J. and {Dey}, A. and {Dey}, B. and {Dhungana}, G. and {Ding}, J. and {Ding}, Z. and {Doel}, P. and {Doshi}, R. and {Douglass}, K. and {Edge}, A. and {Eftekharzadeh}, S. and {Eisenstein}, D.~J. and {Elliott}, A. and {Ereza}, J. and {Escoffier}, S. and {Fagrelius}, P. and {Fan}, X. and {Fanning}, K. and {Fawcett}, V.~A. and {Ferraro}, S. and {Flaugher}, B. and {Font-Ribera}, A. and {Forero-Romero}, J.~E. and {Forero-S{\'a}nchez}, D. and {Frenk}, C.~S. and {G{\"a}nsicke}, B.~T. and {Garc{\'\i}a}, L. {\'A}. and {Garc{\'\i}a-Bellido}, J. and {Garcia-Quintero}, C. and {Garrison}, L.~H. and {Gil-Mar{\'\i}n}, H. and {Golden-Marx}, J. and {Gontcho A Gontcho}, S. and {Gonzalez-Morales}, A.~X. and {Gonzalez-Perez}, V. and {Gordon}, C. and {Graur}, O. and {Green}, D. and {Gruen}, D. and {Guy}, J. and {Hadzhiyska}, B. and {Hahn}, C. and {Han}, J.~J. and {Hanif}, M.~M.~S. and {Herrera-Alcantar}, H.~K. and {Honscheid}, K. and {Hou}, J. and {Howlett}, C. and {Huterer}, D. and {Ir{\v{s}}i{\v{c}}}, V. and {Ishak}, M. and {Jacques}, A. and {Jana}, A. and {Jiang}, L. and {Jimenez}, J. and {Jing}, Y.~P. and {Joudaki}, S. and {Joyce}, R. and {Jullo}, E. and {Juneau}, S. and {Kara{\c{c}}ayl{\i}}, N.~G. and {Karim}, T. and {Kehoe}, R. and {Kent}, S. and {Khederlarian}, A. and {Kim}, S. and {Kirkby}, D. and {Kisner}, T. and {Kitaura}, F. and {Kizhuprakkat}, N. and {Kneib}, J. and {Koposov}, S.~E. and {Kov{\'a}cs}, A. and {Kremin}, A. and {Krolewski}, A. and {L'Huillier}, B. and {Lahav}, O. and {Lambert}, A. and {Lamman}, C. and {Lan}, T. -W. and {Landriau}, M. and {Lang}, D. and {Lange}, J.~U. and {Lasker}, J. and {Leauthaud}, A. and {Le Guillou}, L. and {Levi}, M.~E. and {Li}, T.~S. and {Linder}, E. and {Lyons}, A. and {Magneville}, C. and {Manera}, M. and {Manser}, C.~J. and {Margala}, D. and {Martini}, P. and {McDonald}, P. and {Medina}, G.~E. and {Medina-Varela}, L. and {Meisner}, A. and {Mena-Fern{\'a}ndez}, J. and {Meneses-Rizo}, J. and {Mezcua}, M. and {Miquel}, R. and {Montero-Camacho}, P. and {Moon}, J. and {Moore}, S. and {Moustakas}, J. and {Mueller}, E. and {Mundet}, J. and {Mu{\~n}oz-Guti{\'e}rrez}, A. and {Myers}, A.~D. and {Nadathur}, S. and {Napolitano}, L. and {Neveux}, R. and {Newman}, J.~A. and {Nie}, J. and {Nikutta}, R. and {Niz}, G. and {Norberg}, P. and {Noriega}, H.~E. and {Paillas}, E. and {Palanque-Delabrouille}, N. and {Palmese}, A. and {Pan}, Z. and {Parkinson}, D. and {Penmetsa}, S. and {Percival}, W.~J. and {P{\'e}rez-Fern{\'a}ndez}, A. and {P{\'e}rez-R{\`a}fols}, I. and {Pieri}, M. and {Poppett}, C. and {Porredon}, A. and {Pothier}, S. and {Prada}, F. and {Pucha}, R. and {Raichoor}, A. and {Ram{\'\i}rez-P{\'e}rez}, C. and {Ramirez-Solano}, S. and {Rashkovetskyi}, M. and {Ravoux}, C. and {Rocher}, A. and {Rockosi}, C. and {Ross}, A.~J. and {Rossi}, G. and {Ruggeri}, R. and {Ruhlmann-Kleider}, V. and {Sabiu}, C.~G. and {Said}, K. and {Saintonge}, A. and {Samushia}, L. and {Sanchez}, E. and {Saulder}, C. and {Schaan}, E. and {Schlafly}, E.~F. and {Schlegel}, D. and {Scholte}, D. and {Schubnell}, M. and {Seo}, H. and {Shafieloo}, A. and {Sharples}, R. and {Sheu}, W. and {Silber}, J. and {Sinigaglia}, F. and {Siudek}, M. and {Slepian}, Z. and {Smith}, A. and {Soumagnac}, M.~T. and {Sprayberry}, D. and {Stephey}, L. and {Su{\'a}rez-P{\'e}rez}, J. and {Sun}, Z. and {Tan}, T. and {Tarl{\'e}}, G. and {Tojeiro}, R. and {Ure{\~n}a-L{\'o}pez}, L.~A. and {Vaisakh}, R. and {Valcin}, D. and {Valdes}, F. and {Valluri}, M. and {Vargas-Maga{\~n}a}, M. and {Variu}, A. and {Verde}, L. and {Walther}, M. and {Wang}, B. and {Wang}, M.~S. and {Weaver}, B.~A. and {Weaverdyck}, N. and {Wechsler}, R.~H. and {White}, M. and {Xie}, Y. and {Yang}, J. and {Y{\`e}che}, C. and {Yu}, J. and {Yuan}, S. and {Zhang}, H. and {Zhang}, Z. and {Zhao}, C. and {Zheng}, Z. and {Zhou}, R. and {Zhou}, Z. and {Zou}, H. and {Zou}, S. and {Zu}, Y.},
        title = "{The Early Data Release of the Dark Energy Spectroscopic Instrument}",
      journal = {\aj},
         year = 2024,
        month = aug,
       volume = {168},
       number = {2},
          eid = {58},
        pages = {58},
          doi = {10.3847/1538-3881/ad3217},
archivePrefix = {arXiv},
       eprint = {2306.06308},
 primaryClass = {astro-ph.CO},
       adsurl = {https://ui.adsabs.harvard.edu/abs/2024AJ....168...58D}
}

@ARTICLE{g10cosmosDavies2015,
       author = {{Davies}, L.~J.~M. and {Driver}, S.~P. and {Robotham}, A.~S.~G. and {Baldry}, I.~K. and {Lange}, R. and {Liske}, J. and {Meyer}, M. and {Popping}, A. and {Wilkins}, S.~M. and {Wright}, A.~H.},
        title = "{Galaxy And Mass Assembly (GAMA): curation and reanalysis of 16.6k redshifts in the G10/COSMOS region}",
      journal = {\mnras},
         year = 2015,
        month = feb,
       volume = {447},
       number = {1},
        pages = {1014-1027},
          doi = {10.1093/mnras/stu2515},
archivePrefix = {arXiv},
       eprint = {1409.3574},
 primaryClass = {astro-ph.GA},
       adsurl = {https://ui.adsabs.harvard.edu/abs/2015MNRAS.447.1014D}
}

@ARTICLE{goodsPopesso2009,
       author = {{Popesso}, P. and {Dickinson}, M. and {Nonino}, M. and {Vanzella}, E. and {Daddi}, E. and {Fosbury}, R.~A.~E. and {Kuntschner}, H. and {Mainieri}, V. and {Cristiani}, S. and {Cesarsky}, C. and {Giavalisco}, M. and {Renzini}, A. and {GOODS Team}},
        title = "{The great observatories origins deep survey. VLT/VIMOS spectroscopy in the GOODS-south field}",
      journal = {\aap},
         year = 2009,
        month = feb,
       volume = {494},
       number = {2},
        pages = {443-460},
          doi = {10.1051/0004-6361:200809617},
archivePrefix = {arXiv},
       eprint = {0802.2930},
 primaryClass = {astro-ph},
       adsurl = {https://ui.adsabs.harvard.edu/abs/2009A&A...494..443P}
}

@ARTICLE{acesCooper2012,
       author = {{Cooper}, Michael C. and {Griffith}, Roger L. and {Newman}, Jeffrey A. and {Coil}, Alison L. and {Davis}, Marc and {Dutton}, Aaron A. and {Faber}, S.~M. and {Guhathakurta}, Puragra and {Koo}, David C. and {Lotz}, Jennifer M. and {Weiner}, Benjamin J. and {Willmer}, Christopher N.~A. and {Yan}, Renbin},
        title = "{The DEEP3 Galaxy Redshift Survey: the impact of environment on the size evolution of massive early-type galaxies at intermediate redshift}",
      journal = {\mnras},
         year = 2012,
        month = feb,
       volume = {419},
       number = {4},
        pages = {3018-3027},
          doi = {10.1111/j.1365-2966.2011.19938.x},
archivePrefix = {arXiv},
       eprint = {1109.5698},
 primaryClass = {astro-ph.CO},
       adsurl = {https://ui.adsabs.harvard.edu/abs/2012MNRAS.419.3018C}
}

@ARTICLE{wigglezDrinkwater2010,
       author = {{Drinkwater}, Michael J. and {Jurek}, Russell J. and {Blake}, Chris and {Woods}, David and {Pimbblet}, Kevin A. and {Glazebrook}, Karl and {Sharp}, Rob and {Pracy}, Michael B. and {Brough}, Sarah and {Colless}, Matthew and {Couch}, Warrick J. and {Croom}, Scott M. and {Davis}, Tamara M. and {Forbes}, Duncan and {Forster}, Karl and {Gilbank}, David G. and {Gladders}, Michael and {Jelliffe}, Ben and {Jones}, Nick and {Li}, I. -Hui and {Madore}, Barry and {Martin}, D. Christopher and {Poole}, Gregory B. and {Small}, Todd and {Wisnioski}, Emily and {Wyder}, Ted and {Yee}, H.~K.~C.},
        title = "{The WiggleZ Dark Energy Survey: survey design and first data release}",
      journal = {\mnras},
         year = 2010,
        month = jan,
       volume = {401},
       number = {3},
        pages = {1429-1452},
          doi = {10.1111/j.1365-2966.2009.15754.x},
archivePrefix = {arXiv},
       eprint = {0911.4246},
 primaryClass = {astro-ph.CO},
       adsurl = {https://ui.adsabs.harvard.edu/abs/2010MNRAS.401.1429D}
}

@ARTICLE{vvdsLeFevre2013,
       author = {{Le F{\`e}vre}, O. and {Cassata}, P. and {Cucciati}, O. and {Garilli}, B. and {Ilbert}, O. and {Le Brun}, V. and {Maccagni}, D. and {Moreau}, C. and {Scodeggio}, M. and {Tresse}, L. and {Zamorani}, G. and {Adami}, C. and {Arnouts}, S. and {Bardelli}, S. and {Bolzonella}, M. and {Bondi}, M. and {Bongiorno}, A. and {Bottini}, D. and {Cappi}, A. and {Charlot}, S. and {Ciliegi}, P. and {Contini}, T. and {de la Torre}, S. and {Foucaud}, S. and {Franzetti}, P. and {Gavignaud}, I. and {Guzzo}, L. and {Iovino}, A. and {Lemaux}, B. and {L{\'o}pez-Sanjuan}, C. and {McCracken}, H.~J. and {Marano}, B. and {Marinoni}, C. and {Mazure}, A. and {Mellier}, Y. and {Merighi}, R. and {Merluzzi}, P. and {Paltani}, S. and {Pell{\`o}}, R. and {Pollo}, A. and {Pozzetti}, L. and {Scaramella}, R. and {Tasca}, L. and {Vergani}, D. and {Vettolani}, G. and {Zanichelli}, A. and {Zucca}, E.},
        title = "{The VIMOS VLT Deep Survey final data release: a spectroscopic sample of 35 016 galaxies and AGN out to z \raisebox{-0.5ex}\textasciitilde 6.7 selected with 17.5 {\ensuremath{\leq}} i$_{AB}$ {\ensuremath{\leq}} 24.75}",
      journal = {\aap},
         year = 2013,
        month = nov,
       volume = {559},
          eid = {A14},
        pages = {A14},
          doi = {10.1051/0004-6361/201322179},
archivePrefix = {arXiv},
       eprint = {1307.6518},
 primaryClass = {astro-ph.CO},
       adsurl = {https://ui.adsabs.harvard.edu/abs/2013A&A...559A..14L}
}

@ARTICLE{sdssabolfathi2018,
       author = {{Abolfathi}, Bela and {Aguado}, D.~S. and {Aguilar}, Gabriela and {Allende Prieto}, Carlos and {Almeida}, Andres and {Ananna}, Tonima Tasnim and {Anders}, Friedrich and {Anderson}, Scott F. and {Andrews}, Brett H. and {Anguiano}, Borja and {Arag{\'o}n-Salamanca}, Alfonso and {Argudo-Fern{\'a}ndez}, Maria and {Armengaud}, Eric and {Ata}, Metin and {Aubourg}, Eric and {Avila-Reese}, Vladimir and {Badenes}, Carles and {Bailey}, Stephen and {Balland}, Christophe and {Barger}, Kathleen A. and {Barrera-Ballesteros}, Jorge and {Bartosz}, Curtis and {Bastien}, Fabienne and {Bates}, Dominic and {Baumgarten}, Falk and {Bautista}, Julian and {Beaton}, Rachael and {Beers}, Timothy C. and {Belfiore}, Francesco and {Bender}, Chad F. and {Bernardi}, Mariangela and {Bershady}, Matthew A. and {Beutler}, Florian and {Bird}, Jonathan C. and {Bizyaev}, Dmitry and {Blanc}, Guillermo A. and {Blanton}, Michael R. and {Blomqvist}, Michael and {Bolton}, Adam S. and {Boquien}, M{\'e}d{\'e}ric and {Borissova}, Jura and {Bovy}, Jo and {Bradna Diaz}, Christian Andres and {Brandt}, William Nielsen and {Brinkmann}, Jonathan and {Brownstein}, Joel R. and {Bundy}, Kevin and {Burgasser}, Adam J. and {Burtin}, Etienne and {Busca}, Nicol{\'a}s G. and {Ca{\~n}as}, Caleb I. and {Cano-D{\'\i}az}, Mariana and {Cappellari}, Michele and {Carrera}, Ricardo and {Casey}, Andrew R. and {Cervantes Sodi}, Bernardo and {Chen}, Yanping and {Cherinka}, Brian and {Chiappini}, Cristina and {Choi}, Peter Doohyun and {Chojnowski}, Drew and {Chuang}, Chia-Hsun and {Chung}, Haeun and {Clerc}, Nicolas and {Cohen}, Roger E. and {Comerford}, Julia M. and {Comparat}, Johan and {Correa do Nascimento}, Janaina and {da Costa}, Luiz and {Cousinou}, Marie-Claude and {Covey}, Kevin and {Crane}, Jeffrey D. and {Cruz-Gonzalez}, Irene and {Cunha}, Katia and {da Silva Ilha}, Gabriele and {Damke}, Guillermo J. and {Darling}, Jeremy and {Davidson}, James W., Jr. and {Dawson}, Kyle and {de Icaza Lizaola}, Miguel Angel C. and {de la Macorra}, Axel and {de la Torre}, Sylvain and {De Lee}, Nathan and {de Sainte Agathe}, Victoria and {Deconto Machado}, Alice and {Dell'Agli}, Flavia and {Delubac}, Timoth{\'e}e and {Diamond-Stanic}, Aleksandar M. and {Donor}, John and {Downes}, Juan Jos{\'e} and {Drory}, Niv and {du Mas des Bourboux}, H{\'e}lion and {Duckworth}, Christopher J. and {Dwelly}, Tom and {Dyer}, Jamie and {Ebelke}, Garrett and {Davis Eigenbrot}, Arthur and {Eisenstein}, Daniel J. and {Elsworth}, Yvonne P. and {Emsellem}, Eric and {Eracleous}, Michael and {Erfanianfar}, Ghazaleh and {Escoffier}, Stephanie and {Fan}, Xiaohui and {Fern{\'a}ndez Alvar}, Emma and {Fernandez-Trincado}, J.~G. and {Fernando Cirolini}, Rafael and {Feuillet}, Diane and {Finoguenov}, Alexis and {Fleming}, Scott W. and {Font-Ribera}, Andreu and {Freischlad}, Gordon and {Frinchaboy}, Peter and {Fu}, Hai and {G{\'o}mez Maqueo Chew}, Yilen and {Galbany}, Llu{\'\i}s and {Garc{\'\i}a P{\'e}rez}, Ana E. and {Garcia-Dias}, R. and {Garc{\'\i}a-Hern{\'a}ndez}, D.~A. and {Garma Oehmichen}, Luis Alberto and {Gaulme}, Patrick and {Gelfand}, Joseph and {Gil-Mar{\'\i}n}, H{\'e}ctor and {Gillespie}, Bruce A. and {Goddard}, Daniel and {Gonz{\'a}lez Hern{\'a}ndez}, Jonay I. and {Gonzalez-Perez}, Violeta and {Grabowski}, Kathleen and {Green}, Paul J. and {Grier}, Catherine J. and {Gueguen}, Alain and {Guo}, Hong and {Guy}, Julien and {Hagen}, Alex and {Hall}, Patrick and {Harding}, Paul and {Hasselquist}, Sten and {Hawley}, Suzanne and {Hayes}, Christian R. and {Hearty}, Fred and {Hekker}, Saskia and {Hernandez}, Jesus and {Hernandez Toledo}, Hector and {Hogg}, David W. and {Holley-Bockelmann}, Kelly and {Holtzman}, Jon A. and {Hou}, Jiamin and {Hsieh}, Bau-Ching and {Hunt}, Jason A.~S. and {Hutchinson}, Timothy A. and {Hwang}, Ho Seong and {Jimenez Angel}, Camilo Eduardo and {Johnson}, Jennifer A. and {Jones}, Amy and {J{\"o}nsson}, Henrik and {Jullo}, Eric and {Khan}, Fahim Sakil and {Kinemuchi}, Karen and {Kirkby}, David and {Kirkpatrick}, Charles C., IV and {Kitaura}, Francisco-Shu and {Knapp}, Gillian R. and {Kneib}, Jean-Paul and {Kollmeier}, Juna A. and {Lacerna}, Ivan and {Lane}, Richard R. and {Lang}, Dustin and {Law}, David R. and {Le Goff}, Jean-Marc and {Lee}, Young-Bae and {Li}, Hongyu and {Li}, Cheng and {Lian}, Jianhui and {Liang}, Yu and {Lima}, Marcos and {Lin}, Lihwai and {Long}, Dan and {Lucatello}, Sara and {Lundgren}, Britt and {Mackereth}, J. Ted and {MacLeod}, Chelsea L. and {Mahadevan}, Suvrath and {Maia}, Marcio Antonio Geimba and {Majewski}, Steven and {Manchado}, Arturo and {Maraston}, Claudia and {Mariappan}, Vivek and {Marques-Chaves}, Rui and {Masseron}, Thomas and {Masters}, Karen L. and {McDermid}, Richard M. and {McGreer}, Ian D. and {Melendez}, Matthew and {Meneses-Goytia}, Sofia and {Merloni}, Andrea and {Merrifield}, Michael R. and {Meszaros}, Szabolcs and {Meza}, Andres and {Minchev}, Ivan and {Minniti}, Dante and {Mueller}, Eva-Maria and {Muller-Sanchez}, Francisco and {Muna}, Demitri and {Mu{\~n}oz}, Ricardo R. and {Myers}, Adam D. and {Nair}, Preethi and {Nandra}, Kirpal and {Ness}, Melissa and {Newman}, Jeffrey A. and {Nichol}, Robert C. and {Nidever}, David L. and {Nitschelm}, Christian and {Noterdaeme}, Pasquier and {O'Connell}, Julia and {Oelkers}, Ryan James and {Oravetz}, Audrey and {Oravetz}, Daniel and {Ort{\'\i}z}, Erik Aquino and {Osorio}, Yeisson and {Pace}, Zach and {Padilla}, Nelson and {Palanque-Delabrouille}, Nathalie and {Palicio}, Pedro Alonso and {Pan}, Hsi-An and {Pan}, Kaike and {Parikh}, Taniya and {P{\^a}ris}, Isabelle and {Park}, Changbom and {Peirani}, Sebastien and {Pellejero-Ibanez}, Marcos and {Penny}, Samantha and {Percival}, Will J. and {Perez-Fournon}, Ismael and {Petitjean}, Patrick and {Pieri}, Matthew M. and {Pinsonneault}, Marc and {Pisani}, Alice and {Prada}, Francisco and {Prakash}, Abhishek and {Queiroz}, Anna B{\'a}rbara de Andrade and {Raddick}, M. Jordan and {Raichoor}, Anand and {Barboza Rembold}, Sandro and {Richstein}, Hannah and {Riffel}, Rogemar A. and {Riffel}, Rog{\'e}rio and {Rix}, Hans-Walter and {Robin}, Annie C. and {Rodr{\'\i}guez Torres}, Sergio and {Rom{\'a}n-Z{\'u}{\~n}iga}, Carlos and {Ross}, Ashley J. and {Rossi}, Graziano and {Ruan}, John and {Ruggeri}, Rossana and {Ruiz}, Jose and {Salvato}, Mara and {S{\'a}nchez}, Ariel G. and {S{\'a}nchez}, Sebasti{\'a}n F. and {Sanchez Almeida}, Jorge and {S{\'a}nchez-Gallego}, Jos{\'e} R. and {Santana Rojas}, Felipe Antonio and {Santiago}, Bas{\'\i}lio Xavier and {Schiavon}, Ricardo P. and {Schimoia}, Jaderson S. and {Schlafly}, Edward and {Schlegel}, David and {Schneider}, Donald P. and {Schuster}, William J. and {Schwope}, Axel and {Seo}, Hee-Jong and {Serenelli}, Aldo and {Shen}, Shiyin and {Shen}, Yue and {Shetrone}, Matthew and {Shull}, Michael and {Silva Aguirre}, V{\'\i}ctor and {Simon}, Joshua D. and {Skrutskie}, Mike and {Slosar}, An{\v{z}}e and {Smethurst}, Rebecca and {Smith}, Verne and {Sobeck}, Jennifer and {Somers}, Garrett and {Souter}, Barbara J. and {Souto}, Diogo and {Spindler}, Ashley and {Stark}, David V. and {Stassun}, Keivan and {Steinmetz}, Matthias and {Stello}, Dennis and {Storchi-Bergmann}, Thaisa and {Streblyanska}, Alina and {Stringfellow}, Guy S. and {Su{\'a}rez}, Genaro and {Sun}, Jing and {Szigeti}, Laszlo and {Taghizadeh-Popp}, Manuchehr and {Talbot}, Michael S. and {Tang}, Baitian and {Tao}, Charling and {Tayar}, Jamie and {Tembe}, Mita and {Teske}, Johanna and {Thakar}, Aniruddha R. and {Thomas}, Daniel and {Tissera}, Patricia and {Tojeiro}, Rita and {Tremonti}, Christy and {Troup}, Nicholas W. and {Urry}, Meg and {Valenzuela}, O. and {van den Bosch}, Remco and {Vargas-Gonz{\'a}lez}, Jaime and {Vargas-Maga{\~n}a}, Mariana and {Vazquez}, Jose Alberto and {Villanova}, Sandro and {Vogt}, Nicole and {Wake}, David and {Wang}, Yuting and {Weaver}, Benjamin Alan and {Weijmans}, Anne-Marie and {Weinberg}, David H. and {Westfall}, Kyle B. and {Whelan}, David G. and {Wilcots}, Eric and {Wild}, Vivienne and {Williams}, Rob A. and {Wilson}, John and {Wood-Vasey}, W.~M. and {Wylezalek}, Dominika and {Xiao}, Ting and {Yan}, Renbin and {Yang}, Meng and {Ybarra}, Jason E. and {Y{\`e}che}, Christophe and {Zakamska}, Nadia and {Zamora}, Olga and {Zarrouk}, Pauline and {Zasowski}, Gail and {Zhang}, Kai and {Zhao}, Cheng and {Zhao}, Gong-Bo and {Zheng}, Zheng and {Zheng}, Zheng and {Zhou}, Zhi-Min and {Zhu}, Guangtun and {Zinn}, Joel C. and {Zou}, Hu},
        title = "{The Fourteenth Data Release of the Sloan Digital Sky Survey: First Spectroscopic Data from the Extended Baryon Oscillation Spectroscopic Survey and from the Second Phase of the Apache Point Observatory Galactic Evolution Experiment}",
      journal = {\apjs},
         year = 2018,
        month = apr,
       volume = {235},
       number = {2},
          eid = {42},
        pages = {42},
          doi = {10.3847/1538-4365/aa9e8a},
archivePrefix = {arXiv},
       eprint = {1707.09322},
 primaryClass = {astro-ph.GA},
       adsurl = {https://ui.adsabs.harvard.edu/abs/2018ApJS..235...42A}
}

@ARTICLE{ozdeslidman2020,
       author = {{Lidman}, C. and {Tucker}, B.~E. and {Davis}, T.~M. and {Uddin}, S.~A. and {Asorey}, J. and {Bolejko}, K. and {Brout}, D. and {Calcino}, J. and {Carollo}, D. and {Carr}, A. and {Childress}, M. and {Hoormann}, J.~K. and {Foley}, R.~J. and {Galbany}, L. and {Glazebrook}, K. and {Hinton}, S.~R. and {Kessler}, R. and {Kim}, A.~G. and {King}, A. and {Kremin}, A. and {Kuehn}, K. and {Lagattuta}, D. and {Lewis}, G.~F. and {Macaulay}, E. and {Malik}, U. and {March}, M. and {Martini}, P. and {M{\"o}ller}, A. and {Mudd}, D. and {Nichol}, R.~C. and {Panther}, F. and {Parkinson}, D. and {Pursiainen}, M. and {Sako}, M. and {Swann}, E. and {Scalzo}, R. and {Scolnic}, D. and {Sharp}, R. and {Smith}, M. and {Sommer}, N.~E. and {Sullivan}, M. and {Webb}, S. and {Wiseman}, P. and {Yu}, Z. and {Yuan}, F. and {Zhang}, B. and {Abbott}, T.~M.~C. and {Aguena}, M. and {Allam}, S. and {Annis}, J. and {Avila}, S. and {Bertin}, E. and {Bhargava}, S. and {Brooks}, D. and {Carnero Rosell}, A. and {Carrasco Kind}, M. and {Carretero}, J. and {Castander}, F.~J. and {Costanzi}, M. and {da Costa}, L.~N. and {De Vicente}, J. and {Doel}, P. and {Eifler}, T.~F. and {Everett}, S. and {Fosalba}, P. and {Frieman}, J. and {Garc{\'\i}a-Bellido}, J. and {Gaztanaga}, E. and {Gruen}, D. and {Gruendl}, R.~A. and {Gschwend}, J. and {Gutierrez}, G. and {Hartley}, W.~G. and {Hollowood}, D.~L. and {Honscheid}, K. and {James}, D.~J. and {Kuropatkin}, N. and {Li}, T.~S. and {Lima}, M. and {Lin}, H. and {Maia}, M.~A.~G. and {Marshall}, J.~L. and {Melchior}, P. and {Menanteau}, F. and {Miquel}, R. and {Palmese}, A. and {Paz-Chinch{\'o}n}, F. and {Plazas}, A.~A. and {Roodman}, A. and {Rykoff}, E.~S. and {Sanchez}, E. and {Santiago}, B. and {Scarpine}, V. and {Schubnell}, M. and {Serrano}, S. and {Sevilla-Noarbe}, I. and {Suchyta}, E. and {Swanson}, M.~E.~C. and {Tarle}, G. and {Tucker}, D.~L. and {Varga}, T.~N. and {Walker}, A.~R. and {Wester}, W. and {Wilkinson}, R.~D. and {DES Collaboration}},
        title = "{OzDES multi-object fibre spectroscopy for the Dark Energy Survey: results and second data release}",
      journal = {\mnras},
         year = 2020,
        month = jul,
       volume = {496},
       number = {1},
        pages = {19-35},
          doi = {10.1093/mnras/staa1341},
archivePrefix = {arXiv},
       eprint = {2006.00449},
 primaryClass = {astro-ph.CO},
       adsurl = {https://ui.adsabs.harvard.edu/abs/2020MNRAS.496...19L}
}

@article{Hasinger_2018,
doi = {10.3847/1538-4357/aabacf},
url = {https://dx.doi.org/10.3847/1538-4357/aabacf},
year = {2018},
month = {may},
publisher = {The American Astronomical Society},
volume = {858},
number = {2},
pages = {77},
author = {G. Hasinger and P. Capak and M. Salvato and A. J. Barger and L. L. Cowie and A. Faisst and S. Hemmati and Y. Kakazu and J. Kartaltepe and D. Masters and B. Mobasher and H. Nayyeri and D. Sanders and N. Z. Scoville and H. Suh and C. Steinhardt and Fengwei Yang},
title = {The DEIMOS 10K Spectroscopic Survey Catalog of the COSMOS Field∗},
journal = {The Astrophysical Journal}
}

@ARTICLE{vipersScodeggio2018,
       author = {{Scodeggio}, M. and {Guzzo}, L. and {Garilli}, B. and {Granett}, B.~R. and {Bolzonella}, M. and {de la Torre}, S. and {Abbas}, U. and {Adami}, C. and {Arnouts}, S. and {Bottini}, D. and {Cappi}, A. and {Coupon}, J. and {Cucciati}, O. and {Davidzon}, I. and {Franzetti}, P. and {Fritz}, A. and {Iovino}, A. and {Krywult}, J. and {Le Brun}, V. and {Le F{\`e}vre}, O. and {Maccagni}, D. and {Ma{\l}ek}, K. and {Marchetti}, A. and {Marulli}, F. and {Polletta}, M. and {Pollo}, A. and {Tasca}, L.~A.~M. and {Tojeiro}, R. and {Vergani}, D. and {Zanichelli}, A. and {Bel}, J. and {Branchini}, E. and {De Lucia}, G. and {Ilbert}, O. and {McCracken}, H.~J. and {Moutard}, T. and {Peacock}, J.~A. and {Zamorani}, G. and {Burden}, A. and {Fumana}, M. and {Jullo}, E. and {Marinoni}, C. and {Mellier}, Y. and {Moscardini}, L. and {Percival}, W.~J.},
        title = "{The VIMOS Public Extragalactic Redshift Survey (VIPERS). Full spectroscopic data and auxiliary information release (PDR-2)}",
      journal = {\aap},
         year = 2018,
        month = jan,
       volume = {609},
          eid = {A84},
        pages = {A84},
          doi = {10.1051/0004-6361/201630114},
archivePrefix = {arXiv},
       eprint = {1611.07048},
 primaryClass = {astro-ph.GA},
       adsurl = {https://ui.adsabs.harvard.edu/abs/2018A&A...609A..84S}
}

@ARTICLE{lagcVanderwel2016,
       author = {{van der Wel}, A. and {Noeske}, K. and {Bezanson}, R. and {Pacifici}, C. and {Gallazzi}, A. and {Franx}, M. and {Mu{\~n}oz-Mateos}, J.~C. and {Bell}, E.~F. and {Brammer}, G. and {Charlot}, S. and {Chauk{\'e}}, P. and {Labb{\'e}}, I. and {Maseda}, M.~V. and {Muzzin}, A. and {Rix}, H. -W. and {Sobral}, D. and {van de Sande}, J. and {van Dokkum}, P.~G. and {Wild}, V. and {Wolf}, C.},
        title = "{The VLT LEGA-C Spectroscopic Survey: The Physics of Galaxies at a Lookback Time of 7 Gyr}",
      journal = {\apjs},
         year = 2016,
        month = apr,
       volume = {223},
       number = {2},
          eid = {29},
        pages = {29},
          doi = {10.3847/0067-0049/223/2/29},
archivePrefix = {arXiv},
       eprint = {1603.05479},
 primaryClass = {astro-ph.GA},
       adsurl = {https://ui.adsabs.harvard.edu/abs/2016ApJS..223...29V}
}

@ARTICLE{gamaDriver2022,
       author = {{Driver}, Simon P. and {Bellstedt}, Sabine and {Robotham}, Aaron S.~G. and {Baldry}, Ivan K. and {Davies}, Luke J. and {Liske}, Jochen and {Obreschkow}, Danail and {Taylor}, Edward N. and {Wright}, Angus H. and {Alpaslan}, Mehmet and {Bamford}, Steven P. and {Bauer}, Amanda E. and {Bland-Hawthorn}, Joss and {Bilicki}, Maciej and {Bravo}, Mat{\'\i}as and {Brough}, Sarah and {Casura}, Sarah and {Cluver}, Michelle E. and {Colless}, Matthew and {Conselice}, Christopher J. and {Croom}, Scott M. and {de Jong}, Jelte and {D'Eugenio}, Franceso and {De Propris}, Roberto and {Dogruel}, Burak and {Drinkwater}, Michael J. and {Dvornik}, Andrej and {Farrow}, Daniel J. and {Frenk}, Carlos S. and {Giblin}, Benjamin and {Graham}, Alister W. and {Grootes}, Meiert W. and {Gunawardhana}, Madusha L.~P. and {Hashemizadeh}, Abdolhosein and {H{\"a}u{\ss}ler}, Boris and {Heymans}, Catherine and {Hildebrandt}, Hendrik and {Holwerda}, Benne W. and {Hopkins}, Andrew M. and {Jarrett}, Tom H. and {Heath Jones}, D. and {Kelvin}, Lee S. and {Koushan}, Soheil and {Kuijken}, Konrad and {Lara-L{\'o}pez}, Maritza A. and {Lange}, Rebecca and {L{\'o}pez-S{\'a}nchez}, {\'A}ngel R. and {Loveday}, Jon and {Mahajan}, Smriti and {Meyer}, Martin and {Moffett}, Amanda J. and {Napolitano}, Nicola R. and {Norberg}, Peder and {Owers}, Matt S. and {Radovich}, Mario and {Raouf}, Mojtaba and {Peacock}, John A. and {Phillipps}, Steven and {Pimbblet}, Kevin A. and {Popescu}, Cristina and {Said}, Khaled and {Sansom}, Anne E. and {Seibert}, Mark and {Sutherland}, Will J. and {Thorne}, Jessica E. and {Tuffs}, Richard J. and {Turner}, Ryan and {van der Wel}, Arjen and {van Kampen}, Eelco and {Wilkins}, Steve M.},
        title = "{Galaxy And Mass Assembly (GAMA): Data Release 4 and the z < 0.1 total and z < 0.08 morphological galaxy stellar mass functions}",
      journal = {\mnras},
         year = 2022,
        month = jun,
       volume = {513},
       number = {1},
        pages = {439-467},
          doi = {10.1093/mnras/stac472},
archivePrefix = {arXiv},
       eprint = {2203.08539},
 primaryClass = {astro-ph.GA},
       adsurl = {https://ui.adsabs.harvard.edu/abs/2022MNRAS.513..439D}
}

@ARTICLE{2023Msngr.190...25D,
       author = {{Duncan}, K. and {Baker}, A. and {Best}, P. and {Blyth}, S. and {Hatch}, N. and {Holwerda}, B. and {Jarvis}, M. and {Maddox}, N. and {Smith}, D.~J.~B. and {Arnaudova}, M. and {Chemin}, L. and {Dav{\'e}}, R. and {Dunlop}, J. and {Frank}, B. and {Gawiser}, E. and {Gloudemans}, A. and {Hale}, C. and {Heywood}, I. and {Kannappan}, S. and {Kondapally}, R. and {McLure}, R. and {Morabito}, L. and {Nesvadba}, N. and {Pan}, H. and {Ponomareva}, A. and {Prescott}, M. and {Roberts}, H. and {R{\"o}ttgering}, H. and {Somerville}, R. and {Tudorache}, M. and {Vaccari}, M. and {Whittam}, I. and {Wu}, J. and {Zwaan}, M.},
        title = "{Optical, Radio Continuum and HI Deep Spectroscopic Survey (ORCHIDSS)}",
      journal = {The Messenger},
         year = 2023,
        month = mar,
       volume = {190},
        pages = {25-27},
          doi = {10.18727/0722-6691/5306},
       adsurl = {https://ui.adsabs.harvard.edu/abs/2023Msngr.190...25D}
}

@ARTICLE{2011MNRAS.416.2640R,
       author = {{Robotham}, A.~S.~G. and {Norberg}, P. and {Driver}, S.~P. and {Baldry}, I.~K. and {Bamford}, S.~P. and {Hopkins}, A.~M. and {Liske}, J. and {Loveday}, J. and {Merson}, A. and {Peacock}, J.~A. and {Brough}, S. and {Cameron}, E. and {Conselice}, C.~J. and {Croom}, S.~M. and {Frenk}, C.~S. and {Gunawardhana}, M. and {Hill}, D.~T. and {Jones}, D.~H. and {Kelvin}, L.~S. and {Kuijken}, K. and {Nichol}, R.~C. and {Parkinson}, H.~R. and {Pimbblet}, K.~A. and {Phillipps}, S. and {Popescu}, C.~C. and {Prescott}, M. and {Sharp}, R.~G. and {Sutherland}, W.~J. and {Taylor}, E.~N. and {Thomas}, D. and {Tuffs}, R.~J. and {van Kampen}, E. and {Wijesinghe}, D.},
        title = "{Galaxy and Mass Assembly (GAMA): the GAMA galaxy group catalogue (G$^{3}$Cv1)}",
      journal = {\mnras},
         year = 2011,
        month = oct,
       volume = {416},
       number = {4},
        pages = {2640-2668},
          doi = {10.1111/j.1365-2966.2011.19217.x},
archivePrefix = {arXiv},
       eprint = {1106.1994},
 primaryClass = {astro-ph.CO},
       adsurl = {https://ui.adsabs.harvard.edu/abs/2011MNRAS.416.2640R}
}

@ARTICLE{2023Msngr.190...31H,
       author = {{Haines}, C. and {Jaff{\'e}}, Y. and {Tejos}, N. and {Monachesi}, A. and {Pompei}, E. and {Finoguenov}, A. and {Sif{\'o}n}, C. and {Lopez}, S. and {Manjunatha}, A.~B. and {Bilton}, L. and {Comparat}, J. and {Cuellar}, R. and {D'Ago}, G. and {Demarco}, R. and {Lima-Dias}, C. and {L{\"o}sch}, E. and {Merluzzi}, P. and {Smith Castelli}, A. and {Sodre}, L. and {Vinicius}, E. and {CHANCES Team}},
        title = "{CHANCES: A CHileAN Cluster galaxy Evolution Survey}",
      journal = {The Messenger},
         year = 2023,
        month = mar,
       volume = {190},
        pages = {31-33},
          doi = {10.18727/0722-6691/5308},
       adsurl = {https://ui.adsabs.harvard.edu/abs/2023Msngr.190...31H}
}

@ARTICLE{2024A&A...685A.127S,
       author = {{Signor}, T. and {Rodighiero}, G. and {Bisigello}, L. and {Bolzonella}, M. and {Caputi}, K.~I. and {Daddi}, E. and {De Lucia}, G. and {Enia}, A. and {Gabarra}, L. and {Gruppioni}, C. and {Humphrey}, A. and {La Franca}, F. and {Mancini}, C. and {Pozzetti}, L. and {Serjeant}, S. and {Spinoglio}, L. and {van Mierlo}, S.~E. and {Andreon}, S. and {Auricchio}, N. and {Baldi}, M. and {Bardelli}, S. and {Battaglia}, P. and {Bender}, R. and {Bodendorf}, C. and {Bonino}, D. and {Branchini}, E. and {Brescia}, M. and {Brinchmann}, J. and {Camera}, S. and {Capobianco}, V. and {Carbone}, C. and {Carretero}, J. and {Casas}, S. and {Castellano}, M. and {Cavuoti}, S. and {Cimatti}, A. and {Cledassou}, R. and {Congedo}, G. and {Conselice}, C.~J. and {Conversi}, L. and {Copin}, Y. and {Corcione}, L. and {Courbin}, F. and {Courtois}, H.~M. and {Da Silva}, A. and {Degaudenzi}, H. and {Di Giorgio}, A.~M. and {Dinis}, J. and {Dubath}, F. and {Dupac}, X. and {Dusini}, S. and {Ealet}, A. and {Farina}, M. and {Farrens}, S. and {Ferriol}, S. and {Fotopoulou}, S. and {Franceschi}, E. and {Galeotta}, S. and {Garilli}, B. and {Gillard}, W. and {Gillis}, B. and {Giocoli}, C. and {Grazian}, A. and {Grupp}, F. and {Guzzo}, L. and {Haugan}, S.~V.~H. and {Hook}, I. and {Hormuth}, F. and {Hornstrup}, A. and {Jahnke}, K. and {K{\"u}mmel}, M. and {Kermiche}, S. and {Kiessling}, A. and {Kilbinger}, M. and {Kitching}, T. and {Kurki-Suonio}, H. and {Ligori}, S. and {Lilje}, P.~B. and {Lindholm}, V. and {Lloro}, I. and {Maino}, D. and {Maiorano}, E. and {Mansutti}, O. and {Marggraf}, O. and {Martinet}, N. and {Marulli}, F. and {Massey}, R. and {Medinaceli}, E. and {Melchior}, M. and {Mellier}, Y. and {Meneghetti}, M. and {Merlin}, E. and {Moresco}, M. and {Moscardini}, L. and {Munari}, E. and {Nichol}, R.~C. and {Niemi}, S. -M. and {Padilla}, C. and {Paltani}, S. and {Pasian}, F. and {Pedersen}, K. and {Pettorino}, V. and {Pires}, S. and {Polenta}, G. and {Poncet}, M. and {Popa}, L.~A. and {Raison}, F. and {Renzi}, A. and {Rhodes}, J. and {Riccio}, G. and {Romelli}, E. and {Roncarelli}, M. and {Rossetti}, E. and {Saglia}, R. and {Sapone}, D. and {Sartoris}, B. and {Schneider}, P. and {Schrabback}, T. and {Secroun}, A. and {Seidel}, G. and {Serrano}, S. and {Sirignano}, C. and {Sirri}, G. and {Stanco}, L. and {Surace}, C. and {Tallada-Cresp{\'\i}}, P. and {Teplitz}, H.~I. and {Tereno}, I. and {Toledo-Moreo}, R. and {Torradeflot}, F. and {Tutusaus}, I. and {Valentijn}, E.~A. and {Vassallo}, T. and {Veropalumbo}, A. and {Wang}, Y. and {Weller}, J. and {Williams}, O.~R. and {Zoubian}, J. and {Zucca}, E. and {Burigana}, C. and {Scottez}, V.},
        title = "{Euclid: Identifying the reddest high-redshift galaxies in the Euclid Deep Fields with gradient-boosted trees}",
      journal = {\aap},
         year = 2024,
        month = may,
       volume = {685},
          eid = {A127},
        pages = {A127},
          doi = {10.1051/0004-6361/202348737},
archivePrefix = {arXiv},
       eprint = {2402.04800},
 primaryClass = {astro-ph.CO},
       adsurl = {https://ui.adsabs.harvard.edu/abs/2024A&A...685A.127S}
}

@ARTICLE{2019A&A...624A..13N,
       author = {{Nakoneczny}, S. and {Bilicki}, M. and {Solarz}, A. and {Pollo}, A. and {Maddox}, N. and {Spiniello}, C. and {Brescia}, M. and {Napolitano}, N.~R.},
        title = "{Catalog of quasars from the Kilo-Degree Survey Data Release 3}",
      journal = {\aap},
         year = 2019,
        month = apr,
       volume = {624},
          eid = {A13},
        pages = {A13},
          doi = {10.1051/0004-6361/201834794},
archivePrefix = {arXiv},
       eprint = {1812.03084},
 primaryClass = {astro-ph.IM},
       adsurl = {https://ui.adsabs.harvard.edu/abs/2019A&A...624A..13N}
}

@ARTICLE{2021A&A...649A..81N,
       author = {{Nakoneczny}, S.~J. and {Bilicki}, M. and {Pollo}, A. and {Asgari}, M. and {Dvornik}, A. and {Erben}, T. and {Giblin}, B. and {Heymans}, C. and {Hildebrandt}, H. and {Kannawadi}, A. and {Kuijken}, K. and {Napolitano}, N.~R. and {Valentijn}, E.},
        title = "{Photometric selection and redshifts for quasars in the Kilo-Degree Survey Data Release 4}",
      journal = {\aap},
         year = 2021,
        month = may,
       volume = {649},
          eid = {A81},
        pages = {A81},
          doi = {10.1051/0004-6361/202039684},
archivePrefix = {arXiv},
       eprint = {2010.13857},
 primaryClass = {astro-ph.CO},
       adsurl = {https://ui.adsabs.harvard.edu/abs/2021A&A...649A..81N}
}

@ARTICLE{Anjitha25,
       author = {{John William}, Anjitha and {Jalan}, Priyanka and {Bilicki}, Maciej and {Hellwing}, Wojciech A. and {Thuruthipilly}, Hareesh and {Nakoneczny}, Szymon J.},
        title = "{Hybrid-z: Enhancing the Kilo-Degree Survey bright galaxy sample photometric redshifts with deep learning}",
      journal = {\aap},
         year = 2025,
        month = jun,
       volume = {698},
          eid = {A276},
        pages = {A276},
          doi = {10.1051/0004-6361/202453576},
archivePrefix = {arXiv},
       eprint = {2501.01942},
 primaryClass = {astro-ph.CO},
       adsurl = {https://ui.adsabs.harvard.edu/abs/2025A&A...698A.276J}
}

@INPROCEEDINGS{2016heas.confE..26V,
       author = {{Vaccari}, M. and {Covone}, G. and {Radovich}, M. and {Grado}, A. and {Limatola}, L. and {Botticella}, M.~T. and {Cappellaro}, E. and {Paolillo}, M. and {Pignata}, G. and {De Cicco}, D. and {Falocco}, S. and {Marchetti}, L. and {Brescia}, M. and {Cavuoti}, S. and {Longo}, G. and {Capaccioli}, M. and {Napolitano}, N. and {Schipani}, P.},
        title = "{The VOICE Survey : VST Optical Imaging of the CDFS and ES1 Fields}",
    booktitle = {The 4th Annual Conference on High Energy Astrophysics in Southern Africa (HEASA 2016)},
         year = 2016,
        month = jan,
          eid = {26},
        pages = {26},
          doi = {10.22323/1.275.0026},
archivePrefix = {arXiv},
       eprint = {1704.01495},
 primaryClass = {astro-ph.GA},
       adsurl = {https://ui.adsabs.harvard.edu/abs/2016heas.confE..26V}
}

@ARTICLE{2012A&A...544A.156M,
       author = {{McCracken}, H.~J. and {Milvang-Jensen}, B. and {Dunlop}, J. and {Franx}, M. and {Fynbo}, J.~P.~U. and {Le F{\`e}vre}, O. and {Holt}, J. and {Caputi}, K.~I. and {Goranova}, Y. and {Buitrago}, F. and {Emerson}, J.~P. and {Freudling}, W. and {Hudelot}, P. and {L{\'o}pez-Sanjuan}, C. and {Magnard}, F. and {Mellier}, Y. and {M{\o}ller}, P. and {Nilsson}, K.~K. and {Sutherland}, W. and {Tasca}, L. and {Zabl}, J.},
        title = "{UltraVISTA: a new ultra-deep near-infrared survey in COSMOS}",
      journal = {\aap},
         year = 2012,
        month = aug,
       volume = {544},
          eid = {A156},
        pages = {A156},
          doi = {10.1051/0004-6361/201219507},
archivePrefix = {arXiv},
       eprint = {1204.6586},
 primaryClass = {astro-ph.CO},
       adsurl = {https://ui.adsabs.harvard.edu/abs/2012A&A...544A.156M}
}

@ARTICLE{2013MNRAS.428.1281J,
       author = {{Jarvis}, Matt J. and {Bonfield}, D.~G. and {Bruce}, V.~A. and {Geach}, J.~E. and {McAlpine}, K. and {McLure}, R.~J. and {Gonz{\'a}lez-Solares}, E. and {Irwin}, M. and {Lewis}, J. and {Yoldas}, A. Kupcu and {Andreon}, S. and {Cross}, N.~J.~G. and {Emerson}, J.~P. and {Dalton}, G. and {Dunlop}, J.~S. and {Hodgkin}, S.~T. and {Le}, F{\`e}vre O. and {Karouzos}, M. and {Meisenheimer}, K. and {Oliver}, S. and {Rawlings}, S. and {Simpson}, C. and {Smail}, I. and {Smith}, D.~J.~B. and {Sullivan}, M. and {Sutherland}, W. and {White}, S.~V. and {Zwart}, J.~T.~L.},
        title = "{The VISTA Deep Extragalactic Observations (VIDEO) survey}",
      journal = {\mnras},
         year = 2013,
        month = jan,
       volume = {428},
       number = {2},
        pages = {1281-1295},
          doi = {10.1093/mnras/sts118},
archivePrefix = {arXiv},
       eprint = {1206.4263},
 primaryClass = {astro-ph.CO},
       adsurl = {https://ui.adsabs.harvard.edu/abs/2013MNRAS.428.1281J}
}

@ARTICLE{Zhou2021,
       author = {{Zhou}, Rongpu and {Newman}, Jeffrey A. and {Mao}, Yao-Yuan and {Meisner}, Aaron and {Moustakas}, John and {Myers}, Adam D. and {Prakash}, Abhishek and {Zentner}, Andrew R. and {Brooks}, David and {Duan}, Yutong and {Landriau}, Martin and {Levi}, Michael E. and {Prada}, Francisco and {Tarle}, Gregory},
        title = "{The clustering of DESI-like luminous red galaxies using photometric redshifts}",
      journal = {\mnras},
         year = 2021,
        month = mar,
       volume = {501},
       number = {3},
        pages = {3309-3331},
          doi = {10.1093/mnras/staa3764},
archivePrefix = {arXiv},
       eprint = {2001.06018},
 primaryClass = {astro-ph.CO},
       adsurl = {https://ui.adsabs.harvard.edu/abs/2021MNRAS.501.3309Z}
}

@ARTICLE{Gerdes2010,
       author = {{Gerdes}, David W. and {Sypniewski}, Adam J. and {McKay}, Timothy A. and {Hao}, Jiangang and {Weis}, Matthew R. and {Wechsler}, Risa H. and {Busha}, Michael T.},
        title = "{ArborZ: Photometric Redshifts Using Boosted Decision Trees}",
      journal = {\apj},
         year = 2010,
        month = jun,
       volume = {715},
       number = {2},
        pages = {823-832},
          doi = {10.1088/0004-637X/715/2/823},
archivePrefix = {arXiv},
       eprint = {0908.4085},
 primaryClass = {astro-ph.CO},
       adsurl = {https://ui.adsabs.harvard.edu/abs/2010ApJ...715..823G}
}

@ARTICLE{Masters2015,
       author = {{Masters}, Daniel and {Capak}, Peter and {Stern}, Daniel and {Ilbert}, Olivier and {Salvato}, Mara and {Schmidt}, Samuel and {Longo}, Giuseppe and {Rhodes}, Jason and {Paltani}, Stephane and {Mobasher}, Bahram and {Hoekstra}, Henk and {Hildebrandt}, Hendrik and {Coupon}, Jean and {Steinhardt}, Charles and {Speagle}, Josh and {Faisst}, Andreas and {Kalinich}, Adam and {Brodwin}, Mark and {Brescia}, Massimo and {Cavuoti}, Stefano},
        title = "{Mapping the Galaxy Color-Redshift Relation: Optimal Photometric Redshift Calibration Strategies for Cosmology Surveys}",
      journal = {\apj},
         year = 2015,
        month = nov,
       volume = {813},
       number = {1},
          eid = {53},
        pages = {53},
          doi = {10.1088/0004-637X/813/1/53},
archivePrefix = {arXiv},
       eprint = {1509.03318},
 primaryClass = {astro-ph.CO},
       adsurl = {https://ui.adsabs.harvard.edu/abs/2015ApJ...813...53M}
}

@ARTICLE{DC3R2_2024,
       author = {{McCullough}, J. and {Gruen}, D. and {Amon}, A. and {Roodman}, A. and {Masters}, D. and {Raichoor}, A. and {Schlegel}, D. and {Canning}, R. and {Castander}, F.~J. and {DeRose}, J. and {Miquel}, R. and {Myles}, J. and {Newman}, J.~A. and {Slosar}, A. and {Speagle}, J. and {Wilson}, M.~J. and {Aguilar}, J. and {Ahlen}, S. and {Bailey}, S. and {Brooks}, D. and {Claybaugh}, T. and {Cole}, S. and {Dawson}, K. and {de la Macorra}, A. and {Doel}, P. and {Forero-Romero}, J.~E. and {Gontcho A Gontcho}, S. and {Guy}, J. and {Kehoe}, R. and {Kremin}, A. and {Landriau}, M. and {Le Guillou}, L. and {Levi}, M. and {Manera}, M. and {Martini}, P. and {Meisner}, A. and {Moustakas}, J. and {Nie}, J. and {Percival}, W.~J. and {Poppett}, C. and {Prada}, F. and {Rezaie}, M. and {Rossi}, G. and {Sanchez}, E. and {Seo}, H. and {Tarl{\'e}}, G. and {Weaver}, B.~A. and {Zhou}, Z. and {Zou}, H. and {DESI Collaboration}},
        title = "{DESI complete calibration of the colour-redshift relation (DC3R2): results from early DESI data}",
      journal = {\mnras},
         year = 2024,
        month = jun,
       volume = {531},
       number = {2},
        pages = {2582-2602},
          doi = {10.1093/mnras/stae1316},
archivePrefix = {arXiv},
       eprint = {2309.13109},
 primaryClass = {astro-ph.CO},
       adsurl = {https://ui.adsabs.harvard.edu/abs/2024MNRAS.531.2582M}
}

@ARTICLE{4C3R2_2023,
       author = {{Gruen}, D. and {McCullough}, J. and {Amon}, A. and {Bernstein}, G. and {van den Busch}, J.~L. and {Canning}, R. and {Castander}, F. and {DeRose}, J. and {Hartley}, W. and {Hildebrandt}, H. and {Kuijken}, K. and {Liske}, J. and {Masters}, D. and {Miquel}, R. and {Yuste}, A.~P. and {Roodman}, A. and {Seitz}, S. and {Saglia}, R. and {Stern}, D. and {Tortorelli}, L. and {Wright}, A.~H.},
        title = "{4MOST Complete Calibration of the Colour-Redshift Relation (4C3R2)}",
      journal = {The Messenger},
         year = 2023,
        month = mar,
       volume = {190},
        pages = {28-30},
          doi = {10.18727/0722-6691/5307},
       adsurl = {https://ui.adsabs.harvard.edu/abs/2023Msngr.190...28G}
}

@ARTICLE{DESI_DR1_2025arXiv250314745D,
      author = {{DESI Collaboration} and {Abdul Karim}, M. and {Adame}, A.~G. and {Aguado}, D. and {Aguilar}, J. and {Ahlen}, S. and {Alam}, S. and {Aldering}, G. and {Alexander}, D.~M. and {Alfarsy}, R. and {Allen}, L. and {Allende Prieto}, C. and {Alves}, O. and {Anand}, A. and {Andrade}, U. and {Armengaud}, E. and {Avila}, S. and {Aviles}, A. and {Awan}, H. and {Bailey}, S. and {Baleato Lizancos}, A. and {Ballester}, O. and {Bault}, A. and {Bautista}, J. and {Bean}, R. and {Behera}, J. and {BenZvi}, S. and {Beraldo e Silva}, L. and {Bermejo-Climent}, J.~R. and {Beutler}, F. and {Bianchi}, D. and {Blake}, C. and {Blum}, R. and {Bolton}, A.~S. and {Bonici}, M. and {Brieden}, S. and {Brodzeller}, A. and {Brooks}, D. and {Buckley-Geer}, E. and {Burtin}, E. and {Bystr{\"o}m}, A. and {Canning}, R. and {Carnero Rosell}, A. and {Carr}, A. and {Carrilho}, P. and {Casas}, L. and {Castander}, F.~J. and {Cereskaite}, R. and {Cervantes-Cota}, J.~L. and {Chaussidon}, E. and {Chaves-Montero}, J. and {Chen}, S. and {Chen}, X. and {Circosta}, C. and {Claybaugh}, T. and {Cole}, S. and {Cooper}, A.~P. and {Cousinou}, M.-C. and {Cuceu}, A. and {Davis}, T.~M. and {Dawson}, K.~S. and {de Belsunce}, R. and {de la Cruz}, R. and {de la Macorra}, A. and {de Mattia}, A. and {Deiosso}, N. and {Della Costa}, J. and {Demina}, R. and {Demirbozan}, U. and {DeRose}, J. and {Dey}, A. and {Dey}, B. and {Ding}, J. and {Ding}, Z. and {Doel}, P. and {Douglass}, K. and {Dowicz}, M. and {Ebina}, H. and {Edelstein}, J. and {Eisenstein}, D.~J. and {Elbers}, W. and {Emas}, N. and {Escoffier}, S. and {Fagrelius}, P. and {Fan}, X. and {Fanning}, K. and {Favole}, G. and {Fawcett}, V.~A. and {Fern{\'a}ndez-Garc{\'\i}a}, E. and {Ferraro}, S. and {Findlay}, N. and {Font-Ribera}, A. and {Forero-Romero}, J.~E. and {Forero-S{\'a}nchez}, D. and {Frenk}, C.~S. and {G{\"a}nsicke}, B.~T. and {Galbany}, L. and {Garc{\'\i}a-Bellido}, J. and {Garcia-Quintero}, C. and {Garrison}, L.~H. and {Gazta{\~n}aga}, E. and {Gil-Mar{\'\i}n}, H. and {Gloudemans}, A. and {Gnedin}, O.~Y. and {Gontcho A Gontcho}, S. and {Gonzalez}, D. and {Gonzalez-Morales}, A.~X. and {Gonzalez-Perez}, V. and {Gordon}, C. and {Graur}, O. and {Green}, D. and {Gruen}, D. and {Gsponer}, R. and {Guandalin}, C. and {Gutierrez}, G. and {Guy}, J. and {Hahn}, C. and {Han}, J.~J. and {Han}, J. and {He}, S. and {Herrera-Alcantar}, H.~K. and {Heydenreich}, S. and {Honscheid}, K. and {Hou}, J. and {Howlett}, C. and {Huterer}, D. and {Ir{\v{s}}i{\v{c}}}, V. and {Ishak}, M. and {Jacques}, A. and {Jiang}, L. and {Jimenez}, J. and {Jing}, Y.~P. and {Joachimi}, B. and {Joudaki}, S. and {Joyce}, R. and {Jullo}, E. and {Juneau}, S. and {Kara{\c{c}}ayl{\i}}, N.~G. and {Karim}, T. and {Kehoe}, R. and {Kent}, S. and {Khederlarian}, A. and {Kirkby}, D. and {Kisner}, T. and {Kitaura}, F.-S. and {Kizhuprakkat}, N. and {Kong}, H. and {Koposov}, S.~E. and {Kremin}, A. and {Krolewski}, A. and {Lahav}, O. and {Lai}, Y. and {Lamman}, C. and {Lan}, T.-W. and {Landriau}, M. and {Lang}, D. and {Lange}, J.~U. and {Lasker}, J. and {Le Goff}, J.~M. and {Le Guillou}, L. and {Leauthaud}, A. and {Levi}, M.~E. and {Li}, S. and {Li}, T.~S. and {Liu}, W. and {Lodha}, K. and {Lokken}, M. and {Luo}, Y. and {Magneville}, C. and {Manera}, M. and {Manser}, C.~J. and {Margala}, D. and {Martini}, P. and {Maus}, M. and {McCullough}, J. and {McDonald}, P. and {Medina}, G.~E. and {Medina-Varela}, L. and {Meisner}, A. and {Mena-Fern{\'a}ndez}, J. and {Menegas}, A. and {Meneses-Rizo}, J. and {Mezcua}, M. and {Miquel}, R. and {Montero-Camacho}, P. and {Moon}, J. and {Moustakas}, J. and {Mu{\~n}oz-Guti{\'e}rrez}, A. and {Mu noz-Santos}, D. and {Myers}, A.~D. and {Myles}, J. and {Nadathur}, S. and {Najita}, J. and {Napolitano}, L. and {Newman}, J.~A. and {Nikakhtar}, F. and {Nikutta}, R. and {Niz}, G. and {Noriega}, H.~E. and {Nugent}, P.},
        title = "{Data Release 1 of the Dark Energy Spectroscopic Instrument}",
      journal = {\aj},
         year = 2026,
        month = may,
       volume = {171},
       number = {5},
          eid = {285},
        pages = {285},
          doi = {10.3847/1538-3881/ae4c43},
archivePrefix = {arXiv},
       eprint = {2503.14745},
 primaryClass = {astro-ph.CO},
       adsurl = {https://ui.adsabs.harvard.edu/abs/2026AJ....171..285D}
}

@ARTICLE{Liske2015,
       author = {{Liske}, J. and {Baldry}, I.~K. and {Driver}, S.~P. and {Tuffs}, R.~J. and {Alpaslan}, M. and {Andrae}, E. and {Brough}, S. and {Cluver}, M.~E. and {Grootes}, M.~W. and {Gunawardhana}, M.~L.~P. and {Kelvin}, L.~S. and {Loveday}, J. and {Robotham}, A.~S.~G. and {Taylor}, E.~N. and {Bamford}, S.~P. and {Bland-Hawthorn}, J. and {Brown}, M.~J.~I. and {Drinkwater}, M.~J. and {Hopkins}, A.~M. and {Meyer}, M.~J. and {Norberg}, P. and {Peacock}, J.~A. and {Agius}, N.~K. and {Andrews}, S.~K. and {Bauer}, A.~E. and {Ching}, J.~H.~Y. and {Colless}, M. and {Conselice}, C.~J. and {Croom}, S.~M. and {Davies}, L.~J.~M. and {De Propris}, R. and {Dunne}, L. and {Eardley}, E.~M. and {Ellis}, S. and {Foster}, C. and {Frenk}, C.~S. and {H{\"a}u{\ss}ler}, B. and {Holwerda}, B.~W. and {Howlett}, C. and {Ibarra}, H. and {Jarvis}, M.~J. and {Jones}, D.~H. and {Kafle}, P.~R. and {Lacey}, C.~G. and {Lange}, R. and {Lara-L{\'o}pez}, M.~A. and {L{\'o}pez-S{\'a}nchez}, {\'A}. R. and {Maddox}, S. and {Madore}, B.~F. and {McNaught-Roberts}, T. and {Moffett}, A.~J. and {Nichol}, R.~C. and {Owers}, M.~S. and {Palamara}, D. and {Penny}, S.~J. and {Phillipps}, S. and {Pimbblet}, K.~A. and {Popescu}, C.~C. and {Prescott}, M. and {Proctor}, R. and {Sadler}, E.~M. and {Sansom}, A.~E. and {Seibert}, M. and {Sharp}, R. and {Sutherland}, W. and {V{\'a}zquez-Mata}, J.~A. and {van Kampen}, E. and {Wilkins}, S.~M. and {Williams}, R. and {Wright}, A.~H.},
        title = "{Galaxy And Mass Assembly (GAMA): end of survey report and data release 2}",
      journal = {\mnras},
         year = 2015,
        month = sep,
       volume = {452},
       number = {2},
        pages = {2087-2126},
          doi = {10.1093/mnras/stv1436},
archivePrefix = {arXiv},
       eprint = {1506.08222},
 primaryClass = {astro-ph.GA},
       adsurl = {https://ui.adsabs.harvard.edu/abs/2015MNRAS.452.2087L}
}

@ARTICLE{Masters2017,
       author = {{Masters}, Daniel C. and {Stern}, Daniel K. and {Cohen}, Judith G. and {Capak}, Peter L. and {Rhodes}, Jason D. and {Castander}, Francisco J. and {Paltani}, St{\'e}phane},
        title = "{The Complete Calibration of the Color-Redshift Relation (C3R2) Survey: Survey Overview and Data Release 1}",
      journal = {\apj},
         year = 2017,
        month = jun,
       volume = {841},
       number = {2},
          eid = {111},
        pages = {111},
          doi = {10.3847/1538-4357/aa6f08},
archivePrefix = {arXiv},
       eprint = {1704.06665},
 primaryClass = {astro-ph.CO},
       adsurl = {https://ui.adsabs.harvard.edu/abs/2017ApJ...841..111M}
}

@ARTICLE{LeFevre2005,
       author = {{Le F{\`e}vre}, O. and {Vettolani}, G. and {Garilli}, B. and {Tresse}, L. and {Bottini}, D. and {Le Brun}, V. and {Maccagni}, D. and {Picat}, J.~P. and {Scaramella}, R. and {Scodeggio}, M. and {Zanichelli}, A. and {Adami}, C. and {Arnaboldi}, M. and {Arnouts}, S. and {Bardelli}, S. and {Bolzonella}, M. and {Cappi}, A. and {Charlot}, S. and {Ciliegi}, P. and {Contini}, T. and {Foucaud}, S. and {Franzetti}, P. and {Gavignaud}, I. and {Guzzo}, L. and {Ilbert}, O. and {Iovino}, A. and {McCracken}, H.~J. and {Marano}, B. and {Marinoni}, C. and {Mathez}, G. and {Mazure}, A. and {Meneux}, B. and {Merighi}, R. and {Paltani}, S. and {Pell{\`o}}, R. and {Pollo}, A. and {Pozzetti}, L. and {Radovich}, M. and {Zamorani}, G. and {Zucca}, E. and {Bondi}, M. and {Bongiorno}, A. and {Busarello}, G. and {Lamareille}, F. and {Mellier}, Y. and {Merluzzi}, P. and {Ripepi}, V. and {Rizzo}, D.},
        title = "{The VIMOS VLT deep survey. First epoch VVDS-deep survey: 11 564 spectra with 17.5 {\ensuremath{\leq}} IAB {\ensuremath{\leq}} 24, and the redshift distribution over 0 {\ensuremath{\leq}} z {\ensuremath{\leq}} 5}",
      journal = {\aap},
         year = 2005,
        month = sep,
       volume = {439},
       number = {3},
        pages = {845-862},
          doi = {10.1051/0004-6361:20041960},
archivePrefix = {arXiv},
       eprint = {astro-ph/0409133},
 primaryClass = {astro-ph},
       adsurl = {https://ui.adsabs.harvard.edu/abs/2005A&A...439..845L}
}

@ARTICLE{Balestra2010,
       author = {{Balestra}, I. and {Mainieri}, V. and {Popesso}, P. and {Dickinson}, M. and {Nonino}, M. and {Rosati}, P. and {Teimoorinia}, H. and {Vanzella}, E. and {Cristiani}, S. and {Cesarsky}, C. and {Fosbury}, R.~A.~E. and {Kuntschner}, H. and {Rettura}, A.},
        title = "{The Great Observatories Origins Deep Survey. VLT/VIMOS spectroscopy in the GOODS-south field: Part II}",
      journal = {\aap},
         year = 2010,
        month = mar,
       volume = {512},
          eid = {A12},
        pages = {A12},
          doi = {10.1051/0004-6361/200913626},
archivePrefix = {arXiv},
       eprint = {1001.1115},
 primaryClass = {astro-ph.CO},
       adsurl = {https://ui.adsabs.harvard.edu/abs/2010A&A...512A..12B}
}

@article{Lundberg2020,
  title = {From local explanations to global understanding with explainable AI for trees},
  volume = {2},
  ISSN = {2522-5839},
  url = {http://dx.doi.org/10.1038/s42256-019-0138-9},
  DOI = {10.1038/s42256-019-0138-9},
  number = {1},
  journal = {Nature Machine Intelligence},
  publisher = {Springer Science and Business Media LLC},
  author = {Lundberg,  Scott M. and Erion,  Gabriel and Chen,  Hugh and DeGrave,  Alex and Prutkin,  Jordan M. and Nair,  Bala and Katz,  Ronit and Himmelfarb,  Jonathan and Bansal,  Nisha and Lee,  Su-In},
  year = {2020},
  month = jan,
  pages = {56–67}
}

@ARTICLE{Wu2022,
       author = {{Wu}, John F. and {Peek}, J.~E.~G. and {Tollerud}, Erik J. and {Mao}, Yao-Yuan and {Nadler}, Ethan O. and {Geha}, Marla and {Wechsler}, Risa H. and {Kallivayalil}, Nitya and {Weiner}, Benjamin J.},
        title = "{Extending the SAGA Survey (xSAGA). I. Satellite Radial Profiles as a Function of Host-galaxy Properties}",
      journal = {\apj},
         year = 2022,
        month = mar,
       volume = {927},
       number = {1},
          eid = {121},
        pages = {121},
          doi = {10.3847/1538-4357/ac4eea},
archivePrefix = {arXiv},
       eprint = {2112.01542},
 primaryClass = {astro-ph.GA},
       adsurl = {https://ui.adsabs.harvard.edu/abs/2022ApJ...927..121W}
}

@ARTICLE{Chauhan2019,
       author = {{Chauhan}, Garima and {Lagos}, Claudia del P. and {Obreschkow}, Danail and {Power}, Chris and {Oman}, Kyle and {Elahi}, Pascal J.},
        title = "{The H I velocity function: a test of cosmology or baryon physics?}",
      journal = {\mnras},
         year = 2019,
        month = oct,
       volume = {488},
       number = {4},
        pages = {5898-5915},
          doi = {10.1093/mnras/stz2069},
archivePrefix = {arXiv},
       eprint = {1906.06130},
 primaryClass = {astro-ph.GA},
       adsurl = {https://ui.adsabs.harvard.edu/abs/2019MNRAS.488.5898C}
}

@ARTICLE{Lagos2019,
       author = {{Lagos}, Claudia del P. and {Robotham}, Aaron S.~G. and {Trayford}, James W. and {Tobar}, Rodrigo and {Bravo}, Mat{\'\i}as and {Bellstedt}, Sabine and {Davies}, Luke J.~M. and {Driver}, Simon P. and {Elahi}, Pascal J. and {Obreschkow}, Danail and {Power}, Chris},
        title = "{From the far-ultraviolet to the far-infrared - galaxy emission at 0 {\ensuremath{\leq}} z {\ensuremath{\leq}} 10 in the SHARK semi-analytic model}",
      journal = {\mnras},
         year = 2019,
        month = nov,
       volume = {489},
       number = {3},
        pages = {4196-4216},
          doi = {10.1093/mnras/stz2427},
archivePrefix = {arXiv},
       eprint = {1908.03423},
 primaryClass = {astro-ph.GA},
       adsurl = {https://ui.adsabs.harvard.edu/abs/2019MNRAS.489.4196L}
}

@ARTICLE{2019ApJ...873..111I, author = {{Ivezi{\'c}}, {\v Z}. and {Kahn}, S.~M. and {Tyson}, J.~A. and {Abel}, B. and {Acosta}, E. and {Allsman}, R. and {Alonso}, D. and {AlSayyad}, Y. and {Anderson}, S.~F. and {Andrew}, J. and et al.}, title = "{LSST: From Science Drivers to Reference Design and Anticipated Data Products}", journal = {\apj}, archivePrefix = "arXiv", eprint = {0805.2366}, year = 2019, month = mar, volume = 873, eid = {111}, pages = {111}, doi = {10.3847/1538-4357/ab042c}, adsurl = {http://adsabs.harvard.edu/abs/2019ApJ...873..111I} }

\begin{appendix}

\section{Errors on classification metrics} 
\label{appendix_err}
Here we describe our method for estimating the errors on the classification metrics, based on 5-fold cross-validation. Here we will discuss purity as an example.

As defined in Equation \ref{eq:pur}, purity is computed using the number of true positives (0,0) and false positives (1,0). Given that we employ 5-fold cross-validation, as described in Section \ref{sec:training}, applying these models to the test set produces an array of five values, denoted as ${TP_{i}}$ and ${FP_{i}}$, where $i$ ranges from 1 to 5.

The classification performance metrics introduced in Section~\ref{Sec:metrics} (Eqs. \ref{eq:acc}-\ref{eq:f1}) are derived from the means of the true positives (0,0), false positives (1,0), false negatives (0,1), and true negatives (1,1) obtained from the test performance of the 5-fold models. We define these means as $\overline{TP}$, $\overline{FP}$, $\overline{FN}$, and $\overline{TN}$. 

Consequently, the purity equation (Eq. \ref{eq:pur}) can be rewritten as:
\begin{equation}
    \text{Purity}, P = \frac{\overline{TP}}{\overline{TP} + \overline{FP}}
    \label{eq:pur_mean}
\end{equation}

The Poissonian errors associated with (0,0), (1,0), (0,1), and (1,1) are given by $\sqrt{\overline{TP}}$, $\sqrt{\overline{FP}}$, $\sqrt{\overline{FN}}$, and $\sqrt{\overline{TN}}$, respectively.

The corresponding uncertainty in purity using standard error propagation is then computed as:
\begin{equation}
    \sigma_P = \left [ \left (\dfrac{dP}{d(TP)} \sigma_{TP} \right)  ^2 + \left (\dfrac{dP}{d(FP)} \sigma_{FP} \right )^2 + 2 \left (\dfrac{dP}{d(TP)} \right ) \left (\dfrac{dP}{d(FP)} \right)\rho_{TP,FP}, \sigma_{TP},\sigma_{FP} \right]^{1/2}
    \label{eq:pur_err}
\end{equation}

Here, $\rho_{TP, FP}$ is the Pearson correlation coefficient between the true positives (TP; (0,0)) and false positives (FP; (1,0)) across the five realisations. It accounts for the correlation between TP and FP induced by the stochasticity of the ML training, and is computed as:
\begin{equation}
    \rho_{TP,FP} = \frac{\sum_{i=1}^{5}({FP_{i}} - \overline{FP})(TP_{i} - \overline{TP})}{\sqrt{\sum({FP_{i}} - \overline{FP})^2}\sqrt{\sum({TP_{i}} - \overline{TP})^2}}
    \label{eq:pearson_purr}
\end{equation}

The partial derivatives $\dfrac{dP}{d(TP)}$ and $\dfrac{dP}{d(FP)}$ are given by:
\begin{equation}
    \dfrac{dP}{d(TP)} = \frac{\overline{FP}}{(\overline{TP}+\overline{FP})^2}
    \label{eq:dPdTP}
\end{equation}
\begin{equation}
    \dfrac{dP}{d(FP)} = -\frac{\overline{TP}}{(\overline{TP}+\overline{FP})^2}
    \label{eq:dPdFP}
\end{equation}

Finally, the scatter in the purity metric, as presented in Table \ref{metrics}, is determined using Equations \ref{eq:pur_err}-\ref{eq:dPdFP}. Similarly, the uncertainties for Completeness, F1 score, and Accuracy are worked out following the same approach.

\section{Performance plots for DESI DR1 test set}
\label{app:DESI_performance}

Here we present the dependence of completeness and purity (where relevant) on the spectroscopic redshifts, $Z$-band magnitude and $g-r$ color, in a similar manner as in Sec.~\ref{sec:performance} but for the DESI DR1 test set.

\begin{figure*}
\centering
    \includegraphics[width=0.75\textwidth]{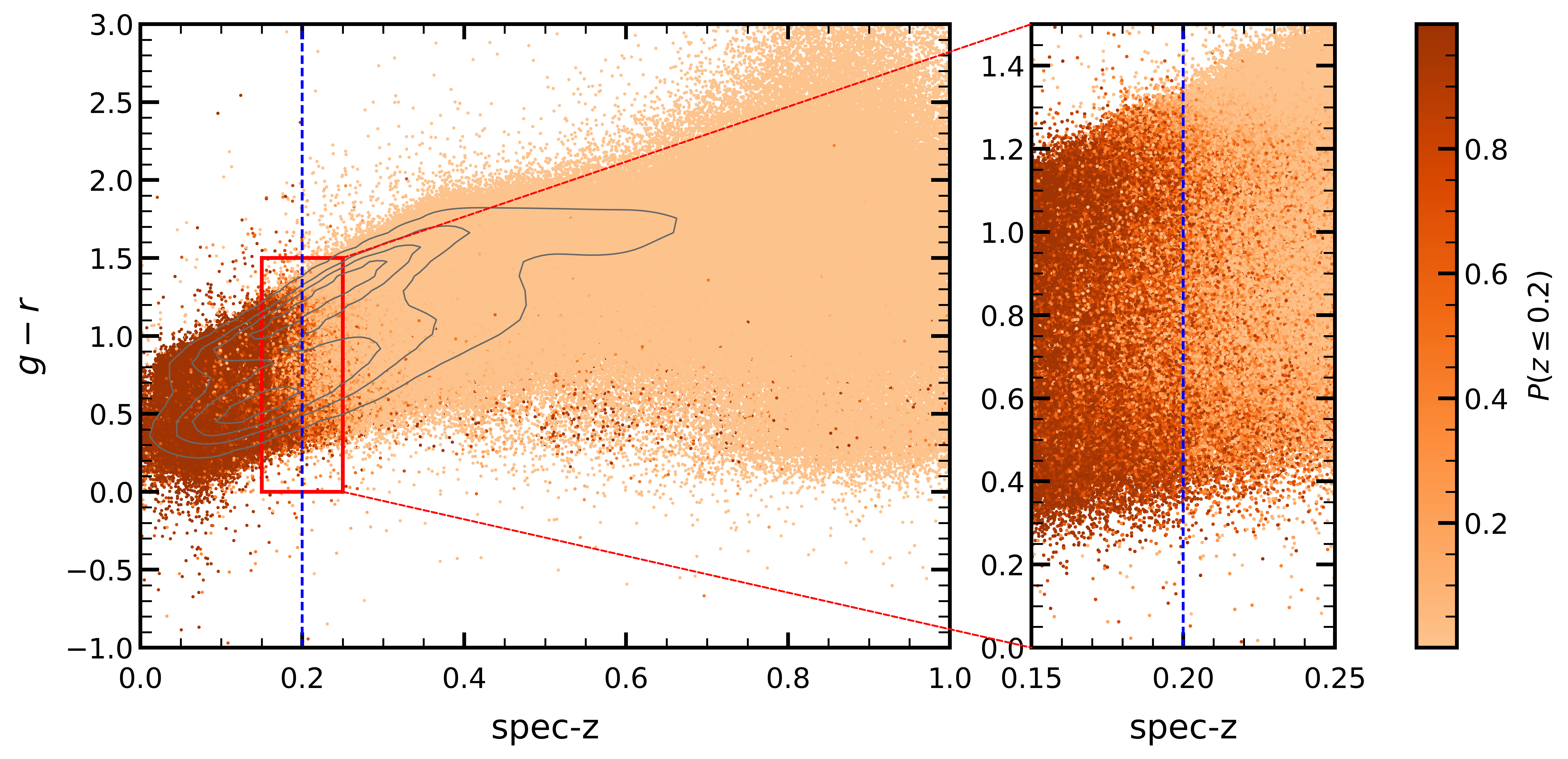}
    \caption{Distribution of the galaxies from the independent test set consisting of DESI DR1 on the true redshift -- observed $g-r$ color plane, color coded by the probability of being a WAVES-Wide target as assigned by our classifier. Dark indicates a high probability of having $z<0.2$, while light means assignment as non-target. The vertical dashed line shows the WW division threshold. Grey contours indicate the region of the largest density of galaxies in the diagram. On the right-hand side, we enlarge the area close to the threshold to highlight the region where most of the misclassification happens.}
    \label{fig:gr_z_prob_desi}
\end{figure*}

Figure \ref{fig:gr_z_prob_desi} shows the performance of the XGB classifier mapped on the $g-r$ color (the most important feature for classification) and redshift plane for the DESI test set, similarly to Fig.~\ref{fig:gr_z_prob} for the random test set. Here, 7.9\%  of the galaxies with $z<0.2$ are labeled non-targets (i.e. false negatives) and show up as dark brown dots on the right side of the $z=0.2$ line in the plot. In DESI DR1 test set, 2.4\% of the class 1 (higher redshift) galaxies are incorrectly assigned as having $z \leq  0.2$, i.e they are false positives.

\begin{figure*}
\centering
    \includegraphics[width=0.6\textwidth]{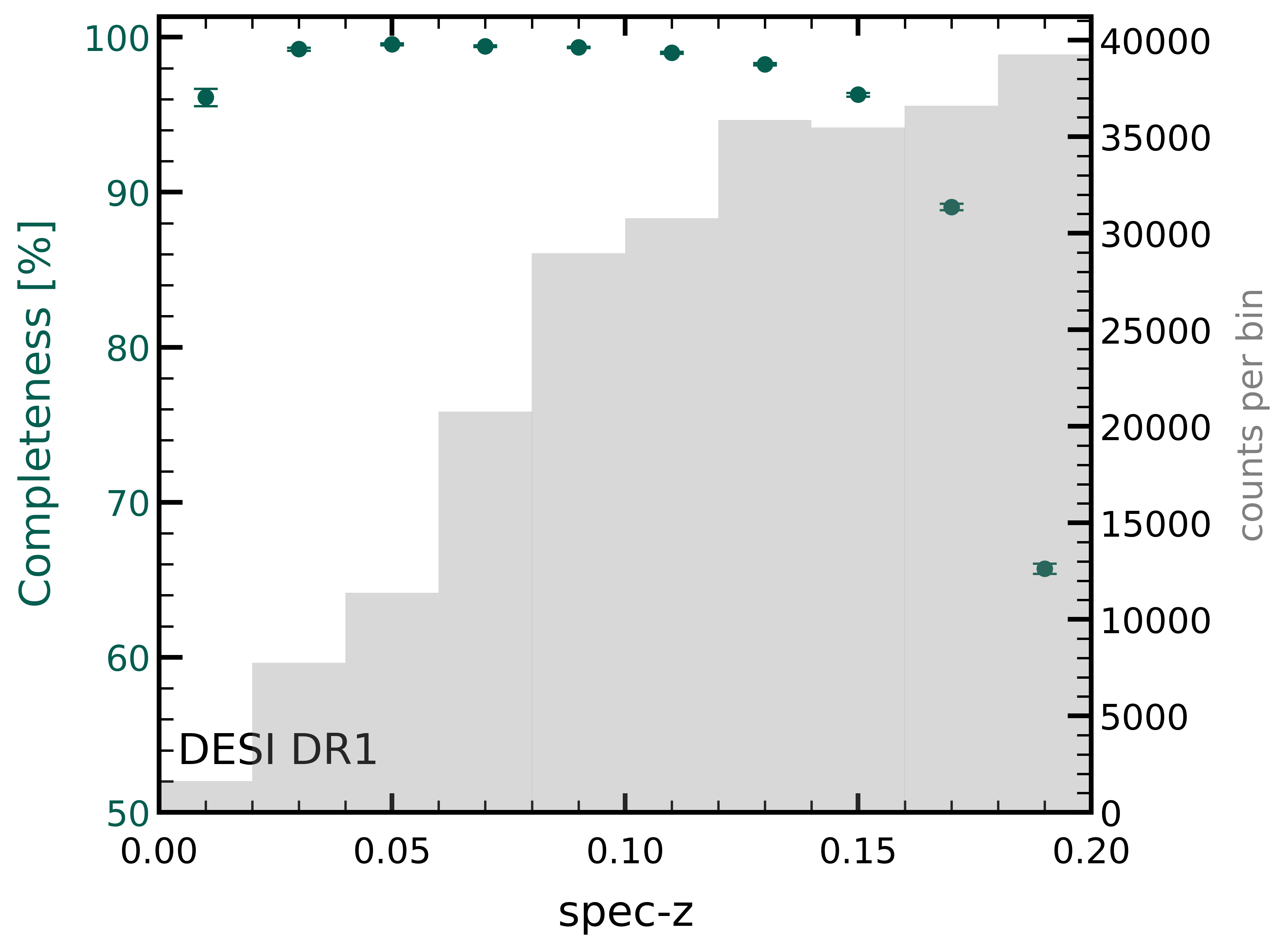}
    \caption{Completeness as a function of true redshift for the desired target (Class ‘0') galaxies, computed per redshift bin of width 0.02, for the DESI DR1 test set. The scatter on the metric is from the Poissonian error on the number of correctly identified (0,0) and missed (0,1) galaxies. The histogram in grey shows the distribution of DESI DR1 spectroscopic redshifts in the test set.}
    \label{fig:dcompdz_DESI}
\end{figure*}

\begin{figure}
    \centering
    \includegraphics[width=0.6\textwidth]{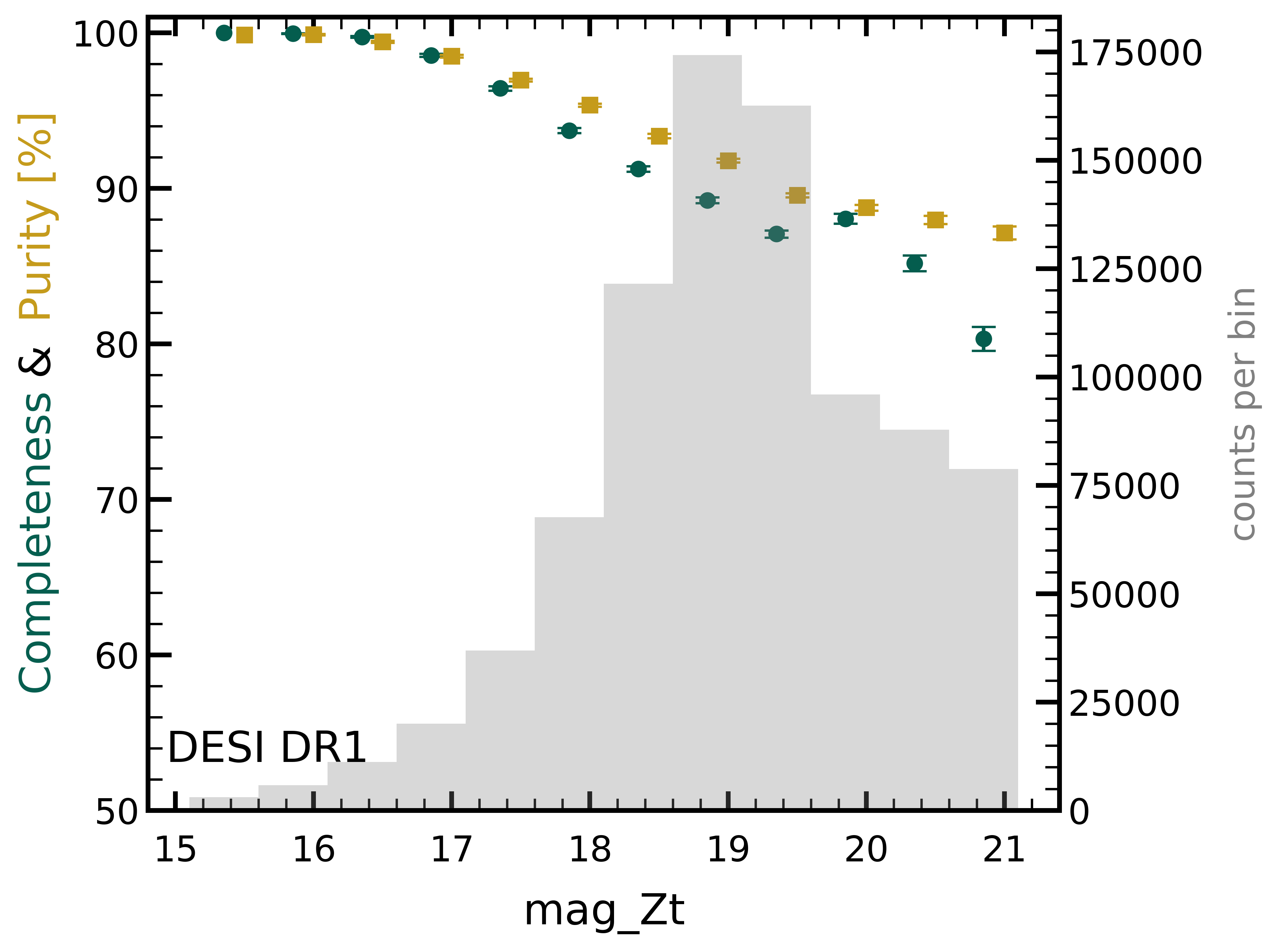}
    \caption{Completeness (dark green) and purity (olive) as a function of $Z$-band magnitude for sources in the DESI DR1 test set, computed per magnitude bin of width 0.5. For better visualization, olive dots representing purity are shifted by 0.15 along the x-axis. The histogram in grey shows the distribution of Z band magnitude in the test set.}
    \label{fig:dcompdmagz_DESI}
\end{figure}

\begin{figure}
\centering
     
    \includegraphics[width=0.6\textwidth]{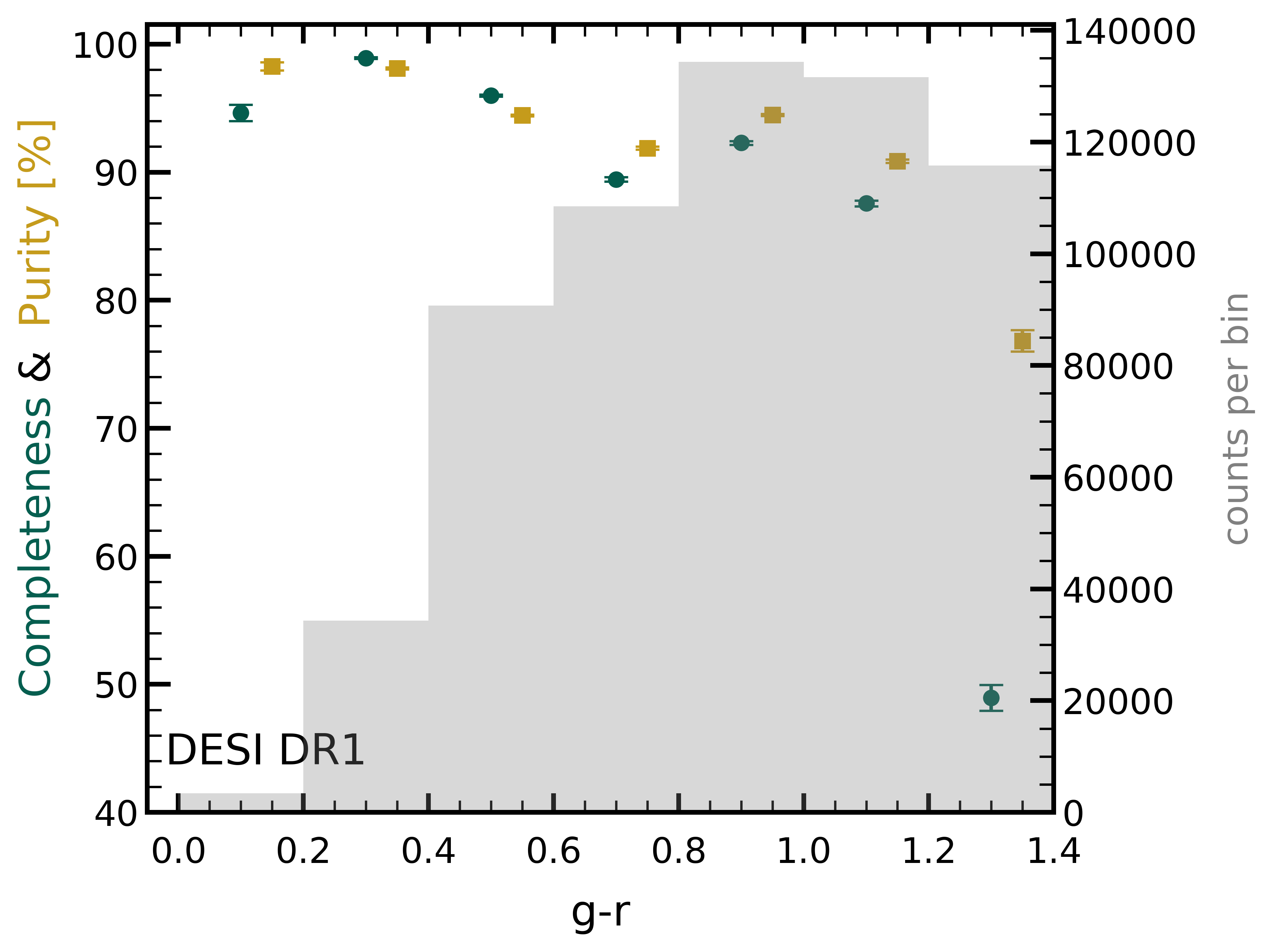}
    \caption{Completeness (dark green) and purity (olive) as a function of the $\text{g-r}$ color for sources in the DESI DR1 test set, computed per bin width of 0.2. For better visualization, olive dots representing purity are shifted by 0.05 along the x-axis. The histogram in grey shows the color distribution of sources in the test set.}
    \label{fig:dcompcolor_DESI}
\end{figure}

The trend of classification metrics per redshift, magnitude and $g - r$ bins for DESI DR1 test set galaxies are shown in Figs.~\ref{fig:dcompdz_DESI}, \ref{fig:dcompdmagz_DESI} and \ref{fig:dcompcolor_DESI}, respectively. It is evident from these figures that completeness and purity deteriorate for redshifts close to the redshift limit and for fainter magnitudes, similar to trends shown in Figs.~\ref{fig:dcompdz} and \ref{fig:dcompdmagz} for the random test set. However, the metrics are slightly better at these ends for DESI DR1. This is not reflected in the metrics present in Table \ref{metrics}, because the number of galaxies in these bins is very small in the random test set as compared to DESI DR1.

\end{appendix}

\end{document}